\documentclass[prd,amsmath,showpacs,preprintnumbers,
               floatfix,twocolumn,letterpaper,superscriptaddress,h-physrev4]{revtex4}
\usepackage{graphicx}
\usepackage{amssymb}
\newcommand{\nc}{\newcommand}
\nc{\lb}{\langle}
\nc{\rk}{\rangle}
\nc{\Blb}{\Big\langle}
\nc{\Brk}{\Big\rangle}

\nc{\mi}{\!\!\mid\!\!}
\nc{\ra}{\rightarrow}
\nc{\Ra}{\Rightarrow}
\nc {\cd}{\partial}
\nc {\sla}{\slashed}
\nc{\ro}{\mathrm}
\nc{\ca}{\mathcal}
\nc{\bo}{\mathbf}
\nc{\Tr}{\ro{Tr}\,}
\nc{\Str}{\ro{Str}}
\nc{\realtrace}{\ro{Re\; Tr}}
\nc{\maxrealtrace}{\ro{max\, Re\; Tr}}
\nc{\ud}{\ro{d}}
\nc{\nn}{\nonumber}

\nc{\pb}{\bar{\psi}}
\nc{\p}{\psi}
\nc{\Pb}{\bar{\Psi}}
\nc{\vp}{\vec{\pi}}
\nc{\vap}{\varphi}
\nc{\vt}{\vec{\tau}}
\nc{\si}{\sigma}
\nc{\Si}{\Sigma}
\nc{\tSi}{\tilde{\Si}}
\nc{\tsi}{\tilde{\si}}
\nc{\g}{\gamma}
\nc{\G}{\Gamma}
\nc{\la}{\lambda}
\nc{\La}{\Lambda}
\nc{\ep}{\epsilon}
\nc{\de}{\delta}
\nc{\De}{\Delta}
\nc{\cL}{\ca{L}}
\nc{\cLe}{\ca{L}_{\ro{eff}}}
\nc {\ti}{\tilde}
\nc{\f}{\frac}
\nc{\da}{\dagger}
\nc{\SU}{\ro{SU}}
\nc{\om}{\omega}
\nc{\Om}{\Omega}

\nc{\darrow}{\stackrel{\leftrightarrow}{\cd}}
\nc{\darrows}{\stackrel{\leftrightarrow}{\sla{\cd}}}
\nc{\Darrows}{\stackrel{\leftrightarrow}{\sla{D}}}

\nc {\eqb}{\begin{equation}}
\nc {\eqe}{\end{equation}}
\nc {\eqab}{\begin{eqnarray}}
\nc {\eqae}{\end{eqnarray}}

\nc{\note}[2]{{\bf #1: #2}}

\begin{document}

\title{Power Counting Regime of Chiral Effective Field Theory and Beyond}

\author{J. M. M. Hall} 

\affiliation{Special Research Centre for the Subatomic Structure of
  Matter (CSSM), School of Chemistry \& Physics, University of
  Adelaide, SA 5005, Australia}

\author{D. B. Leinweber} 

\affiliation{Special Research Centre for the Subatomic Structure of
  Matter (CSSM), School of Chemistry \& Physics, University of
  Adelaide, SA 5005, Australia}

\author{R. D. Young} 

\affiliation{Special Research Centre for the Subatomic Structure of
  Matter (CSSM), School of Chemistry \& Physics, University of
  Adelaide, SA 5005, Australia}
\affiliation{ Argonne National Laboratory, 9700 S. Cass Avenue,
  Argonne, IL 60439, USA}

\preprint{ADP-10-03/T699}

\begin{abstract}
Chiral effective field theory ($\chi$EFT) complements numerical
simulations of quantum chromodynamics (QCD) on a space-time lattice. 
It provides a
model-independent formalism for connecting lattice simulation results
at finite volume and a variety of quark masses to the physical
world.
The asymptotic nature of the chiral expansion places the focus on the
first few terms of the expansion.  Thus, knowledge of the
power-counting regime (PCR) of $\chi$EFT, where higher-order terms of
the expansion may be regarded as negligible, is as important as
knowledge of the expansion itself.
Through the consideration of a variety of renormalization schemes and
associated parameters, techniques to identify the PCR
where results are independent of the renormalization scheme are established. 
 The
nucleon mass is considered as a benchmark for illustrating this
general approach.
Because the PCR is small, the numerical simulation
results are also examined 
to search for the possible presence of an intrinsic scale
which may be used in a nonperturbative manner to describe lattice
simulation results outside of the PCR. Positive results
that improve on the current optimistic application of 
 chiral perturbation theory ($\chi$PT) beyond the PCR are reported.

\end{abstract}


\maketitle


\section{Introduction}

The low energy chiral effective field theory ($\chi$EFT) of quantum 
chromodynamics (QCD) 
provides a model-independent approach for understanding the
consequences of dynamical chiral-symmetry breaking in the chiral
properties of hadrons.  Non-analytic contributions in the quark mass
are generated by the pseudo-Goldstone meson dressings of hadrons
through meson-loop integrals.
Chiral perturbation theory ($\chi$PT) provides a formal approach to
counting the powers of low energy momenta and quark masses such that
an ordered expansion in powers of the quark mass $m_q \propto m_\pi^2$
is constructed.  $\chi$PT indicates that, in general, the most singular 
nonanalytic contributions to hadron properties lie in the one-loop 
`meson cloud' of the hadron.  For example, the leading nonanalytic
behavior of a baryon mass is proportional to $m_q^{3/2}$ or $m_\pi^3$.
More generally, baryon masses can be written as an ordered expansion
in quark mass or $m_\pi^2$. 

 To establish a model-independent
framework in $\chi$PT, the expansion must display the properties of a
convergent series for the terms considered.  There is a power-counting
regime (PCR) where the quark mass is small, and higher-order terms in
the expansion are negligible beyond the order calculated.  Within the
PCR, the truncation of the chiral expansion is reliable to a
prescribed precision.

The asymptotic nature of the chiral expansion places the focus on the
first few terms of the expansion.  A survey of the literature for the
baryon sector of $\chi$EFT illustrates the rarity of calculations
beyond one-loop \cite{McGovern:1998tm,McGovern:2006fm,Schindler:2006ha},
 and there are no two-loop calculations which
incorporate the effects of placing a baryon in a finite volume.  With
only a few terms of the expansion known for certain, knowledge of the PCR of
$\chi$EFT  is as important as knowledge of the expansion itself. It is within
the PCR that higher-order terms of the expansion may be regarded
as negligible.

Numerical simulations of QCD on a space-time lattice are complemented
by $\chi$EFT through the provision of a model-independent formalism
for connecting lattice simulation results to the physical world.
Simulations at finite volume and a variety of quark masses are related
to the infinite volume and physical quark masses through
this formalism.  However, the formalism is accurate only if one
works within the PCR of the truncated expansion.  Present practice in
the field is best described as optimistic.  Truncated expansions are
regularly applied to a wide range of quark (or pion) masses with
little regard to a rigorous determination of the PCR.

When considering nucleons, there is some evidence that the PCR may be
 quite small; constrained by $m_\pi \lesssim 200$ MeV at $1\%$
accuracy at the chiral order $\ca{O}(m_\pi^4\,\log m_\pi)$
\cite{Beane:2004ks} \cite{Leinweber:2005xz}.
  This estimate of the PCR of $\chi$PT was
identified using specific finite-range regularization (FRR) techniques
to analyze lattice QCD data.   Using FRR, the regime is manifest when the 
quark-mass dependence of the nucleon mass is independent of the
renormalization scheme parameter.

A chief focus of this paper is to establish a rigorous approach to
 determining the PCR of a truncated chiral expansion quantitatively.
Through the consideration of a variety of renormalization schemes and
associated parameters, new techniques to identify the PCR are established. 
The PCR is the regime where results are scheme independent.
The nucleon mass is considered as a benchmark for illustrating this
general approach.
%
Here, the chiral expansion is examined, focusing on
 the individual low energy coefficients of the chiral
expansion.  This approach provides a determination of the PCR for a
truncated expansion in $\chi$EFT.
As discussed in detail in the following section, the PCR is indeed small for
the nucleon mass.  Other observables are expected to show a similar if
not smaller PCR.  Thus most of today's lattice simulation results lie
outside the PCR, and the truncated chiral expansions have been used to
extrapolate from outside the PCR.  The
low energy coefficients determined by applying the truncated expansion
outside the PCR will take on unphysical values, as they accommodate
important but otherwise missing contributions from non-negligible
higher-order terms.

While continued advances in numerical simulations of lattice QCD will
be vital to some extent in resolving this problem, the physical value
of the strange-quark mass presents a challenge that will not diminish
with supercomputing advances.  If one were to include the effects of
kaons, vital to understanding strangeness in the nucleon for example,
then one must either calculate to significantly higher order in the
expansion of $\chi$PT, or develop new non-perturbative approaches
which utilize the non-perturbative information expressed in the
lattice simulation results.  Since the former is likely to be
compromised by the asymptotic nature of the expansion, 
attention is given to the latter approach.

Thus the second focus of this paper is to examine the numerical
simulation results, to identify the possible presence of an intrinsic
scale. This may then be used to address lattice simulation results outside
of the PCR in a non-perturbative manner.  Of course, the
non-perturbative formalism must incorporate the exact perturbative results
of $\chi$PT in the PCR.  Positive results are reported that improve on
the current optimistic application of $\chi$PT 
outside of the PCR.

The outline of the presentation is as follows. 
Sec.~\ref{sect:eft} reviews chiral effective
field theory and the process of regularization and
renormalization.
The adoption of finite-range regularization (FRR)
 provides a wide range of schemes
and scales, which overlap with the more popular massless
renormalization schemes as the finite-range regulator parameter is
taken to infinity. Sec.~\ref{sect:intr} investigates FRR in the context
of a particular model. By generating a set of pseudodata and analysing it 
with a
variety of renormalization schemes, a robust method for determining
the PCR is obtained, along with an optimal 
  renormalization scale to use beyond the PCR.
Finally, Sec.~\ref{sect:scales} includes the analysis of three sets of 
lattice results for the nucleon mass, 
utilizing the tools developed in the previous section.
 Conclusions are summarized in Sec.~\ref{sect:conc}.

\section{Effective Field Theory}
\label{sect:eft}

This section begins by briefly reviewing the process of regularization and
renormalization in finite-range regularized (FRR) chiral effective
field theory, providing a range of renormalization schemes
and scales.  A central focus is to search for the dependence of
physical results on the scheme and associated scales, as these will be
an indication that one is applying the chiral expansion outside the
PCR.

The focus is to establish techniques that provide a quantitative test
of whether a given range of $m_\pi$ lies within the PCR.  This is
achieved through an examination of the flow of the low energy
coefficients as a function of the renormalization scheme
parameter(s).  A negligible dependence would confirm that the pion-mass
range is within the PCR.  On the other hand, the properties of the
flow will be used to identify a preferred regularization scheme in a
non-perturbative sense that best describes the results beyond the PCR.

\subsection{Renormalization in FRR $\chi$EFT}
\label{sect:renorm}
\begin{figure}[tp]
\centering
\includegraphics[width=0.45\hsize]{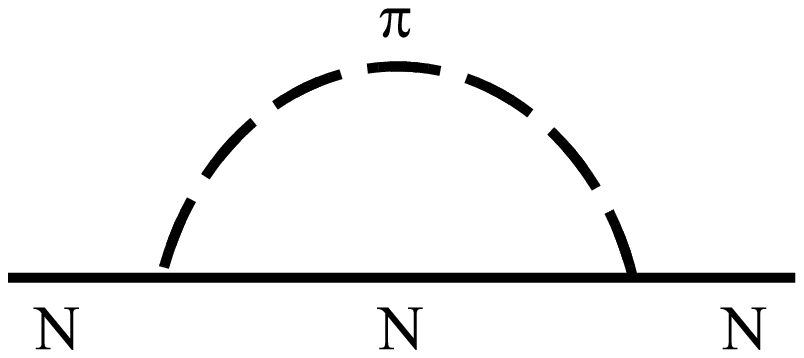}
\vspace{-6pt}
\caption{The pion loop contribution to the self-energy
    of the nucleon, providing the leading nonanalytic contribution to
    the nucleon mass.  All charge conserving transitions are implicit.}
\label{fig:nucSEpiN}
\centering
\includegraphics[width=0.45\hsize]{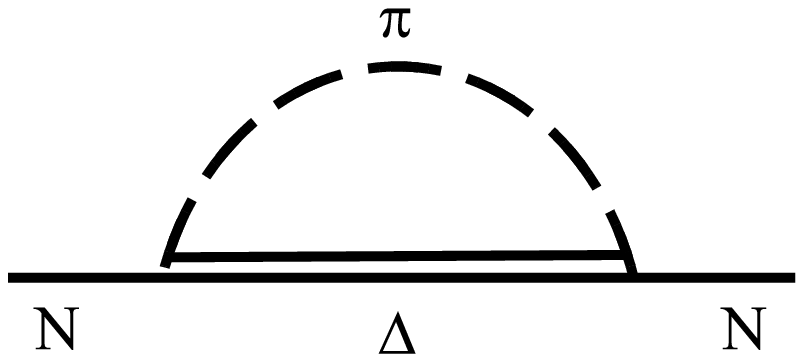}
\vspace{-6pt}
\caption{The pion loop contribution to the self-energy
    of the nucleon allowing transitions to the near-by and
    strongly-coupled decuplet baryons.}
\label{fig:nucSEpiD}
\centering
\includegraphics[width=0.45\hsize]{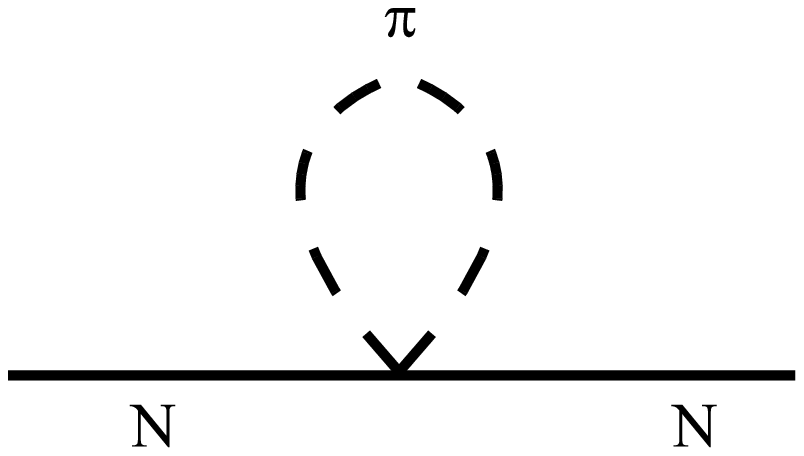}
\vspace{-6pt}
\caption{Tadpole contributions to the nucleon self energy.}
\label{fig:nucSEtad}
\end{figure}

Using the standard Gell-Mann--Oakes--Renner relation connecting
quark and pion masses, $m_q \propto m_\pi^2$ \cite{GellMann:1968rz},
the formal chiral expansion of the nucleon can be written as a
polynomial expansion in $m_\pi^2$ plus the meson-loop integral
contributions: 
\eqab
M_N &=& \{a_0 + a_2 m_\pi^2 + a_4 m_\pi^4 + \ca{O}(m_\pi^6)\} \nn\\
&+& \Si_N + \Si_\De + \Si_{tad}\,.
\label{eqn:mNresid}
\eqae
The pion cloud corrections are considered in the heavy-baryon limit,
with loop integrals, $\Si_N$, $\Si_\De$ and $\Si_{tad}$, corresponding to
Figures \ref{fig:nucSEpiN} through \ref{fig:nucSEtad}, respectively.
The coefficients $a_i$ of the analytic polynomial, contained in brackets 
$\{\,\}$ in Eq.~(\ref{eqn:mNresid}), are related to the
low energy constants of $\chi$PT.  In this investigation, they will be
determined by fitting to lattice QCD data.  These
coefficients will be referred to 
as the \emph{residual series} coefficients.  These bare
coefficients undergo renormalization due to contributions from the
loop integrals $\Si_N$, $\Si_\De$ and $\Si_{tad}$. 

Under the most general considerations, each loop integral, when
evaluated, produces an analytic polynomial in $m_\pi^2$ and
nonanalytic terms:
\eqab
\label{eqn:sigmaN}
\Si_N &=& b_0^N + b_2^N m_\pi^2 + \chi_N m_\pi^3 + b_4^N m_\pi^4 +
 \ca{O}(m_\pi^5)\,,\\
\label{eqn:sigmaD}
\Si_\De &=& b_0^\De + b_2^\De m_\pi^2  + b_4^\De m_\pi^4 \nn\\
&&+\f{3}{4\pi\De} 
\chi_\De m_\pi^4\,\log\f{m_\pi}{\mu} + \ca{O}(m_\pi^5)\,,\\
\label{eqn:sigmaT}
\Si_{tad} &=& b_2^{t'} m_\pi^2  +  b_4^{t'} m_\pi^4 + \chi_t'
 m_\pi^4\,\log\f{m_\pi}{\mu} + \ca{O}(m_\pi^5) .
\eqae
Here $\De$ is the delta-nucleon mass splitting in the chiral limit,
taken to be $292$ MeV.  $\chi_N$, $\chi_\Delta$ and $\chi_t^\prime$
denote the model independent chiral coefficients of the terms that are
nonanalytic in the quark mass.  The $b_i$ coefficients are
renormalization scheme dependent as are the $a_i$ coefficients.
It can be noted that the tadpole loop contribution $\Si_{tad}$ does not
produce a $b_0^t$ term because it enters with a leading factor of
$m_\pi^2$, as discussed in Section \ref{subsect:loops}.  The primes on
the coefficients $b_2^{t'}$ and $\chi_t'$ here simply indicate that
they will be used later in a slightly different form.

The process of renormalization in FRR $\chi$EFT proceeds by combining
the renormalization-scheme dependent coefficients to provide the
physical low energy coefficients, which are denoted as $c_i$.  Thus, the
nucleon mass expansion takes on the standard form:
\eqab
M_N &=& c_0 + c_2 m_\pi^2 + \chi_N m_\pi^3 +
c_4 m_\pi^4 \nn\\
&+& \!\left(\!\!-\f{3}{4\pi\De}\chi_\De + c_2\chi_t
 \!\right)\!m_\pi^4\log\f{m_\pi}{\mu} + \ca{O}(m_\pi^5)\,.
\label{eqn:mNexpansion}
\eqae
By comparing Eqs.~(\ref{eqn:mNresid}) through (\ref{eqn:mNexpansion}),
 the following renormalization procedure is obtained:
\eqab
\label{eqn:c0norm}
c_0 &=& a_0 + b_0^N + b_0^\De\,,\\
\label{eqn:c2norm}
c_2 &=& a_2 + b_2^N + b_2^\De + b_2^{t'}\,,\\
\label{eqn:c4norm}
c_4 &=& a_4 + b_4^N + b_4^\De + b_4^{t'}\,, \mbox{\,\,etc.}
\eqae
The coefficients $c_i$ are scheme independent quantities, and
 this property will be demonstrated when determined within the PCR.  The
value of $c_0$ is the nucleon mass in the chiral limit ($m_\pi^2 =
0$), and $c_2$ is related to the so-called sigma term of explicit
chiral symmetry breaking
\cite{Leinweber:2000sa,Wright:2000gg,Holl:2005st}.  The nonanalytic
terms $m_\pi^3$ and $m_\pi^4\,\log m_\pi/\mu$ have known,
constant coefficients denoted by $\chi_N$, $\chi_\De$ and $\chi_t$.
The value of $c_4$ is scale dependent, such that the total 
$m_\pi^4$ term in Eq.~(\ref{eqn:mNexpansion}), including the logarithm,
 is independent of the scale 
$\mu$. It can be noted
 that the nucleon mass itself is completely independent of the
choice of $\mu$. For the numerical analysis, 
$\mu$ is set equal to $1\,{\rm GeV}$.

Of course, EFT loop calculations are commonly divergent without some
regularization method.  Since the effective field theory is only
applicable for low energies, hard momenta contributions to loop
calculations may be eliminated.  However, the traditional schemes
including dimensional regularization (DR) often do not involve an
explicit scale dependence when evaluating loop diagrams.  Without any
momentum cutoff, the $b_i$ coefficients from each loop integral become
either infinite or vanish, and the $c_i$ coefficients from Eq.
(\ref{eqn:mNexpansion}) undergo an infinite renormalization or none at
all:
\eqab
\label{eqn:c0normdr}
c_0 &=& a_0 - \infty \,, \\
\label{eqn:c2normdr}
c_2 &=& a_2 + \infty \,,\\
\label{eqn:c4normdr}
c_4 &=& a_4 + 0 \,, \mbox{\,\,etc.}
\eqae
Since the $c_i$ coefficients are finite after renormalization,
the $a_i$ coefficients must have been infinite, with the opposite sign
of the $b_i$ coefficients.  As emphasized above, both the $a_i$ and
$b_i$ coefficients are scheme dependent.  The infinities are absorbed
in constructing
 the $c_i$ coefficients and thus subtracted from the chiral
 expansion.  This minimal subtraction scheme with no explicit scale
dependence makes DR quite suitable for elementary fields, where the
{\em absence} of new degrees of freedom at higher energies is assumed.
However, for EFTs there is an energy scale beyond
which the effective fields are no longer the relevant degrees of
freedom. When one integrates loop contributions over this high energy
domain, there is no guarantee that one can efficiently subtract the
model-dependent, ultraviolet physics with a finite number of
counter-terms (unless in the PCR).
As a result, the chiral expansion typically only shows reliable
convergence properties over a narrow range of pion mass.

Indeed this problem of beginning with rapidly varying loop
contributions, which must then be removed with a finite number of
counter-terms, can easily be overcome. The hard momentum contributions
to the meson-loop diagrams can be suppressed via the introduction of a
regulator.  
As such, the coefficients of the residual expansion are
likely to be smaller, and the utility of the expansion has the potential to
apply to a broader range of quark or pion masses.  The introduction of
a regulator acts to resum the chiral expansion, with loop integrals
having the general properties described in Eqs.~(\ref{eqn:sigmaN})
through (\ref{eqn:sigmaT}).

The resummation of the chiral series through the introduction of a
regulator (or similar variant) has been studied in various instances
\cite{Donoghue:1998bs,Young:2002cj,Young:2002ib,Borasoy:2002jv,Leinweber:2003dg,Bernard:2003rp,Djukanovic:2004px,Young:2004tb,Leinweber:2004tc,Leinweber:2005xz}.
The method consists of inserting a regulator function $u(k^2)$
 into the integrand of the meson-loop integrals. The
regulator can take any form, so long as it is normalized to $1$, and approaches
 $0$ sufficiently fast to ensure convergence of the loop.
 Unlike DR, this method
involves an explicit momentum cutoff scale, $\La$. The chiral
expansion can now be written in terms of this cutoff scale:
\eqab
M_N &=& \{a_0^\La + a_2^\La\, m_\pi^2 + a_4^\La\, m_\pi^4 
+ \ca{O}(m_\pi^6)\} \nn\\
&+& \Si_N(m_\pi^2,\La) + \Si_\De(m_\pi^2,\La) + \Si_{tad}(m_\pi^2,\La).\,\,
\label{eqn:mNresidB}
\eqae
The superscript $\La$ denotes the scheme dependence of the
$a_i^\La$ coefficients. The loop integrals are functions of the scale
$\La$ and also $m_\pi^2$. 

Through the introduction of the regulator, the loop integrals are now
low energy contributions, significant for small $m_\pi^2$ and becoming
negligible as $m_\pi^2$ becomes large.  The scheme dependent $a_i^\La$
coefficients undergo a renormalization, as before, via their
combination with the $b_i^\La$ coefficients, whose scheme dependence
is now explicit, reflecting the regularization of the loop integrals:
\eqab
\label{eqn:c0normfrr}
c_0 &=& a_0^\La + b_0^{\La,N} + b_0^{\La,\De}\,,\\
\label{eqn:c2normfrr}
c_2 &=& a_2^\La + b_2^{\La,N} + b_2^{\La,\De} + b_2^{\La,t'}\,,\\
\label{eqn:c4normfrr}
c_4 &=& a_4^\La + b_4^{\La,N} + b_4^{\La,\De} + b_4^{\La,t'}\,, \mbox{\,\,etc.}
\eqae

Dimensional analysis reveals that the coefficients $b_i$ are
proportional to $\Lambda^{(3-i)}$.  Thus it can be realized that as the cutoff
scale $\La$ goes to infinity the FRR expansion reduces to that of
Eq.~(\ref{eqn:mNexpansion}) via 
Eqs.~(\ref{eqn:c0normdr}) through (\ref{eqn:c4normdr}).
 At any finite $\Lambda$, a partial
resummation of higher-order terms is introduced.

Previous studies indicate that extrapolation results show very little 
sensitivity to the precise functional form of the regulator 
\cite{Leinweber:2003dg}.
In this investigation, the
family of smoothly attenuating dipole regulators will be considered.
The general $n-$tuple dipole function takes the following form, for a
cutoff scale of $\La$:
\eqb
u_n(k^2) = {\left(1 + {\f{k^{2n}}{\La^{2n}}} \right)}^{-2}\,.
\eqe
The standard dipole is recovered for $n=1$. The cases $n=2,3$ are the
`double-' and `triple-dipole' regulators, respectively.  In the
following, $u(k^2)$ is used to denote one of these 
regulators. This functional form allows one to interpolate between the
dipole regulator and the step function (which corresponds to $n\to\infty$).


In a study by Bernard \textit{et al.} \cite{Bernard:2003rp}, 
it was suggested that only a 
sharp cutoff FRR scheme is consistent with chiral symmetry.
Djukanovic  
\textit{et al.} \cite{Djukanovic:2004px} 
have demonstrated more general functional forms can be 
generated by proposing a scheme in which the regulator function is 
interpreted as a modification 
to the propagators of the theory, obtained from a new chiral 
symmetry-preserving Lagrangian.
Higher-derivative coupling terms are 
built into the Lagrangian in order to produce a regulator from the 
Feynman Rules in a symmetry preserving manner.

The regulators used in the present investigation
 are introduced in a less systematic fashion, such that chiral symmetry 
is not automatically preserved to the order calculated. 
The higher derivative couplings 
of the regulator induces scheme-dependent nonanalytic terms.
To maintain chiral symmetry, one must introduce the necessary vertex 
corrections.

Alternatively, one can choose the regulator judiciously such that any extra 
scheme-dependent nonanalytic are removed to any chosen order. For example, 
the $n-$tuple dipole regulators generate extra nonanalytic terms 
in the chiral expansion of Eq. (\ref{eqn:mNexpansion}) at  
 higher chiral orders. For a dipole regulator, regulator-dependent 
nonanalytic terms occur at odd 
powers of $m_\pi$, beginning at $\ca{O}(m_\pi^5)$ \footnotemark. 
\footnotetext{While scheme-dependent, it is significant to note that
with a dipole regulator, $\La = 0.8$ GeV, the 
coefficient of the induced 
$m_\pi^5$ term compares favorably with the two-loop calculation 
\cite{McGovern:1998tm,Leinweber:2003dg,Leinweber:2005xz,McGovern:2006fm,Schindler:2006ha}.}
In the case of the double dipole, the nonanalytic terms begin 
at $\ca{O}(m_\pi^7)$, and for the triple dipole the nonanalytic terms begin 
only at $\ca{O}(m_\pi^9)$. 


In a final observation, it is essential to note that the degrees of freedom
present in the residual series coefficients, $a_i^\La$, are sufficient
to eliminate any dependence on the regulator parameter, $\Lambda$, to
the order of the chiral expansion calculated: in this case ${\cal
O}(m_\pi^4)$.  By definition, higher order terms in the FRR expansion
are negligible in the PCR, and therefore FRR $\chi$EFT is
mathematically equivalent to $\chi$PT in the PCR.
Any differences observed in results obtained at the same chiral order
but with different regularization schemes are a direct result of
considering data that lie outside the PCR (provided that the regulator 
$\La$ is not chosen too small such as to introduce an unphysical 
low energy scale).


%
\subsection{Loop Integrals and Definitions}
\label{subsect:loops}
The leading order loop integral contributions to the nucleon mass,
corresponding to the diagrams in Figures
\ref{fig:nucSEpiN} through \ref{fig:nucSEtad}, can be simplified to a
convenient form by taking the  heavy-baryon limit, and
performing the pole integration for $k_0$.
%
Renormalization, as outlined above, is achieved by subtracting the
relevant terms in the Taylor expansion of the loop integrals and
absorbing them into the corresponding low energy constants, $c_i$:
\begin{eqnarray}
\label{eqn:NN}
\tSi_{N} &=& \chi_N\f{1}{2\pi^2}\int\!\!\ud^3 k
\f{k^2 u^2(k^2)}{\om^2(k)} -  b_0^{\La,N} - b_2^{\La,N} m_\pi^2\\
 &=& \chi_N m_\pi^3 + b_4^{\La,N}m_\pi^4 + \ca{O}(m_\pi^5)\,,
\label{eqn:NNexpn}
\end{eqnarray}
\begin{eqnarray}
\label{eqn:NDe}
\tSi_{\De} &=& \chi_\De\f{1}{2\pi^2}\int\!\!\ud^3 k
\f{k^2 u^2(k^2)}{\om(k)\left(\De + \om(k)\right)} \nn\\
&& \quad - b_0^{\La,\De} - b_2^{\La,\De} m_\pi^2 \\
 &=&  b_4^{\La,\De} m_\pi^4- \f{3}{4\pi\De} 
\chi_\De m_\pi^4\,\log\f{m_\pi}{\mu} + \ca{O}(m_\pi^5)\,,\qquad
\label{eqn:NDeexpn}
\end{eqnarray}
\begin{eqnarray}
\label{eqn:tad}
\tSi_{tad} &=&  c_2 m_\pi^2\left(
\chi_t\f{1}{4\pi}\int\!\!\ud^3 k
\f{2 u^2(k^2)}{\om(k)} - b_2^{\La,t} \right)\\
\label{eqn:tadexpnSi}
 &=& c_2 m_\pi^2 \left(b_4^{\La,t} m_\pi^2  + 
\chi_t m_\pi^2\,\log\f{m_\pi}{\mu}
+ \ca{O}(m_\pi^5)\right)\qquad\\
\label{eqn:tadexpnsi}
&=& c_2 m_\pi^2 \tsi_{tad}\,.
\end{eqnarray}
These integrals are expressed in terms of the pion energy, $\om(k) =
\sqrt{k^2 + m_\pi^2}$.  The tilde ($\,\tilde{\,}\,$) denotes that the
integrals are written out in renormalized form to chiral order 
 $\ca{O}(m_\pi^2)$.
  As the $b_i$
coefficients are regulator and scale dependent, this
subtraction removes this dependence. The coefficients $a_0$ and $a_2$ of
the analytic terms in the chiral expansion in
Eq.~(\ref{eqn:mNexpansion}) are now automatically the renormalized
coefficients $c_0$ and $c_2$. This is because the $b_0$ and $b_2$ terms in
Eqs.~(\ref{eqn:c0normfrr}) and (\ref{eqn:c2normfrr}) 
 are removed in the subtraction.
 Note also that the tadpole amplitude in
Eqs.~(\ref{eqn:tadexpnSi}) and (\ref{eqn:tadexpnsi}) contains the
renormalized $c_2$ in its coefficient. The interaction vertex in this
diagram arises from expanding out the pion field in the leading quark
mass insertion.

The constant coefficients $\chi_{N}$, $\chi_\De$ and $\chi_t$ for each
integral are defined in terms of the pion decay constant, which is
taken to be $f_\pi = 93$ MeV, and the axial coupling parameters $D$,
$F$ and $\ca{C}$ which couple the baryons to the pion field. 
The phenomenological values for these couplings are used, applying the $\SU(6)$
flavour-symmetry relations \cite{Jenkins:1991ts,Lebed:1994ga} to yield
$\ca{C} = -2D$, $F = \f{2}{3}D$ and the value $D = 0.76$:
\begin{eqnarray}
\chi_N   &=& -\f{3}{32\pi f_\pi^2}(D+F)^2\,,\\
\chi_\De &=& -\f{3}{32\pi f_\pi^2}\f{8}{9}\ca{C}^2\,,\\
\chi_t   &=& -\f{3}{16\pi^2 f_\pi^2}\,.
\end{eqnarray}
%

%
With the renormalized integrals specified, the FRR
modified version of the chiral expansion in Eq.~(\ref{eqn:mNexpansion})
 takes the form:
\begin{equation}
M_N = c_0 + c_2 m_\pi^2(1+{\tsi}_{tad})
+ a_4^\La m_\pi^4 + {\tSi}_{N} + {\tSi}_{\De}\,.\\
\label{eqn:mNfit}
\end{equation}
The $a_4^\Lambda$ term is left in unrenormalized form
for simplicity. Indeed, the $b_4$ can be evaluated by expanding out
corresponding loop integrals, such as in Ref.~\cite{Young:2002ib}.
However, the focus here is on the behaviour of $c_0$ and $c_2$.

Since the results of lattice simulations reflect the presence of
discrete momentum values associated with the finite volume of the
lattices, the formalism must also take into account these finite
volume effects.  In order to accommodate the effect of the finite
volume, the continuous loop integrals occurring in the meson loop
calculations in infinite volume are transformed into a sum over
discrete momentum values.  The difference between a loop sum and its
corresponding loop integral is the finite volume correction, which
should vanish for all integrals as $m_\pi L$ becomes large
\cite{Beane:2004tw}. 

While Eq.~(\ref{eqn:mNfit}) is useful in describing the pion mass
evolution of the nucleon mass, for the consideration of lattice QCD
results, one also needs to incorporate corrections to allow for the
finite-volume nature of the numerical simulations.  As the pion is the
lightest degree of freedom in the system, it is the leading order 
 pion loop effects
 that are most sensitive to the periodic boundary
conditions.  The corrections can be determined by considering the
transformation of each loop integral in
Eqs.~(\ref{eqn:NN}), (\ref{eqn:NDe}) and (\ref{eqn:tad}), 
into a discrete sum for
lattice volume $V = L_x L_y L_z$ \cite{Armour:2005mk}:
\begin{equation}
\int\!\!\ud^3 k \ra \f{{(2\pi)}^3}{L_x L_y L_z} \sum_{k_x,k_y,k_z}\,.
\end{equation}
Each momentum component is quantized in units of $2\pi/L$, that
is $k_i=n_i2\pi/L$ for integers $n_i$.  The finite volume correction
$\de^\ro{FVC}$ can be written as the difference between the finite sum
and the integral:
\begin{equation}
\de^\ro{FVC}_i \!\!= \f{\chi_i}{2\pi^2}\Big[\f{{(2\pi)}^3}{L_x L_y L_z}
\!\!\sum_{k_x,k_y,k_z}\!\!\!\!\!I_i(\vec{k},m_\pi^2,\La)\, -
  \int\!\!\ud^3 k\,\, I_i(\vec{k},m_\pi^2,\La)\Big],\quad
\end{equation}
where $i=N,\De$, and the integrands are denoted $I_i(\vec{k},m_\pi^2,\La)$.
%
By adding the relevant finite volume correction (FVC) to each loop 
contribution, 
 the finite volume nucleon mass can be parameterized:
\begin{eqnarray}
M_N^V &=&  c_0 + c_2 m_\pi^2(1+{\tsi}_{tad}) + a_4^\La m_\pi^4 \nn\\
&+& ({\tSi}_{N}+\de_{N}^{\ro{FVC}}) +
 ({\tSi}_{\De}+\de_{\De}^{\ro{FVC}})\,.
\label{eqn:mNfitfin}
\end{eqnarray}

\begin{figure}[tp]
\includegraphics[height=0.76\hsize,angle=90]{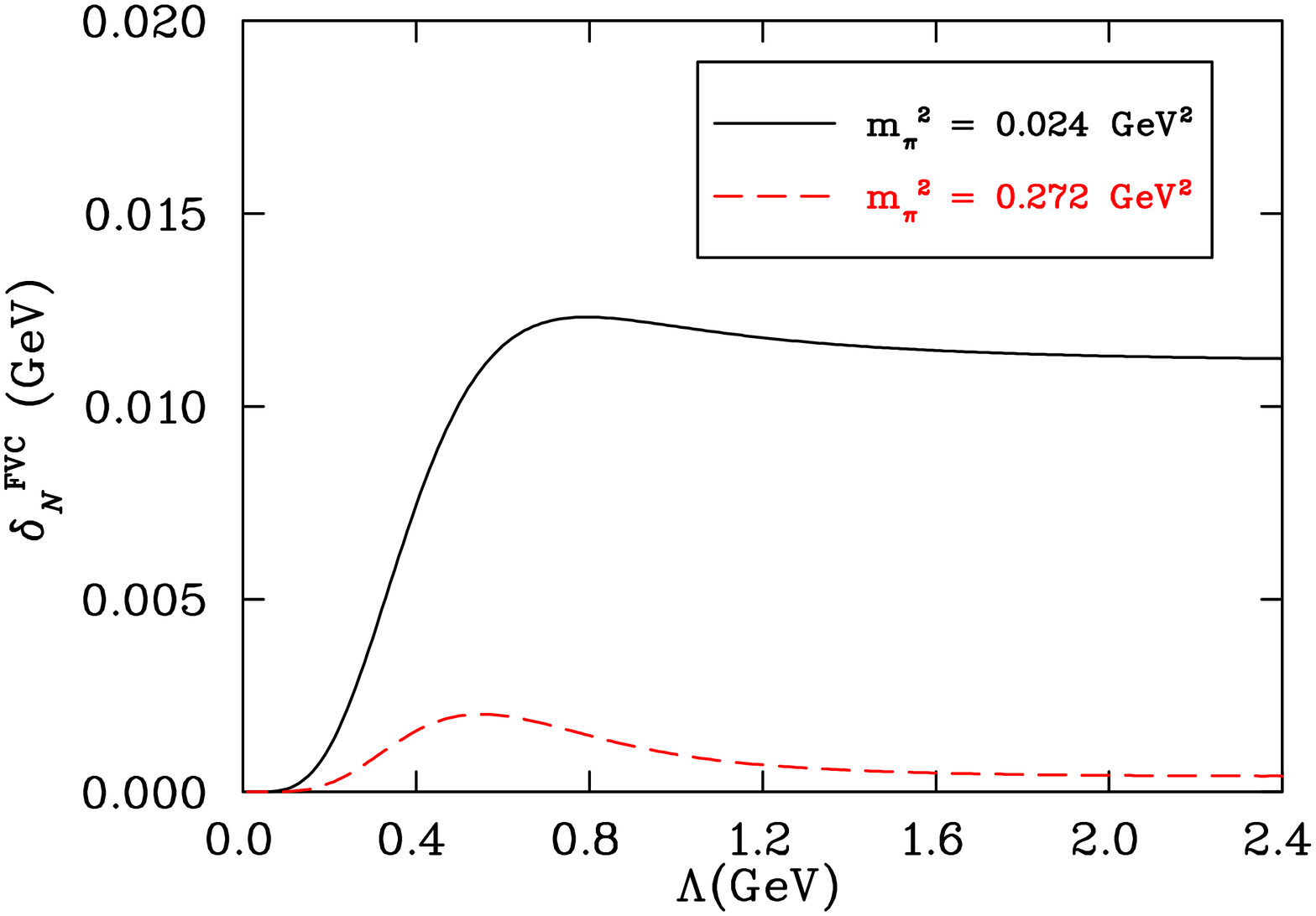}
\vspace{-12pt}
\caption{(color online). Behaviour of the finite volume
    corrections $\de_N^\ro{FVC}$ vs.\ $\La$ on a $2.9$ fm box using a dipole regulator. Results for two different values of $m_\pi^2$ are shown.}
\label{fig:fvcN}
\includegraphics[height=0.76\hsize,angle=90]{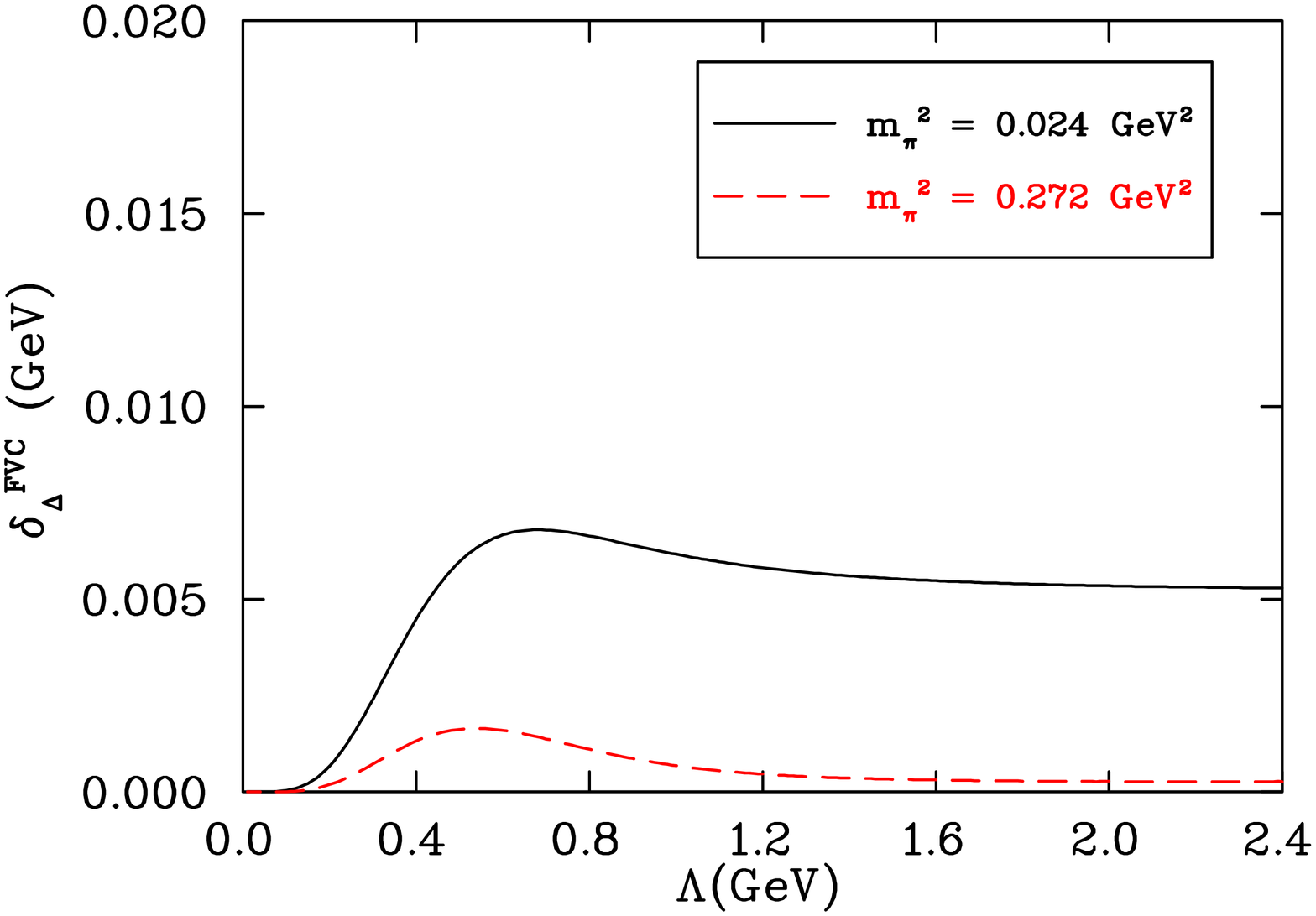}
\vspace{-12pt}
\caption{(color online). Behaviour of finite volume corrections
  $\de_\De^\ro{FVC}$ vs.\ $\La$ on a $2.9$ fm box using a dipole regulator. Results for two different values of $m_\pi^2$ are shown.}
\label{fig:fvcD}
\includegraphics[height=0.76\hsize,angle=90]{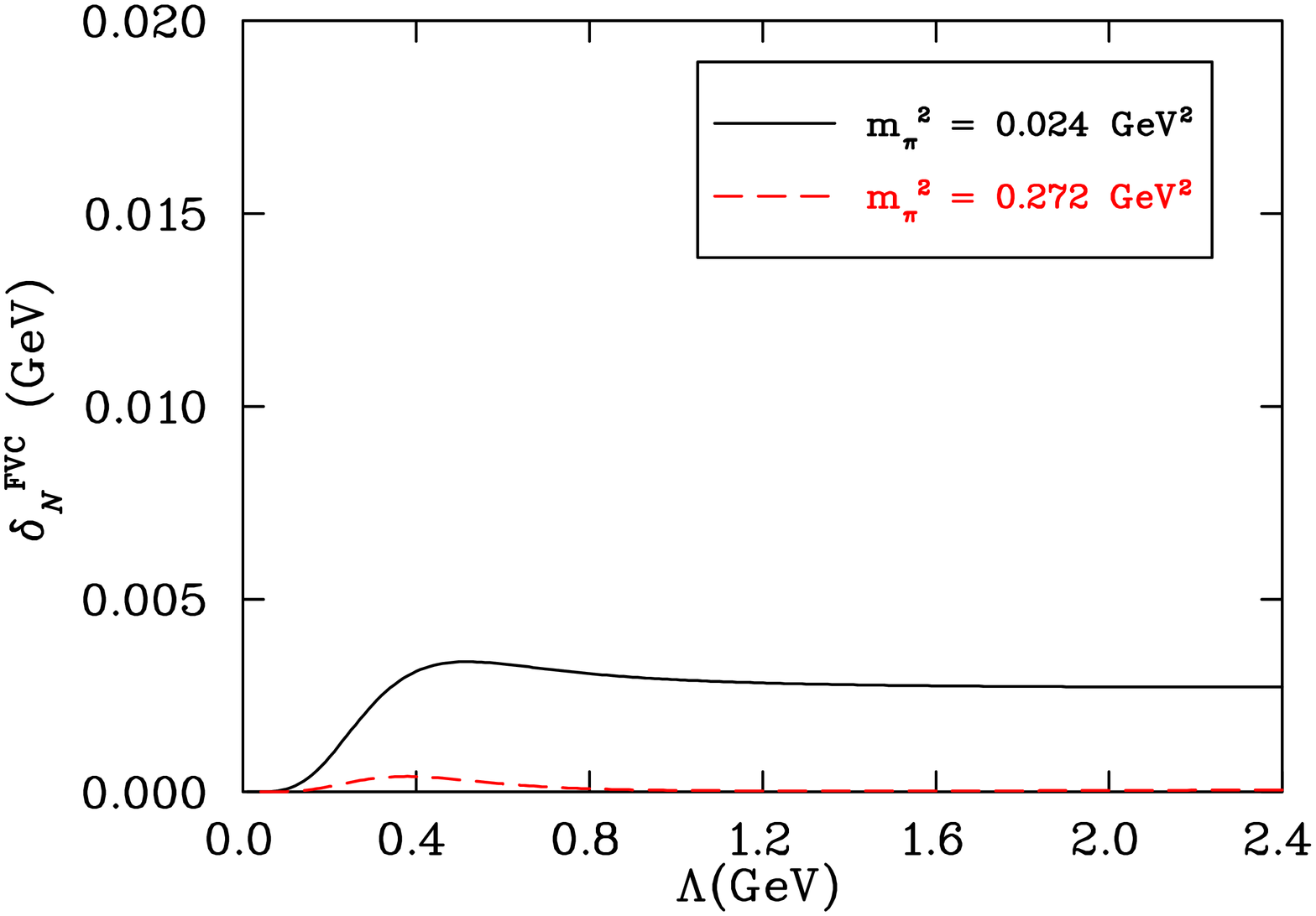}
\vspace{-12pt}
\caption{(color online). Behaviour of finite volume
    corrections $\de_N^\ro{FVC}$ vs.\ $\La$ on a $4.0$ fm box using a dipole regulator. Results for two different values of $m_\pi^2$ are shown.}
\label{fig:fvcN4fm}
\includegraphics[height=0.76\hsize,angle=90]{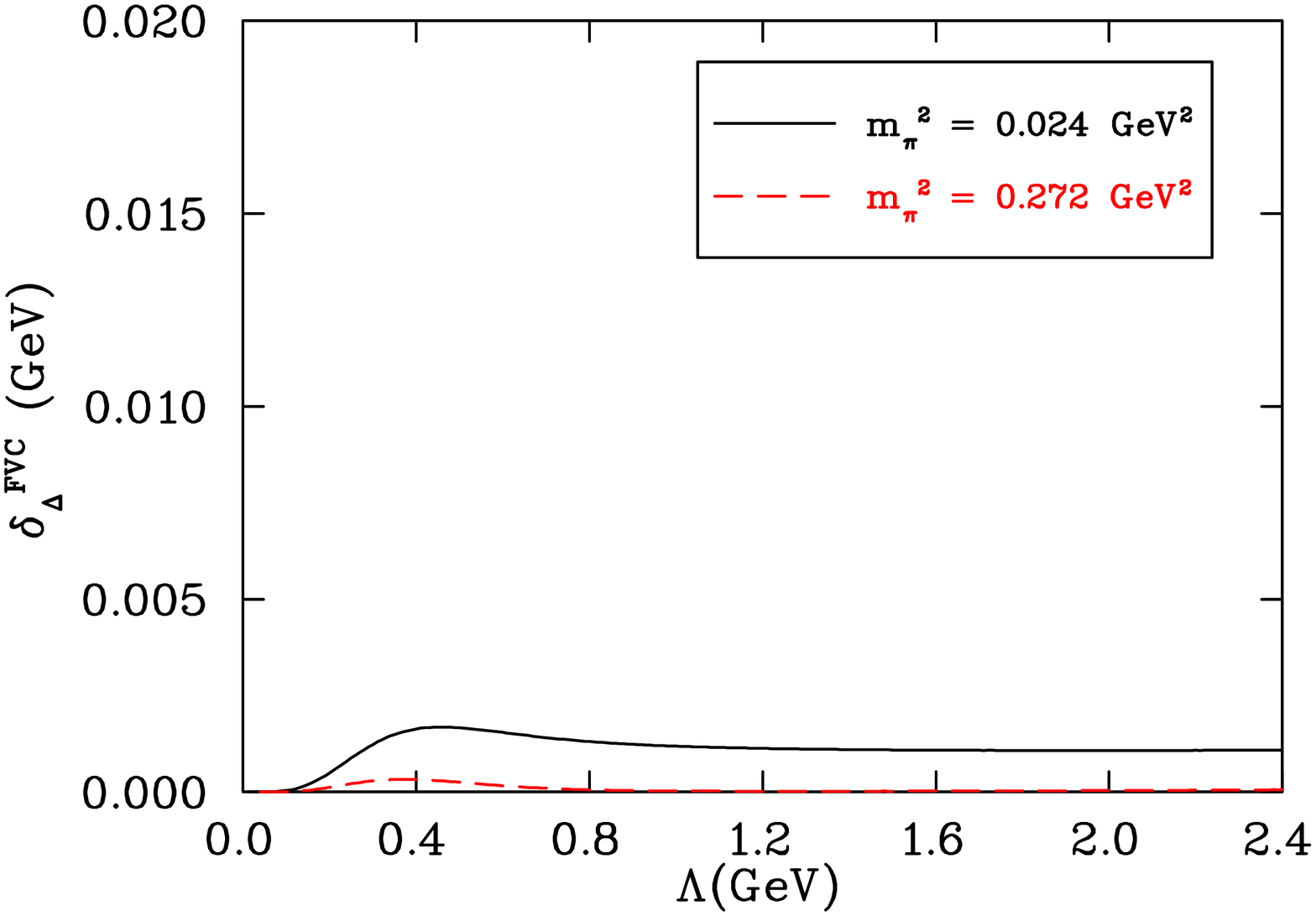}
\vspace{-12pt}
\caption{(color online). Behaviour of finite volume corrections
  $\de_\De^\ro{FVC}$ vs.\ $\La$ on a $4.0$ fm box using a dipole regulator. Results for two different values of $m_\pi^2$ are shown.}
\label{fig:fvcD4fm}
\end{figure}

It is also anticipated that the FVC are
independent of the regularization scale $\Lambda$ in this domain. In Figures
\ref{fig:fvcN} and \ref{fig:fvcD}, the scale dependence of
the finite-volume corrections is shown for a dipole regulator and a 
$2.9$ fm box
 (the same box size used for the PACS-CS data \cite{Aoki:2008sm}).
  It is notable that choosing $\Lambda$
too small suppresses the very infrared physics that one is trying to
describe, and therefore it is sensible to
 be cautious by not selecting a $\Lambda$ that is
too low.
Figures \ref{fig:fvcN4fm} and \ref{fig:fvcD4fm} 
show the behaviour of the FVC 
for a $4.0$ fm box, and the corrections are much smaller as expected. 

For large $\Lambda$ the results saturate to a fixed
result. For the light pion masses, provided $\Lambda\gtrsim
0.8$ GeV, the estimated finite volume corrections are stable.
The asymptotic result is used, which has been demonstrated to be
successful in previous studies \cite{AliKhan:2003cu}.  Numerically, this is
achieved by evaluating the finite volume corrections with a parameter,
$\La' = 2.0$ GeV, $\de_i^\ro{FVC} = \de_i^\ro{FVC}(\La')$. It should be noted
that this is equivalent to the more algebraic approach outlined in
Ref.~\cite{Beane:2004tw}.

%
\section{Intrinsic Scale: Pseudodata}
\label{sect:intr}
\begin{figure}[tp]
\includegraphics[height=0.76\hsize,angle=90]{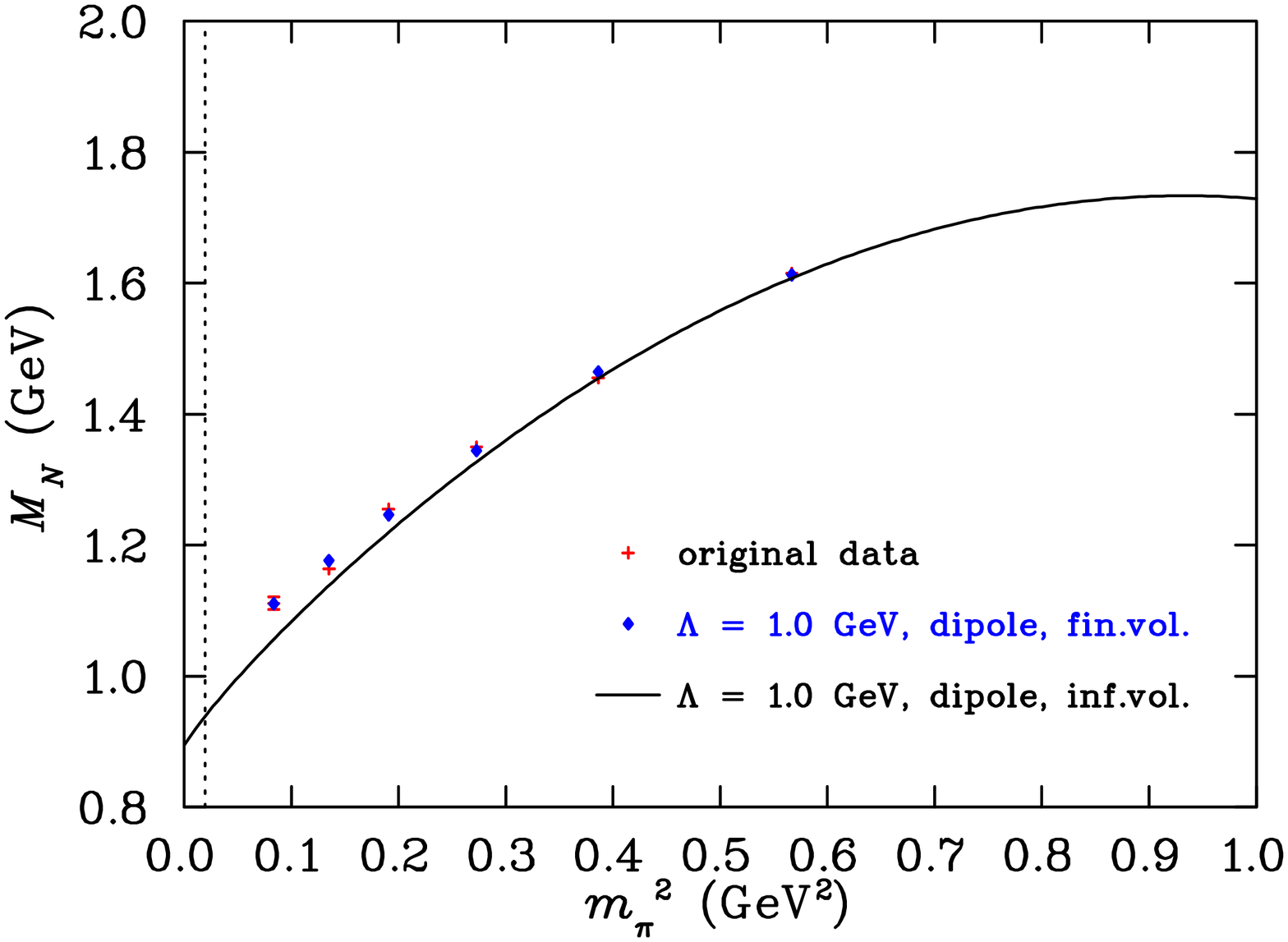}
\vspace{-12pt}
\caption{\footnotesize{(color online). Example dipole extrapolation based on JLQCD data \cite{Ohki:2008ff}, box size: $1.9$ fm.}}
\label{fig:OhkiExt}
\includegraphics[height=0.76\hsize,angle=90]{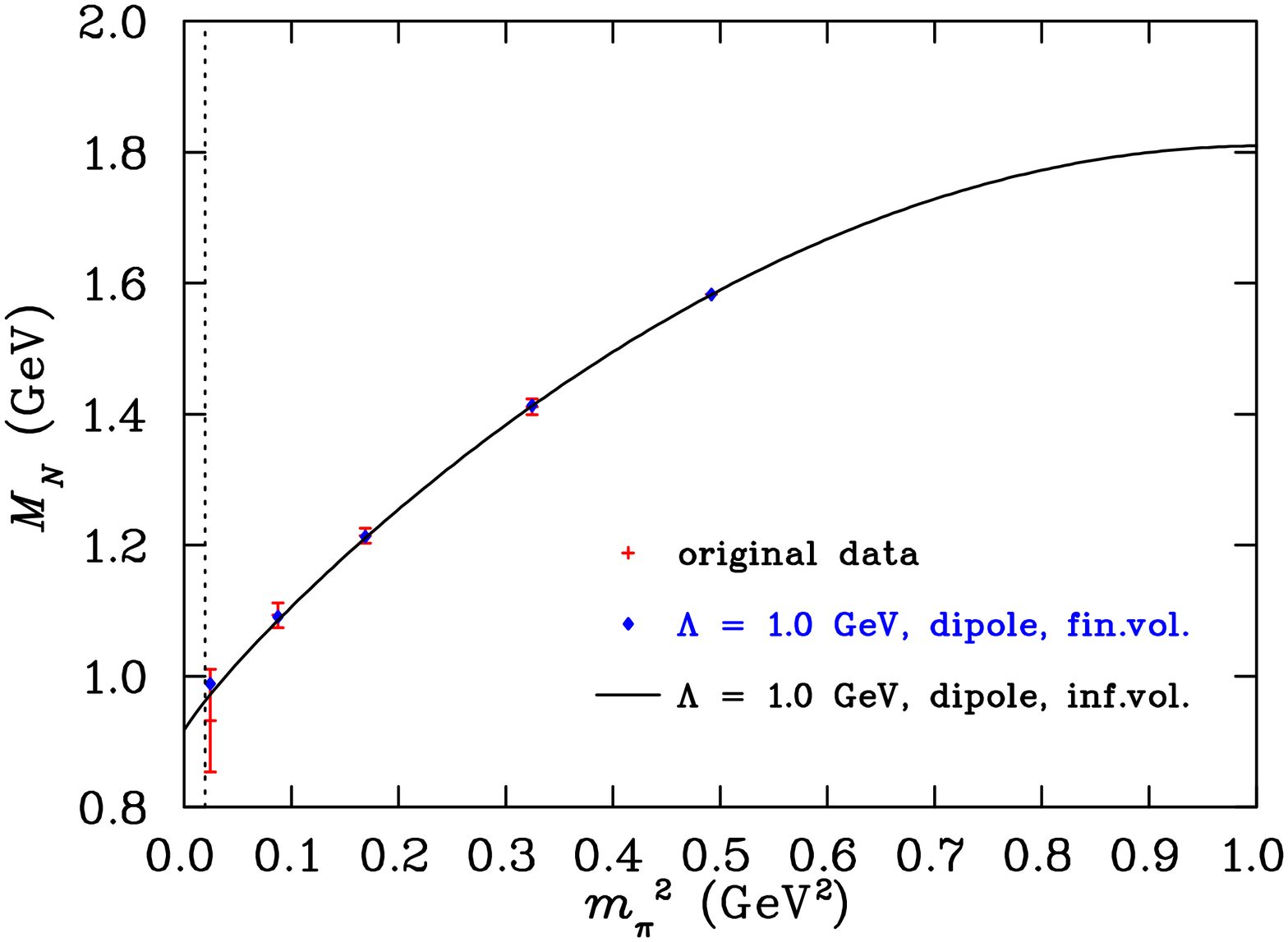}
\vspace{-12pt}
\caption{\footnotesize{(color online). Example dipole extrapolation based on PACS-CS data \cite{Aoki:2008sm}, box size: $2.9$ fm.}}
\label{fig:AokiExt}
\end{figure}

\begin{figure}[tp]
\includegraphics[height=0.76\hsize,angle=90]{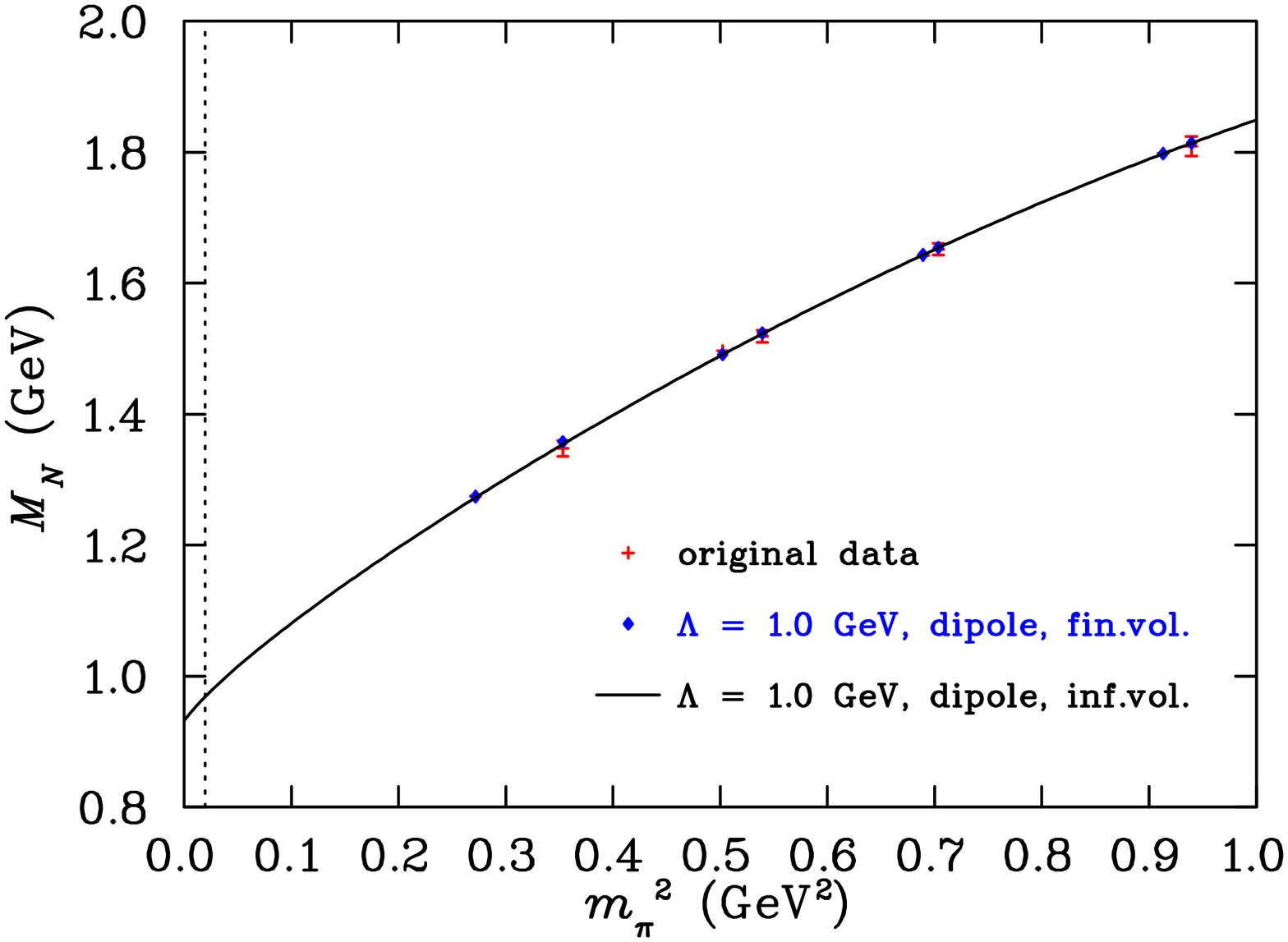}
\vspace{-12pt}
\caption{\footnotesize{(color online). Example dipole extrapolation based on CP-PACS data \cite{AliKhan:2001tx}, box size: $2.3-2.8$ fm.}}
\label{fig:YoungExt}
\includegraphics[height=0.76\hsize,angle=90]{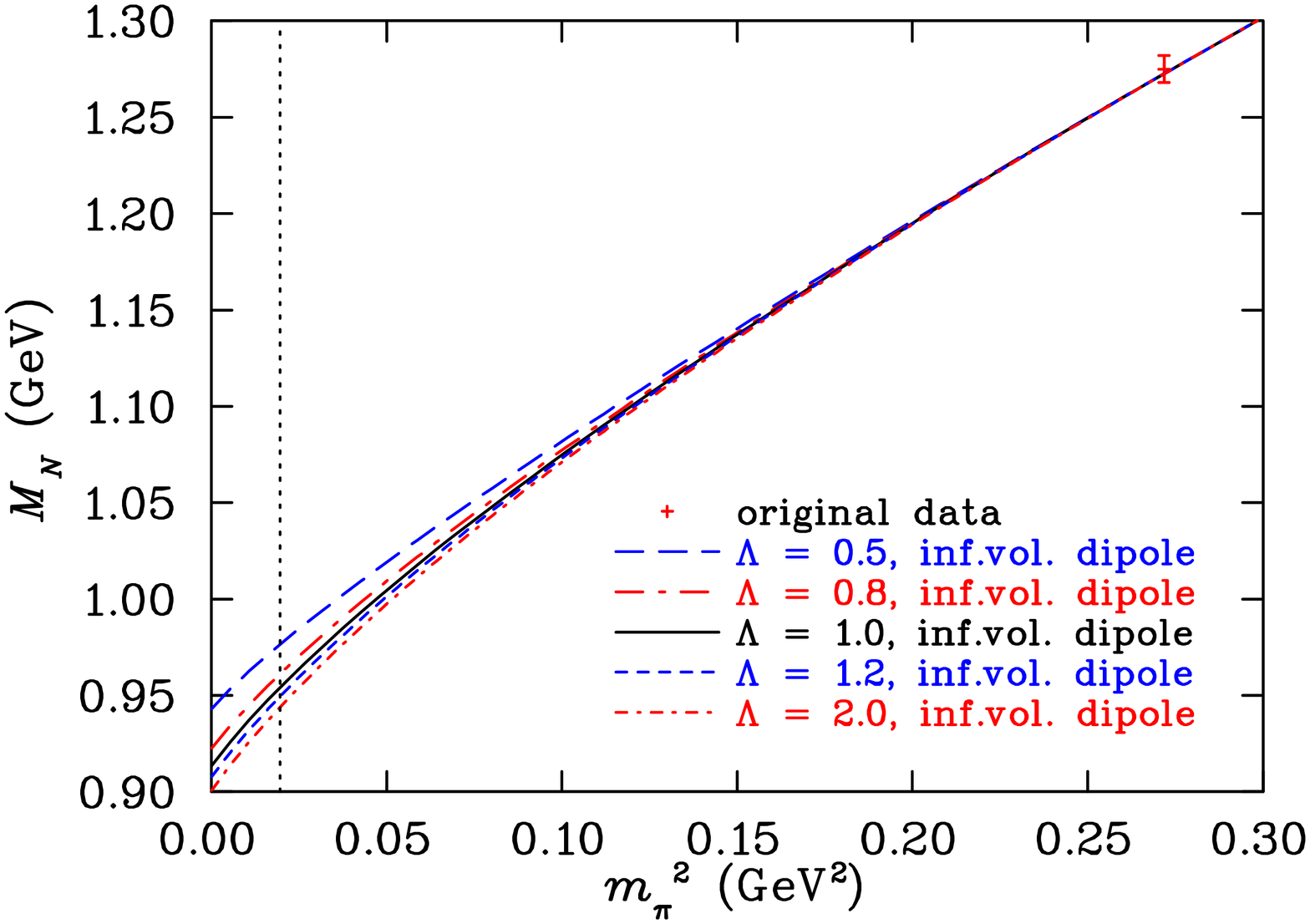}
\vspace{-12pt}
\caption{\footnotesize{(color online). Close zoom of the regulator dependence for dipole extrapolation based on CP-PACS data. Only the data point corresponding to the smallest $m_\pi^2$ value is shown at this scale.}}
\label{fig:YoungExtmulti}
\end{figure}

This $\chi$EFT extrapolation scheme to order 
 $\ca{O}(m_\pi^4\,\ro{log}\,m_\pi)$ 
 will be used in conjunction with 
lattice QCD data from JLQCD, PACS-CS and CP-PACS to predict the nucleon mass
for any value of $m_\pi^2$.
The lattice data used in this analysis can be used to extrapolate 
$M_N$ to the physical point by
taking into account the relevant curvature from the loop integrals
in Eqs. ~(\ref{eqn:NNexpn}), (\ref{eqn:NDeexpn}) and (\ref{eqn:tadexpnSi}).
As an example, a regulator value of $\La = 1.0$ GeV was chosen 
 for Figures \ref{fig:OhkiExt} through \ref{fig:YoungExt}, where
the finite volume corrected EFT appears concordant with previous QCDSF-UKQCD
 results \cite{AliKhan:2003cu}.
If the regulator is changed away from the choice $\La = 1.0$ GeV,
 the extrapolation curve also changes.
This signifies a scheme dependence in the result due to using lattice QCD
data beyond the PCR.

To demonstrate this, consider the infinite volume extrapolation of the
CP-PACS data. The extrapolation is achieved by subtracting the finite volume 
loop integral
contributions defined in 
Eqs.~(\ref{eqn:NN}), (\ref{eqn:NDe}) and (\ref{eqn:tad}) from each data point
and then fitting the result to obtain the coefficients $c_0$, $c_2$ and 
$a_4^\La$ as shown in Eq.~(\ref{eqn:mNfit}). The infinite volume 
loop integrals are then
 added back at any desired value of $m_\pi^2$.

Figure \ref{fig:YoungExtmulti} shows that the curves overlap
exactly when $m_\pi^2$ is large, where the lattice data reside.
 They diverge as the chiral regime is approached. This section addresses
this problem in detail.
%



A particular regularization scale is selected and a dense and
precise data set is generated, 
which smoothly connects with state of the art lattice
simulation results.  If all the data considered lie
within the PCR then the choice of regulator parameter is irrelevant,
and the FRR chiral expansion is mathematically equivalent to scale-invariant 
renormalzation schemes including DR.  However, the purpose here is to
consider an insightful scenario, whereby a set of
\emph{ideal} pseudodata with known low energy coefficients is produced.
  This scenario will form the basis of the investigation of the PCR, and
ultimately the possible existence of an intrinsic scale hidden within
the actual lattice QCD data.

The pseudodata are produced by performing a finite volume extrapolation
such as shown in Figures \ref{fig:OhkiExt} through \ref{fig:YoungExt}.
 The difference is that $100$ infinite volume 
extrapolation points are produced close to the chiral regime.
 The exercise is to pretend that it is actual lattice QCD data.
 Clearly, a regularization scheme must be chosen 
to produce the pseudodata. In this case,
a dipole regulator was chosen
 and pseudodata were created at $\La_\ro{c} \equiv 1.0$ GeV.

\begin{figure}[tp]
\includegraphics[height=0.76\hsize,angle=90]{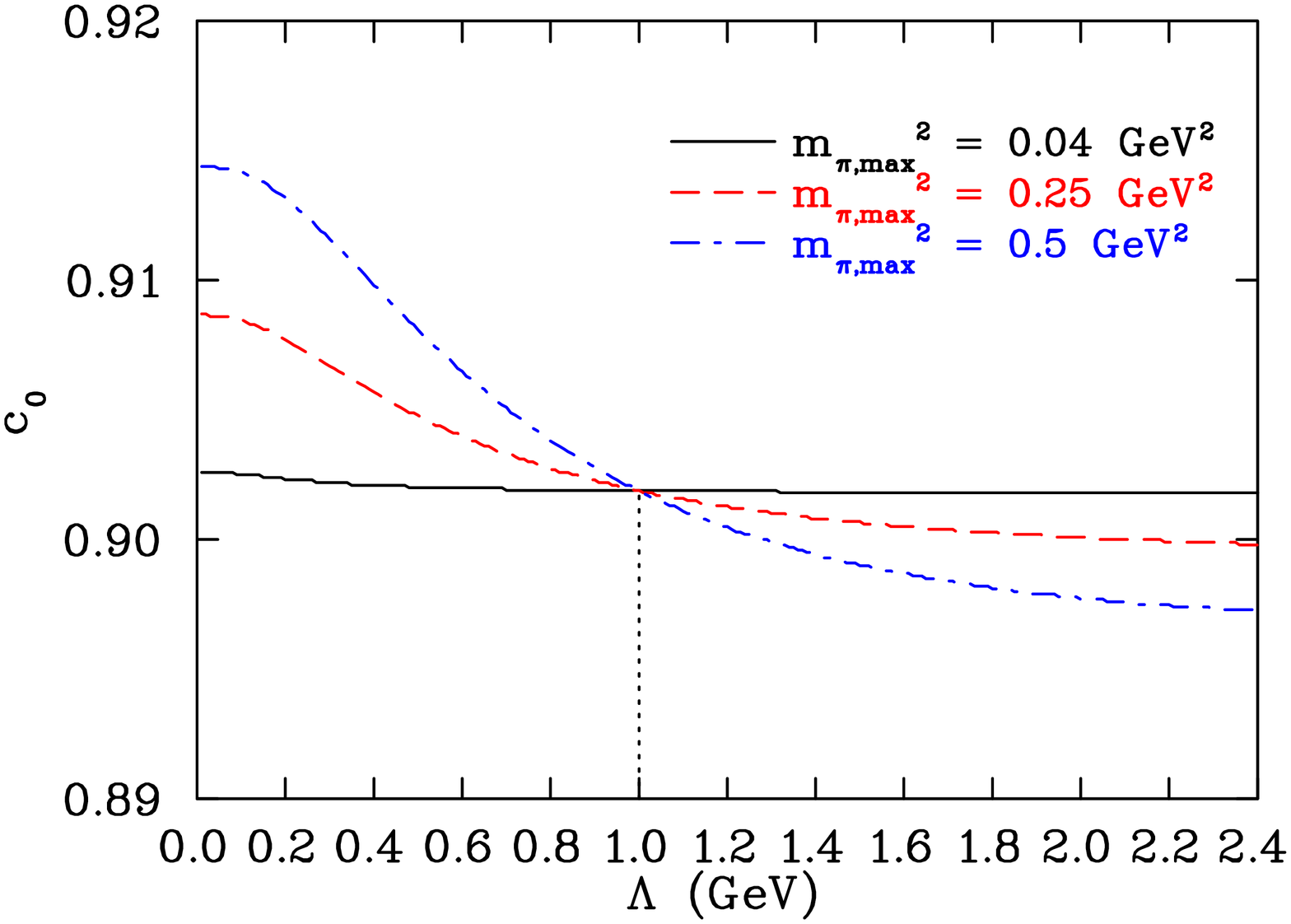}
\vspace{-12pt}
\caption{\footnotesize{(color online). Behaviour of $c_0$ vs.\ regulator parameter $\La$, based on infinite volume pseudodata created with a dipole regulator at $\La_\ro{c} = 1.0$ GeV (based on lightest four
data points from PACS-CS). Each curve uses pseudodata with
 a different upper value of pion mass $m_{\pi,\ro{max}}^2$.}}
\label{fig:pdatac0}
\includegraphics[height=0.76\hsize,angle=90]{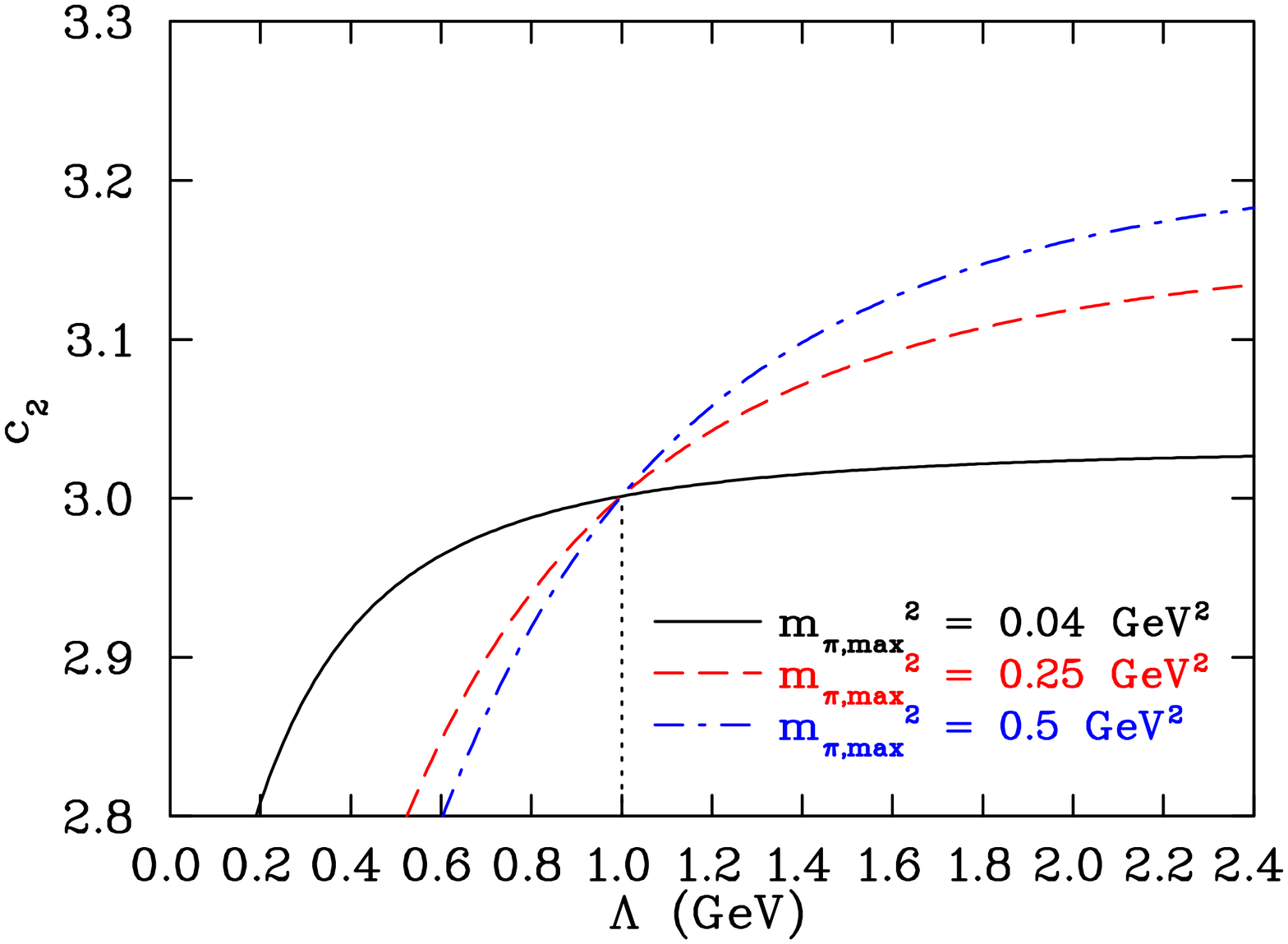}
\vspace{-12pt}
\caption{\footnotesize{(color online). Behaviour of $c_2$ vs.\ $\La$, based on infinite volume
pseudodata created with a dipole regulator at $\La_\ro{c} = 1.0$ GeV (based on lightest four
data points from PACS-CS). Each curve uses pseudodata with
 a different upper value of pion mass $m_{\pi,\ro{max}}^2$.}}
\label{fig:pdatac2}
\end{figure}

The regularization dependence of the extrapolation is characterized by the 
scale dependence of the renormalized constants $c_i$.
Consider how $c_0$ and $c_2$ behave when analyzed with a variety of 
regulator values in Figures \ref{fig:pdatac0} and \ref{fig:pdatac2}.
By choosing to use pseudodata produced at infinite volume, one eliminates 
 the concern that behaviour of the low energy constants across a range of
 regulators and pion masses is a finite volume artefact. The equivalents
of Figures \ref{fig:pdatac0} and \ref{fig:pdatac2} for finite volume
 pseudodata exhibit the same features.

If three pseudodata sets are compared, each with different upper bounds
on the range of $m_\pi^2$ considered in the fit,
 an increasing regulator dependence in $c_0$ and $c_2$ is seen 
further from the PCR.
A steep line indicates a strong scheme dependence in the result, 
and naturally occurs for data samples extending far outside the PCR.
Scheme independence will appear as a completely horizontal graph.
The latter
 is what one expects for a value of $m_{\pi,\ro{max}}^2 < 0.04$ GeV$^2$,
as shown in Figures \ref{fig:pdatac0} and \ref{fig:pdatac2}.
Note that, for each figure,
 all three curves (corresponding to different $m_{\pi,\ro{max}}^2$)
 arrive at stable values
for $c_0$ and $c_2$ on the righthand side of the graph (large $\La$).
To read off the values of $c_0$ and $c_2$ for large $\La$ is 
 tempting but wrong. It is known that the correct values of $c_0$ and $c_2$
are recovered at $\La = 1.0$ GeV, because at that value the pseudodata were 
created.

The analysis of the pseudodata in 
Figures \ref{fig:pdatac0} and \ref{fig:pdatac2} 
shows that as the value of $m_{\pi,\ro{max}}^2$ is changed, 
the correct value of $c_0$ is recovered at exactly $\La = \La_\ro{c}$, where
the curves intersect.
 This is also the intersection point for $c_2$ at $\La = \La_\ro{c}$.
This suggests that when considering lattice QCD results extending outside the
PCR, there may be an optimal finite-range cutoff.
Physically, such a cutoff would be associated with an intrinsic 
scale reflecting the finite
size of the source of the pion dressings.
Mathematically, this optimal cutoff is reflected by an independence of 
 the fit parameters on $m_{\pi,\ro{max}}^2$.

\begin{figure}[tp]
\includegraphics[height=0.76\hsize,angle=90]{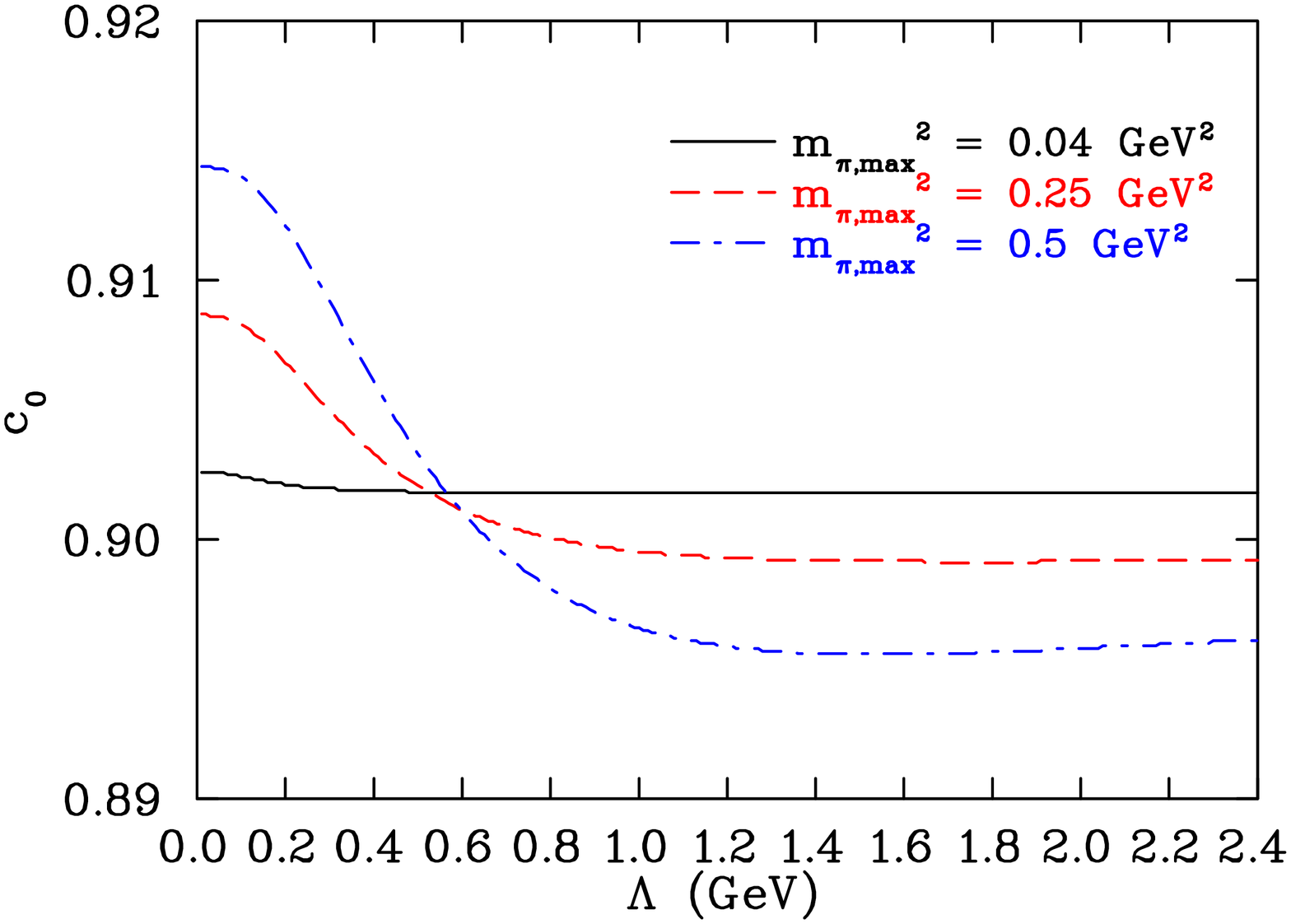}
\vspace{-12pt}
\caption{\footnotesize{(color online). Behaviour of $c_0$ vs.\ $\La$, based on infinite volume pseudodata created with a dipole regulator at $\La_\ro{c} = 1.0$ GeV but subsequently analyzed using a triple-dipole regulator. }}
\label{fig:pdatac0diptrip}
\includegraphics[height=0.76\hsize,angle=90]{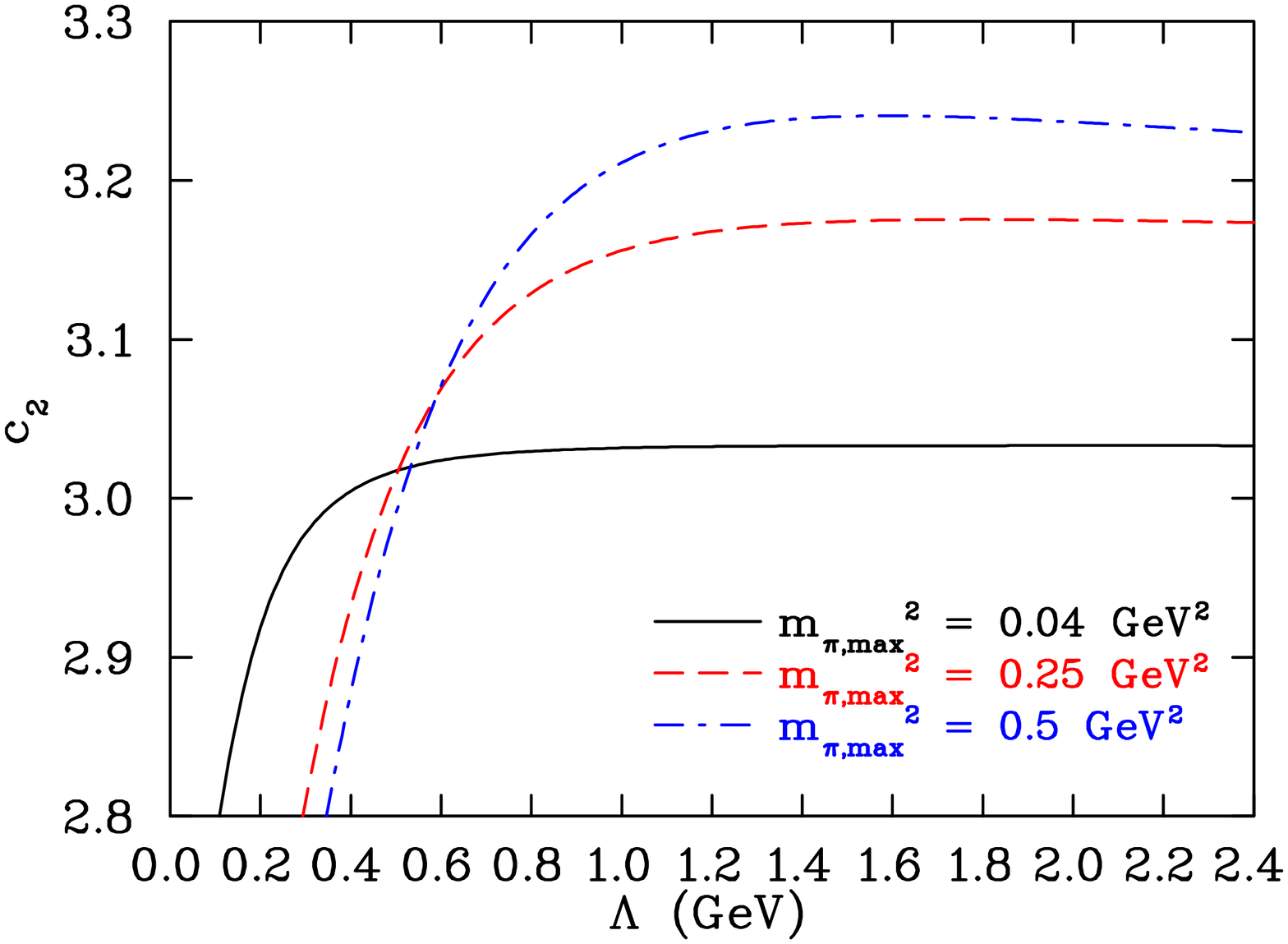}
\vspace{-12pt}
\caption{\footnotesize{(color online). Behaviour of $c_2$ vs.\ $\La$, based on infinite volume pseudodata created with a dipole regulator at $\La_\ro{c} = 1.0$ GeV but subsequently analyzed using a triple-dipole regulator.}}
\label{fig:pdatac2diptrip}
\end{figure}

To illustrate the non-triviality of this scale of curve-intersection, 
the pseudodata were analyzed with a different regulator, e.g. 
a triple-dipole regulator.
Figures \ref{fig:pdatac0diptrip} and \ref{fig:pdatac2diptrip}
show that the scale of the intersection is no longer a clear point, but a 
cluster centred about $0.5$ to $0.6$ GeV. 
%
The triple-dipole will of course predict
a different `best scale', since the shape of the regulator is different from
that of the dipole used to create the pseudodata. The
 essential point of this exercise is that clustering of curve 
intersections identifies a preferred 
renormalization scale that allows one to recover the correct low energy 
coefficients.
In this case, the crossing of the dash and dot-dash curves (from fitting) 
clearly identifies $\La^\ro{scale}_\ro{trip} = 0.6$ GeV 
as a preferred regulator,
 which reflects
the intrinsic scale used to create the data. Table \ref{table:pdata}
 compares the values
for $c_0$ and $c_2$ recovered in this analysis for two different regulators:
the preferred value $\La^\ro{scale}_\ro{trip} = 0.6$ GeV, and a 
large value $\La_\ro{trip} = 2.4$ GeV reflecting
the asymptotic result recovered from DR. The input values of $c_0$ and $c_2$
used to create the pseudodata are also indicated.
\begin{table*}[tp]
  \newcommand\T{\rule{0pt}{2.8ex}}
  \newcommand\B{\rule[-1.4ex]{0pt}{0pt}}
  \begin{center}
    \begin{tabular}{lllll}
      \hline
      \hline
      \T\B parameter \qquad &$\,\,$ input \quad  & $\La^\ro{scale}_\ro{trip} = 0.6$  $\,\,$& $\La_\ro{trip} = 2.4$ $\,\,$& $\La_\ro{trip} = 2.4$   \\
                     \qquad &$\,\,$  & $m_{\pi,\ro{max}}^2 = 0.25$ $\,\,$& $m_{\pi,\ro{max}}^2 = 0.25$  $\,\,$&  $m_{\pi,\ro{max}}^2 = 0.5$  \\
      \hline
      $c_0$   &\T $0.902$ & $0.901$ & $0.899$ & $0.896$ \\
      $c_2$   &\T $3.00$ & $3.07$ & $3.17$ & $3.23$ \\
      \hline
    \end{tabular}
  \end{center}
\vspace{-6pt}
  \caption{\footnotesize{A comparison of the parameters $c_0$ (GeV) and $c_2$ (GeV$^{-1}$) at their input value (pseudodata created with a dipole at $\La_\ro{c} = 1.0$ GeV) with the values when analysed with a triple dipole regulator. Different values of $\La_\ro{trip}$ (GeV) and $m_{\pi,\ro{max}}^2$ (GeV$^2$) are chosen to demonstrate the scheme dependence of $c_0$ and $c_2$ for data extending outside the PCR.}}
  \label{table:pdata}
\end{table*}

Note that the finite-range renormalization scheme
 breaks down if the finite-range regulator is too small. 
This is because $\Lambda$ must be large enough to include
the chiral physics being studied. 
 The exact value of a sensible lower bound in the 
finite-range regulator will depend on the functional form chosen
as regulator. This is estimated for three dipole-like regulators
 in Section (\ref{sect:scales}).

Figure \ref{fig:pdatac2} shows that 
 the renormalization for $c_2$ breaks down for small 
values of the regulator $\La$. FRR breaks down for a value of $\La_\ro{dip}$ 
much
below $0.6$ GeV, because the coefficients $b_i$ of the loop integral expansion
in Eqs.~(\ref{eqn:NNexpn}), (\ref{eqn:NDeexpn}) and (\ref{eqn:tadexpnSi})
 are proportional to $\La^{(3-i)}$. For high-order terms with large $i$, 
the coefficients  will become large when $\La$ is small. This will adversely
 affect 
the convergence properties of the chiral expansion. One obtains a residual
series expansion with good convergence properties only when $\La$ reflects
the intrinsic scale of the source of the pion dressings of the hadron 
in question.

 The pseudodata analysis provides a good indication of a lower bound
 for $\La$ using a dipole 
regulator: $\La_{\ro{dip}} \gtrsim 0.6$ GeV. Similarly, 
Figure \ref{fig:pdatac2diptrip} suggests a lower bound for the triple dipole
 regulator: $\La_{\ro{trip}} \gtrsim 0.3$ GeV. The same analysis can be 
repeated for the double dipole regulator 
to obtain $\La_{\ro{doub}} \gtrsim 0.4$ GeV.

One may also constrain the lowest value
that $\La$ can take by considering phenomenological arguments.
Based on the physical values of the sigma
commutator and the nucleon mass, a pion mass of
 $m_\pi \approx 0.5$ GeV bounds the radius of convergence 
\cite{Borasoy:2002jv,Young:2009ub,Young:2009zb}.
Therefore, in order to ensure the inclusion of important contributions
to the chiral physics, one should choose an energy scale 
$\La_{\ro{sharp}} \sim 0.5$ GeV for a sharp cutoff (step function) 
regulator.  
 To compare this estimate 
for the sharp cutoff to that of dipole-like regulators, 
one can calculate the regulator value required
such that $u^2_{n}(k^2) = 1/2$ when the momentum takes the 
energy scale of $\La_\ro{sharp}$.
This results in a rough estimate for a sensible value 
for the dipole, double dipole and triple dipole.
These values are $\La_{\ro{dip}} \sim 1.1$ GeV, 
$\La_{\ro{doub}} \sim 0.76$ GeV and 
$\La_{\ro{trip}} \sim 0.66$ GeV, respectively.
In any event, a wide range of regulator values will be considered, and
the intersections of the curves for the 
low energy coefficients will be used 
in order to construct fits outside the PCR.
This will be done in order to identify the presence of an intrinsic scale
for the pion source and an associated preferred regularization scale.

%



%
\section{Intrinsic Scale: Lattice Results}
\label{sect:scales}
\subsection{Evidence for an Intrinsic Scale}
\label{subsect:ev}

In the example of the pseudodata, an optimal finite-range cutoff
was obtained from the data themselves. Clearly, the pseudodata
have an \emph{intrinsic scale}: the renormalization scale $\La_\ro{c}$ at 
which they were created.
This test example leads the researcher
 to wonder if actual lattice QCD data 
 have an intrinsic cutoff scale embedded within them. That is, by
 analysing lattice QCD data
in the same way as the pseudodata, can a similar intersection
point be obtained from the renormalization scale flow of $c_0$ and $c_2$?
If so, this indicates that the lattice QCD data contain information
regarding an optimal finite-range regularization scale, 
which can be calculated.

The results for the renormalization 
of $c_0$ and $c_2$ as a 
function of $\La$ are now presented 
for JLQCD \cite{Ohki:2008ff}, PACS-CS \cite{Aoki:2008sm} 
and CP-PACS \cite{AliKhan:2001tx} lattice QCD data. The JLQCD data use 
 overlap fermions in two-flavor QCD, but the lattice box size for each
data point is $\sim 1.9$ fm, smaller than the other two data sets.
 The PACS-CS data use the nonperturbatively $\ca{O}(a)$-improved
 Wilson quark action at a lattice box size of $\sim 2.9$ fm, but the data set
only contains five data points and a large statistical error in the smallest 
 $m_\pi^2$ point. The CP-PACS data use a mean field improved clover quark 
action on lattice box sizes for each data point varying from $\sim 2.2$ fm
 to $\sim 2.8$ fm.

\begin{figure}[tp]
\includegraphics[height=0.76\hsize,angle=90]{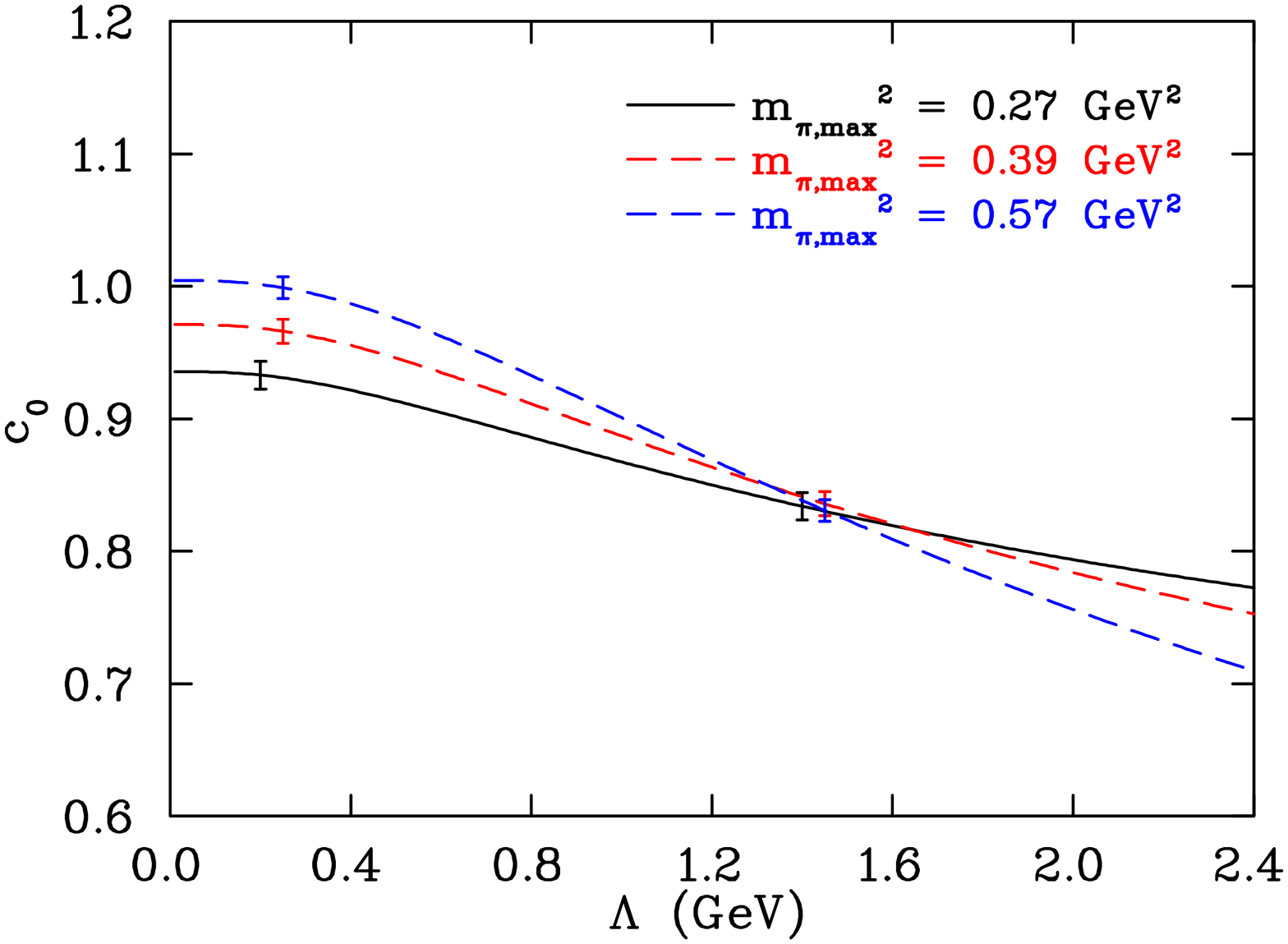}
\vspace{-12pt}
\caption{\footnotesize{(color online). Behaviour of $c_0$ vs.\ $\La$, based on JLQCD data. The chiral expansion is taken to order $\ca{O}(m_\pi^3)$, and a dipole regulator is used. A few points are selected to indicate the general size of the statistical error bars.}}
\label{fig:Ohkic0truncDIP}
\includegraphics[height=0.76\hsize,angle=90]{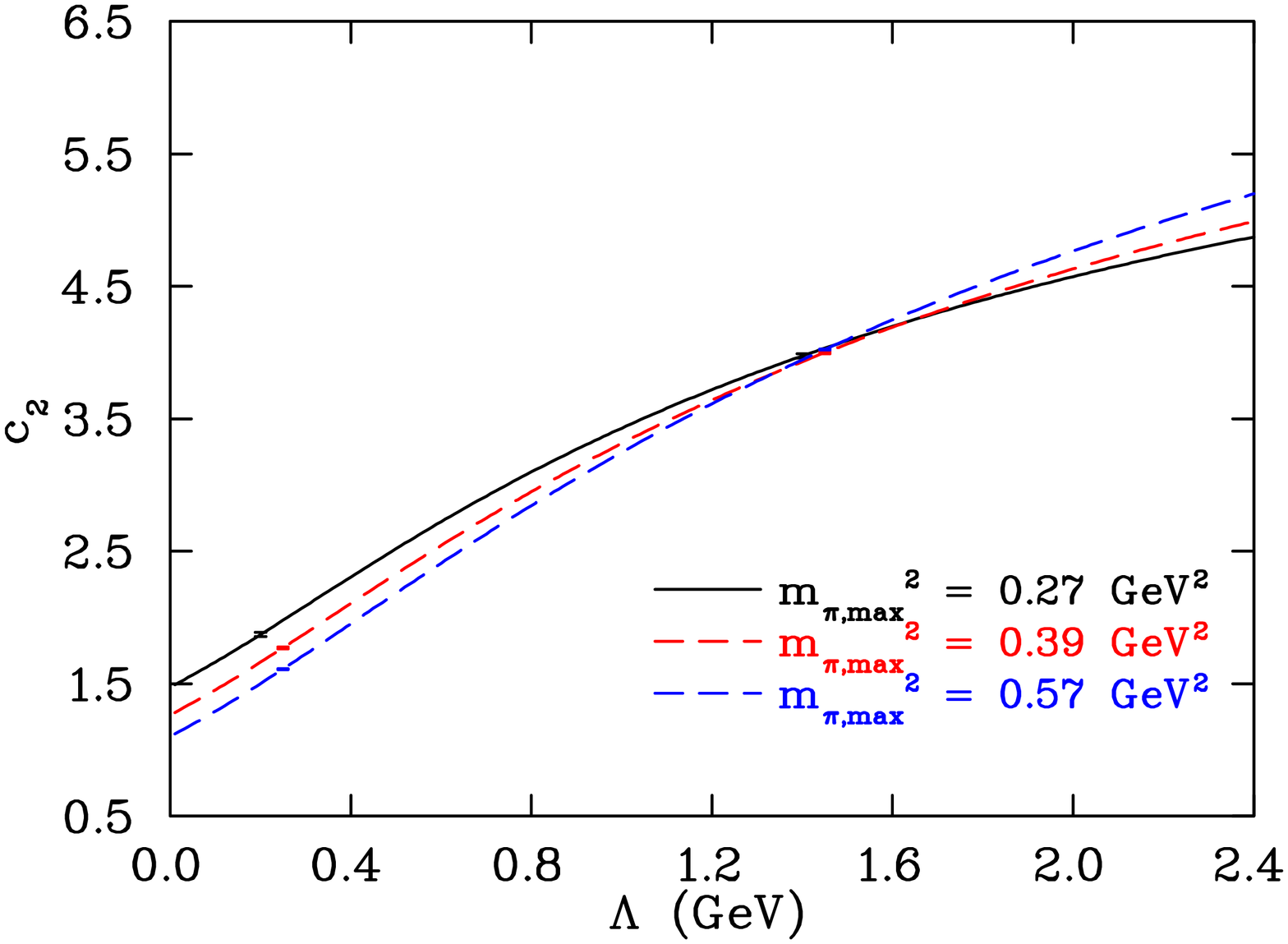}
\vspace{-12pt}
\caption{\footnotesize{(color online). Behaviour of $c_2$ vs.\ $\La$, based on JLQCD data. The chiral expansion is taken to order $\ca{O}(m_\pi^3)$, and a dipole regulator is used. A few points are selected to indicate the general size of the statistical error bars.}}
\label{fig:Ohkic2truncDIP}
\includegraphics[height=0.76\hsize,angle=90]{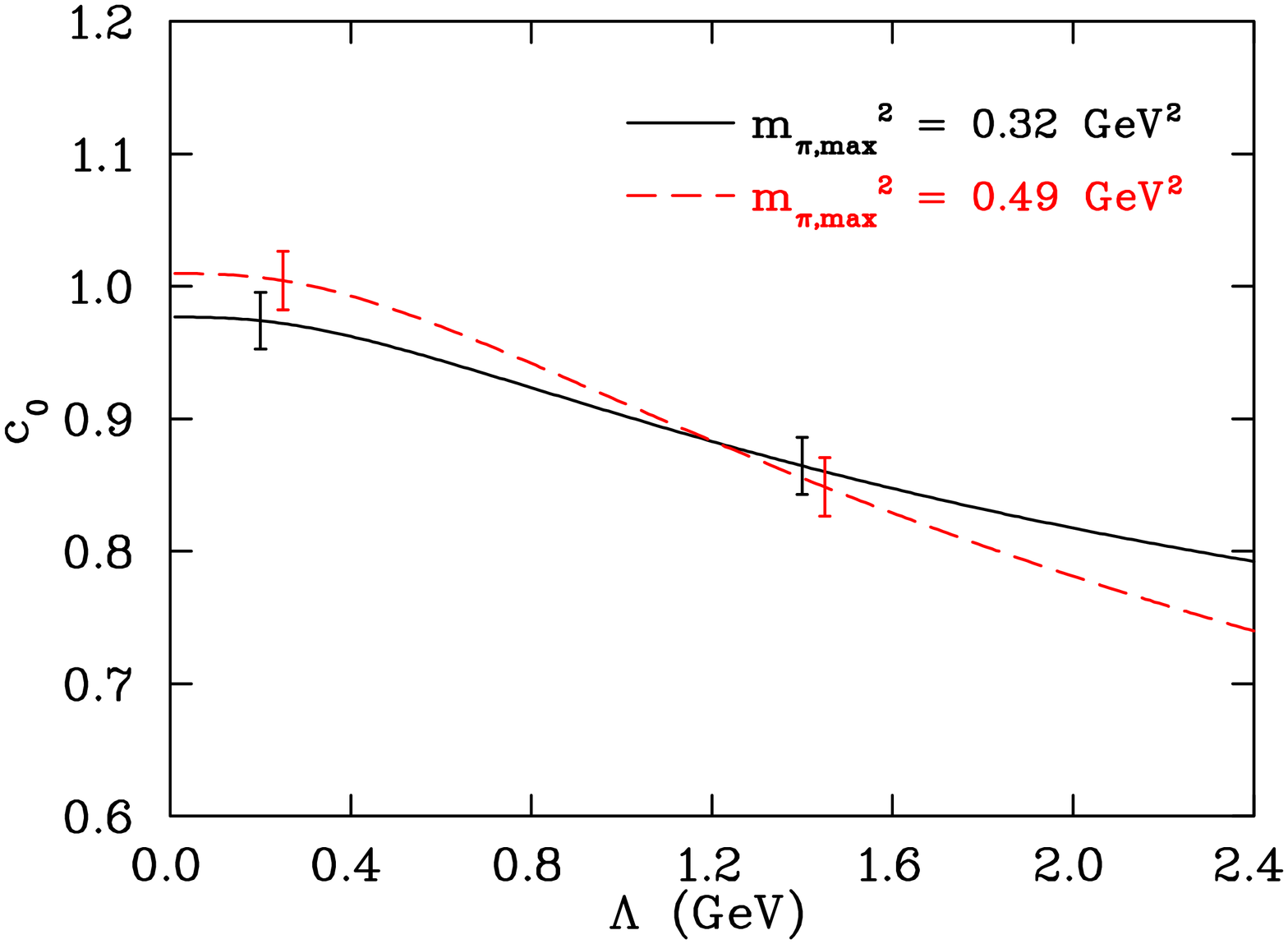}
\vspace{-12pt}
\caption{\footnotesize{(color online). Behaviour of $c_0$ vs.\ $\La$, based on PACS-CS data. The chiral expansion is taken to order $\ca{O}(m_\pi^3)$, and a dipole regulator is used. A few points are selected to indicate the general size of the statistical error bars.}}
\label{fig:Aokic0truncDIP}
\includegraphics[height=0.76\hsize,angle=90]{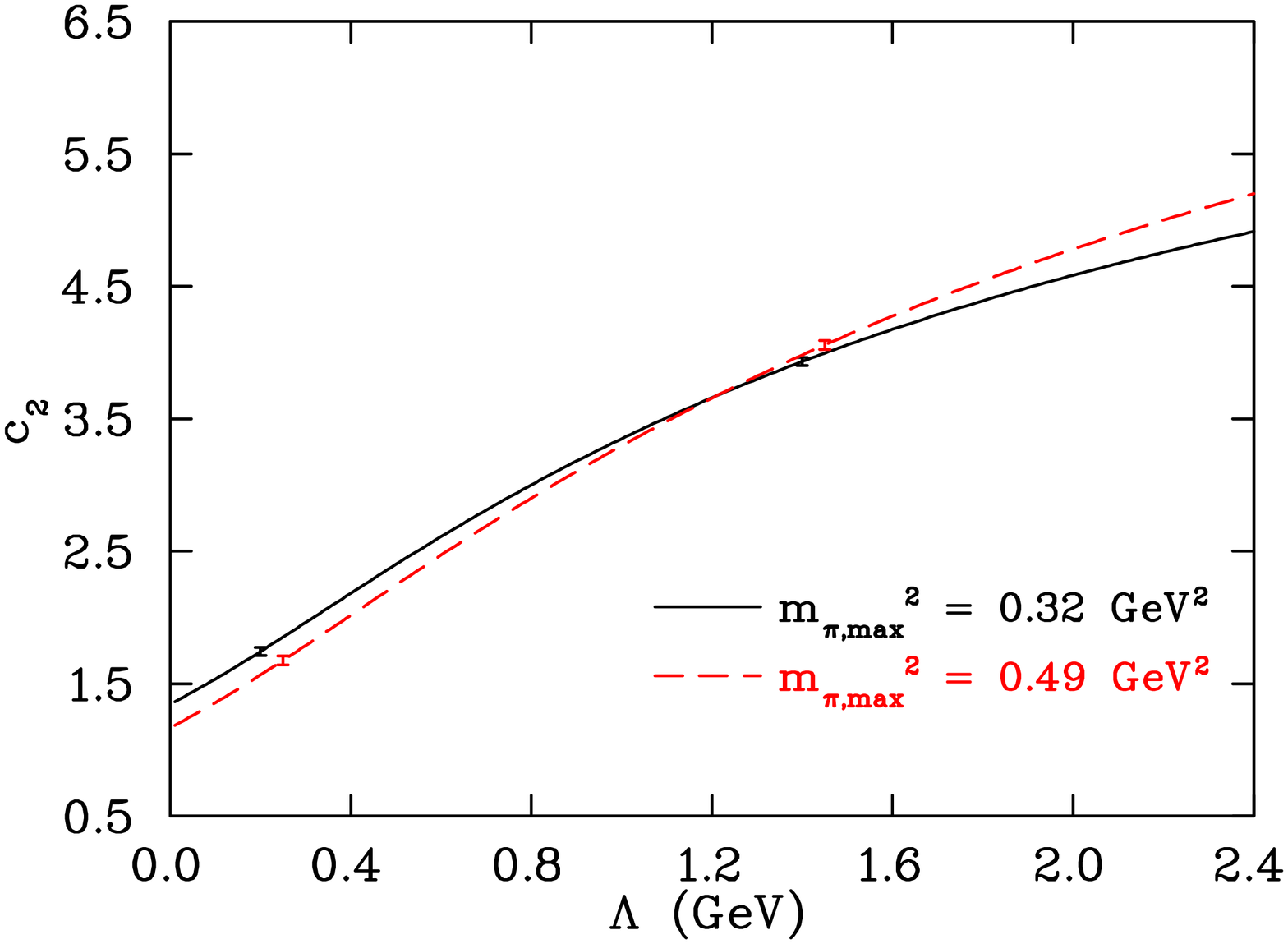}
\vspace{-12pt}
\caption{\footnotesize{(color online). Behaviour of $c_2$ vs.\ $\La$, based on PACS-CS data. The chiral expansion is taken to order $\ca{O}(m_\pi^3)$, and a dipole regulator is used. A few points are selected to indicate the general size of the statistical error bars.}}
\label{fig:Aokic2truncDIP}
\end{figure}

\begin{figure}[tp]
\includegraphics[height=0.76\hsize,angle=90]{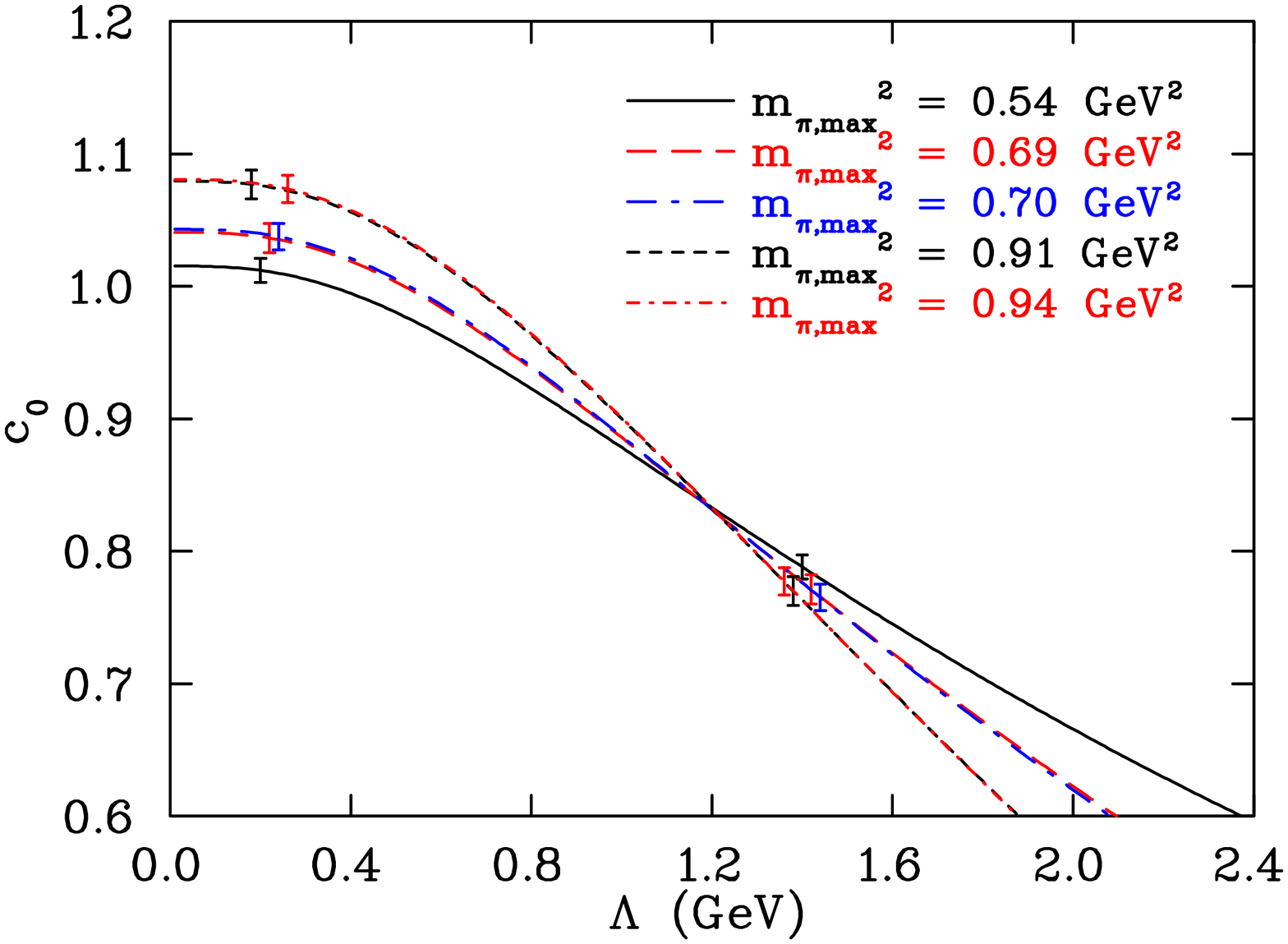}
\vspace{-12pt}
\caption{\footnotesize{(color online). Behaviour of $c_0$ vs.\ $\La$, based on CP-PACS data. The chiral expansion is taken to order $\ca{O}(m_\pi^3)$, and a dipole regulator is used. A few points are selected to indicate the general size of the statistical error bars.}}
\label{fig:Youngc0truncDIP}
\includegraphics[height=0.76\hsize,angle=90]{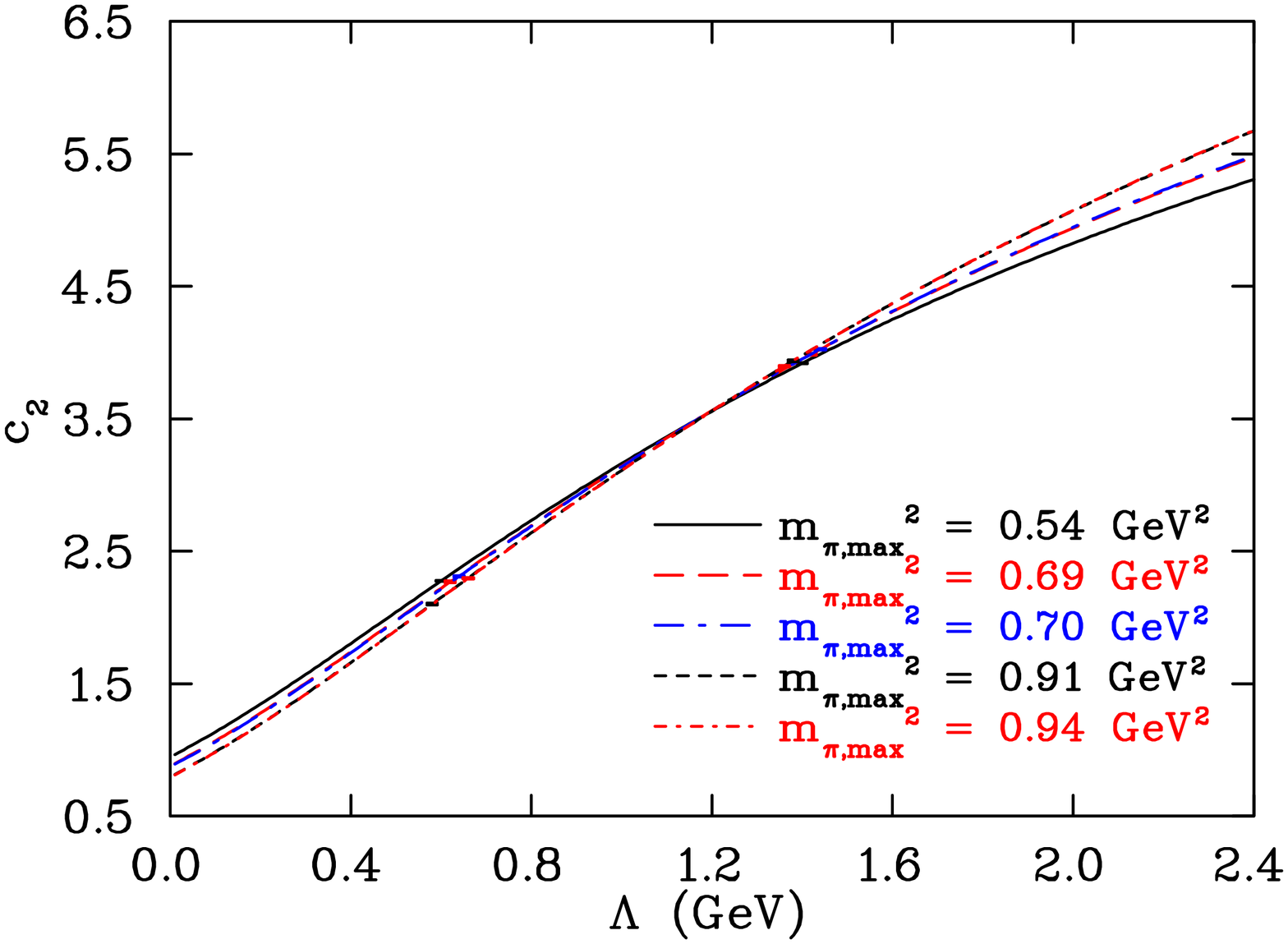}
\vspace{-12pt}
\caption{\footnotesize{(color online). Behaviour of $c_2$ vs.\ $\La$, based on CP-PACS data. The chiral expansion is taken to order $\ca{O}(m_\pi^3)$, and a dipole regulator is used. A few points are selected to indicate the general size of the statistical error bars.}}
\label{fig:Youngc2truncDIP}
\end{figure}

\begin{figure}[tp]
\includegraphics[height=0.76\hsize,angle=90]{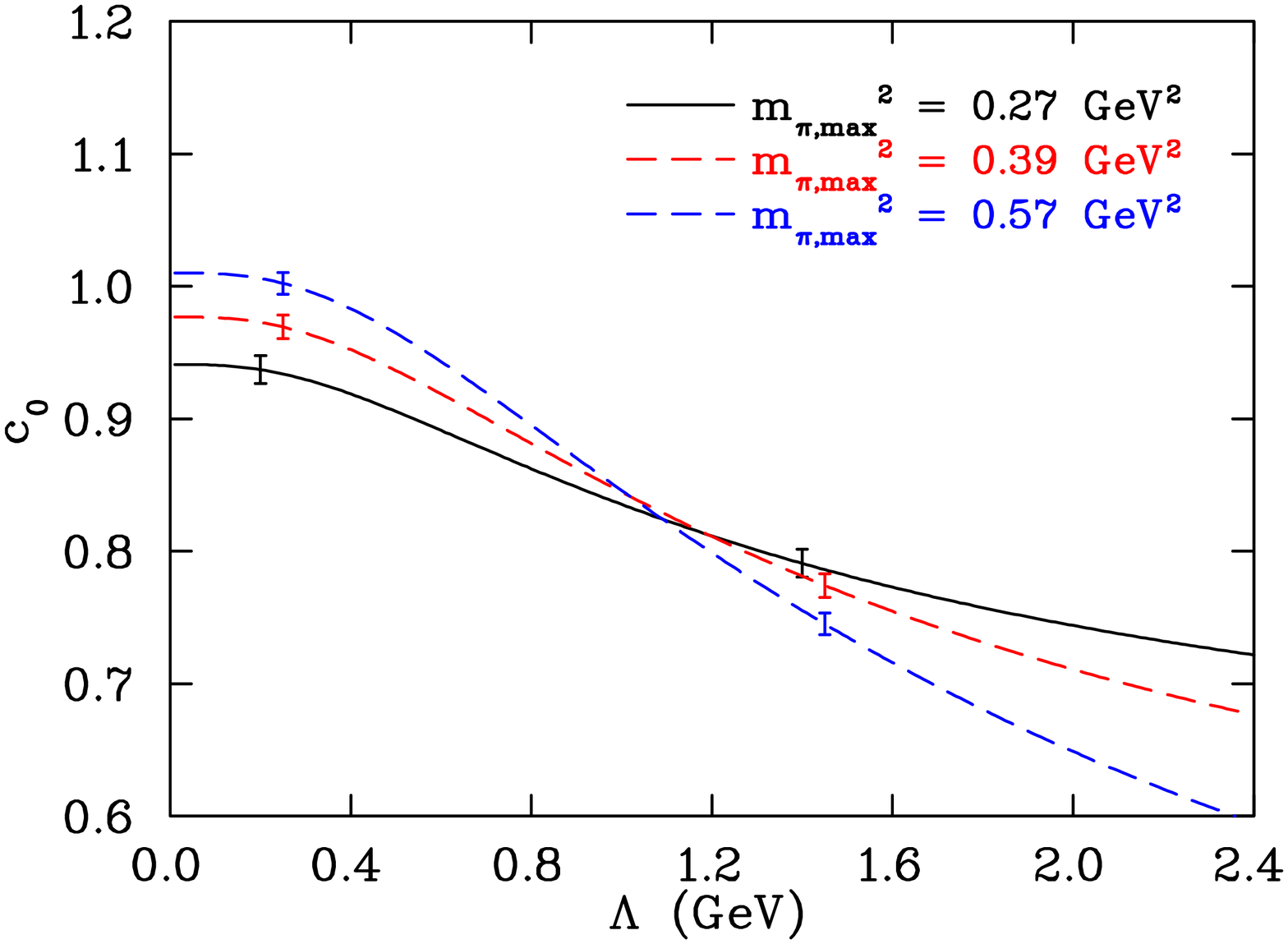}
\vspace{-12pt}
\caption{\footnotesize{(color online). Behaviour of $c_0$ vs.\ $\La$, based on JLQCD data. The chiral expansion is taken to order $\ca{O}(m_\pi^3)$ and a double dipole regulator is used. A few points are selected to indicate the general size of the statistical error bars.}}
\label{fig:Ohkic0truncDOUB}
\includegraphics[height=0.76\hsize,angle=90]{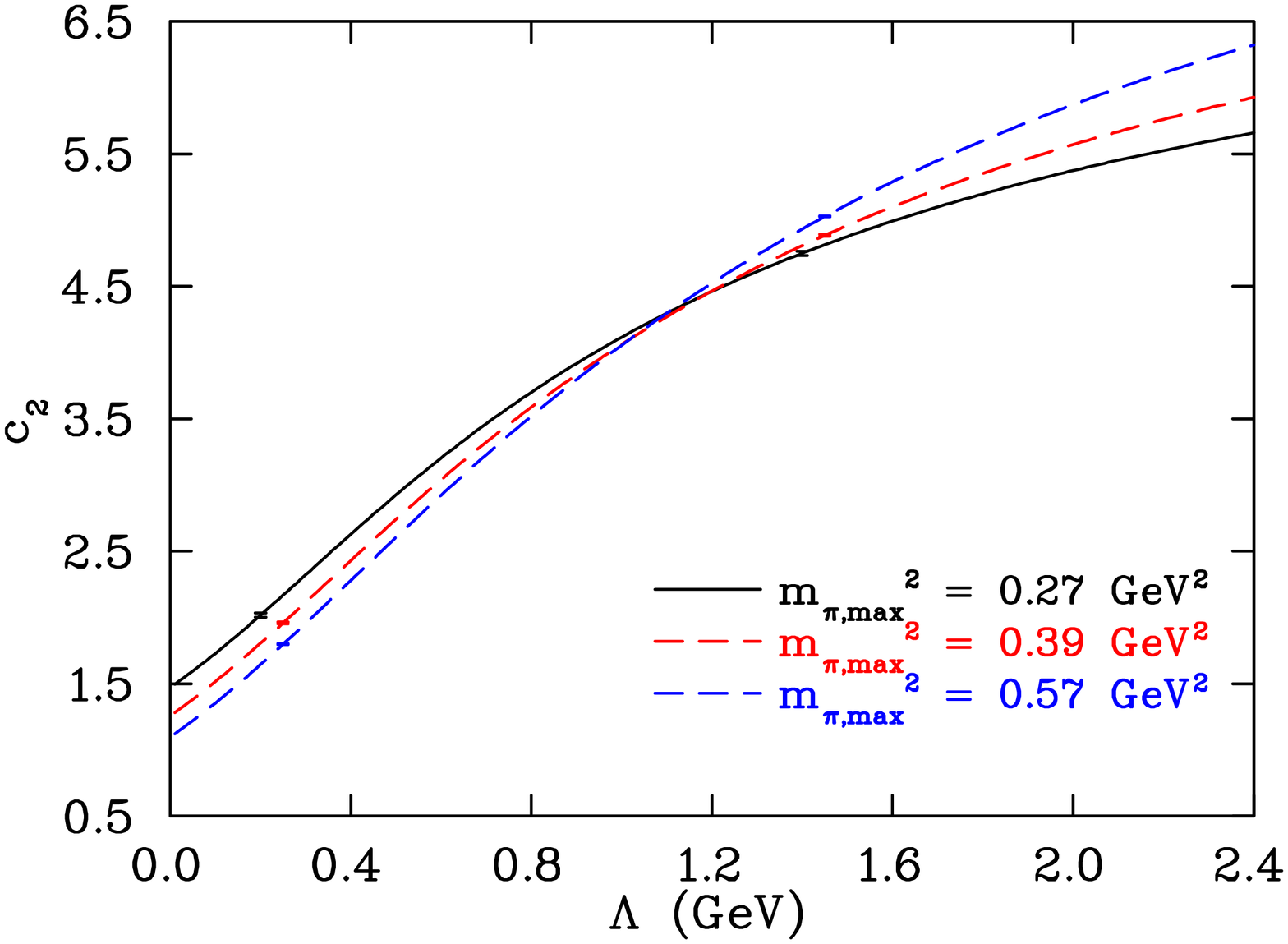}
\vspace{-12pt}
\caption{\footnotesize{(color online). Behaviour of $c_2$ vs.\ $\La$, based on JLQCD data. The chiral expansion is taken to order $\ca{O}(m_\pi^3)$ and a double dipole regulator is used. A few points are selected to indicate the general size of the statistical error bars.}}
\label{fig:Ohkic2truncDOUB}
\includegraphics[height=0.76\hsize,angle=90]{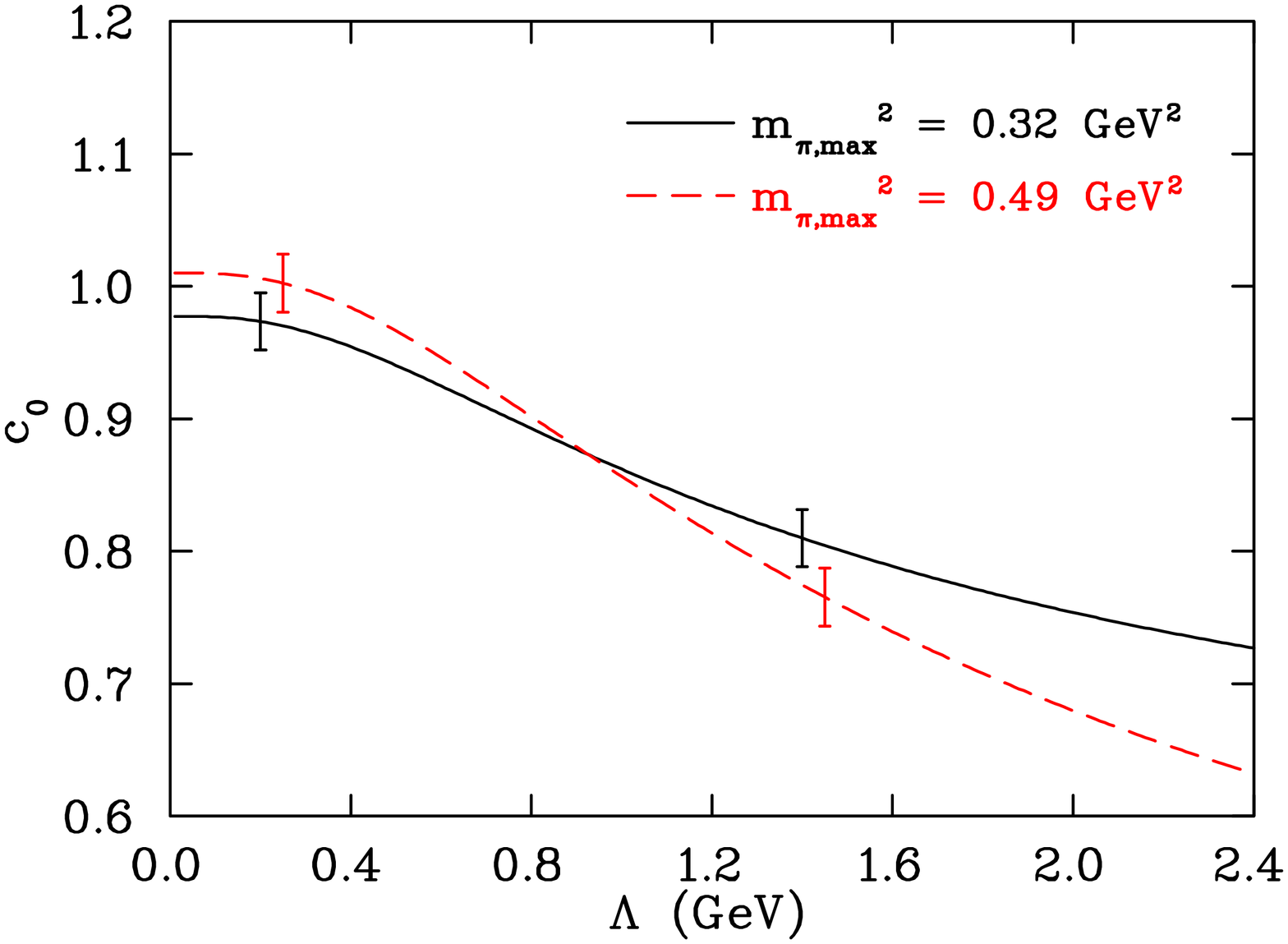}
\vspace{-12pt}
\caption{\footnotesize{(color online). Behaviour of $c_0$ vs.\ $\La$, based on PACS-CS data. The chiral expansion is taken to order $\ca{O}(m_\pi^3)$ and a double dipole regulator is used. A few points are selected to indicate the general size of the statistical error bars.}}
\label{fig:Aokic0truncDOUB}
\includegraphics[height=0.76\hsize,angle=90]{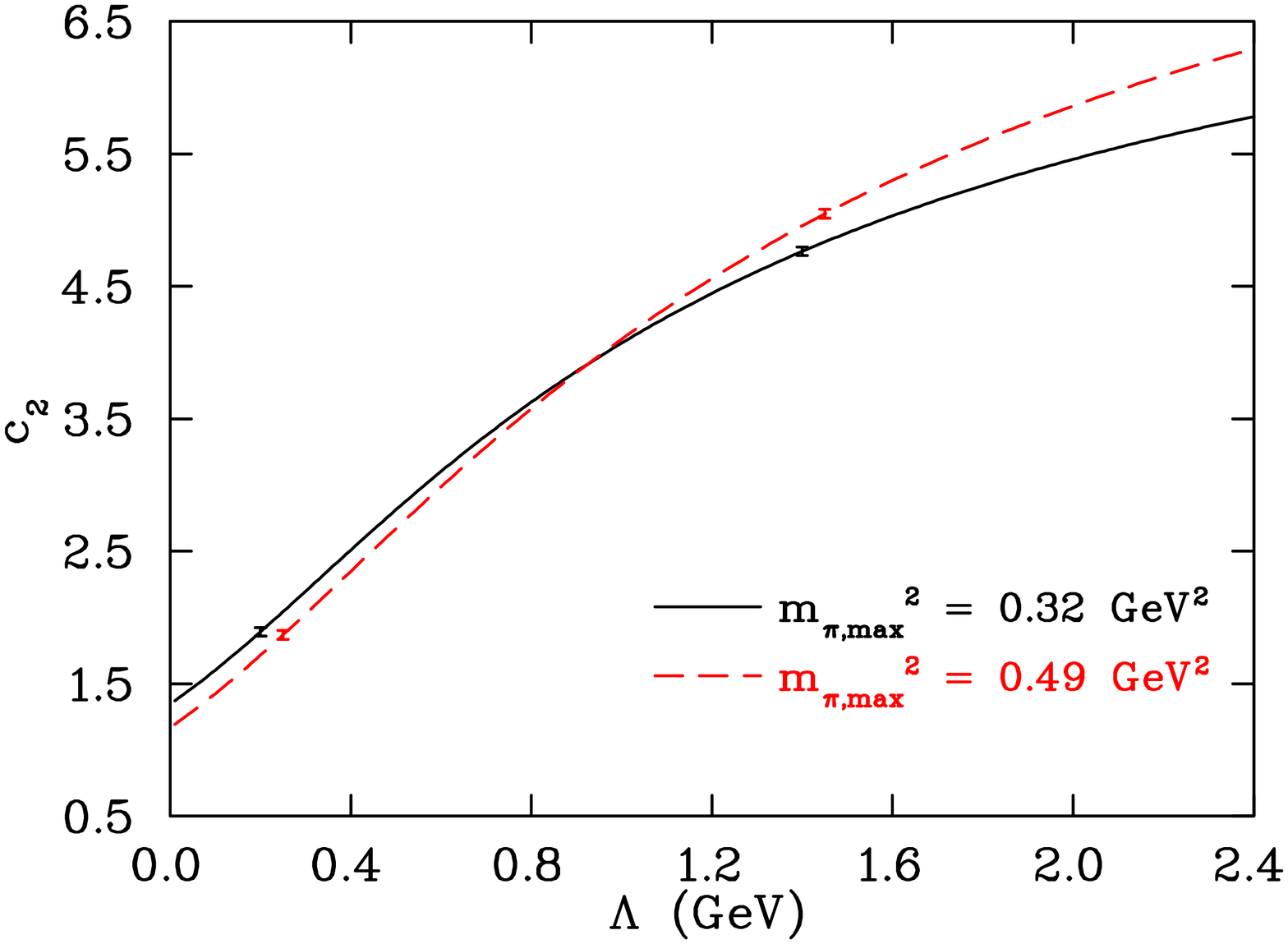}
\vspace{-12pt}
\caption{\footnotesize{(color online). Behaviour of $c_2$ vs.\ $\La$, based on PACS-CS data. The chiral expansion is taken to order $\ca{O}(m_\pi^3)$ and a double dipole regulator is used. A few points are selected to indicate the general size of the statistical error bars.}}
\label{fig:Aokic2truncDOUB}
\end{figure}

\begin{figure}[tp]
\includegraphics[height=0.76\hsize,angle=90]{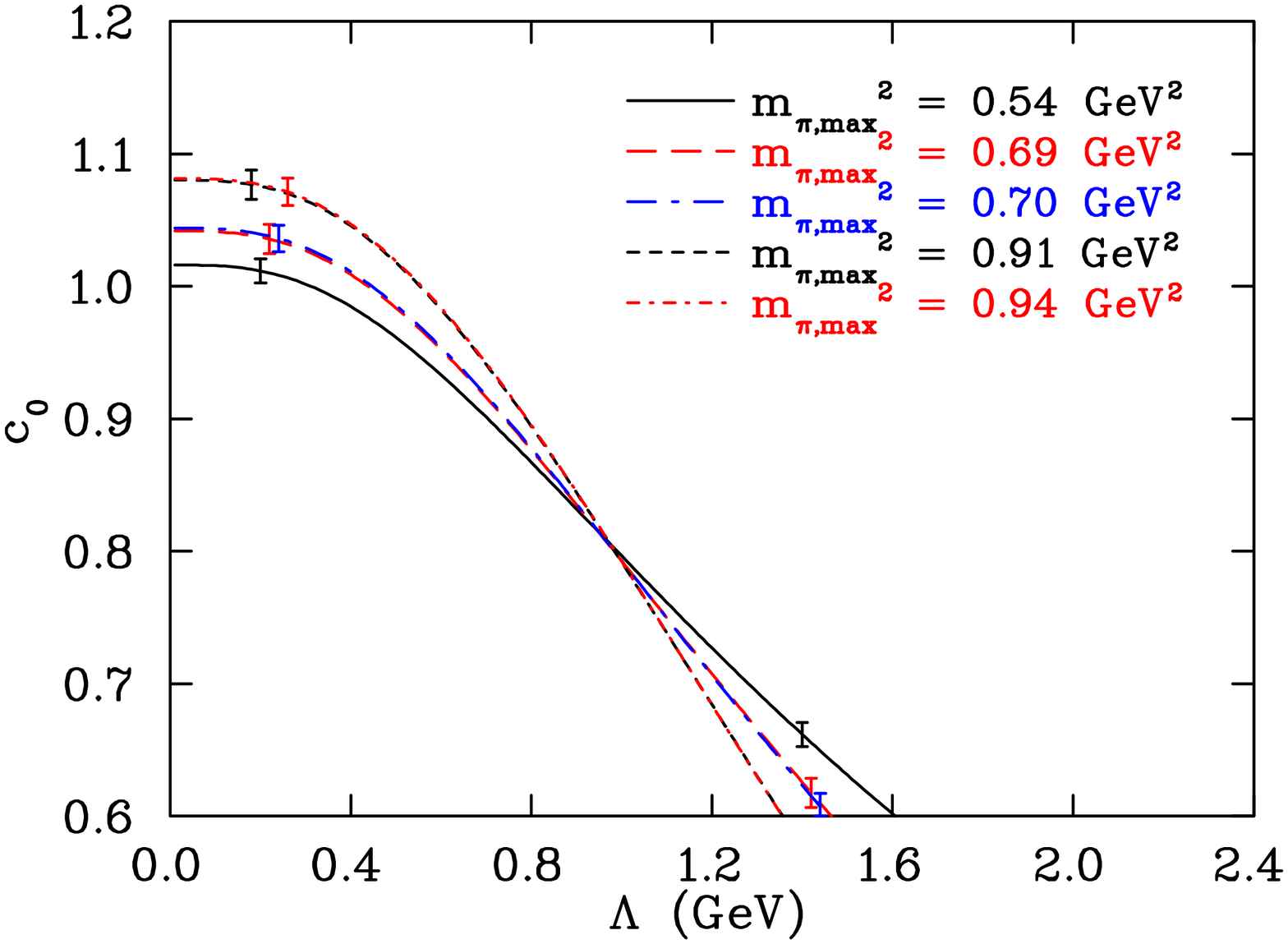}
\vspace{-12pt}
\caption{\footnotesize{(color online). Behaviour of $c_0$ vs.\ $\La$, based on CP-PACS data. The chiral expansion is taken to order $\ca{O}(m_\pi^3)$ and a double dipole regulator is used. A few points are selected to indicate the general size of the statistical error bars.}}
\label{fig:Youngc0truncDOUB}
\includegraphics[height=0.76\hsize,angle=90]{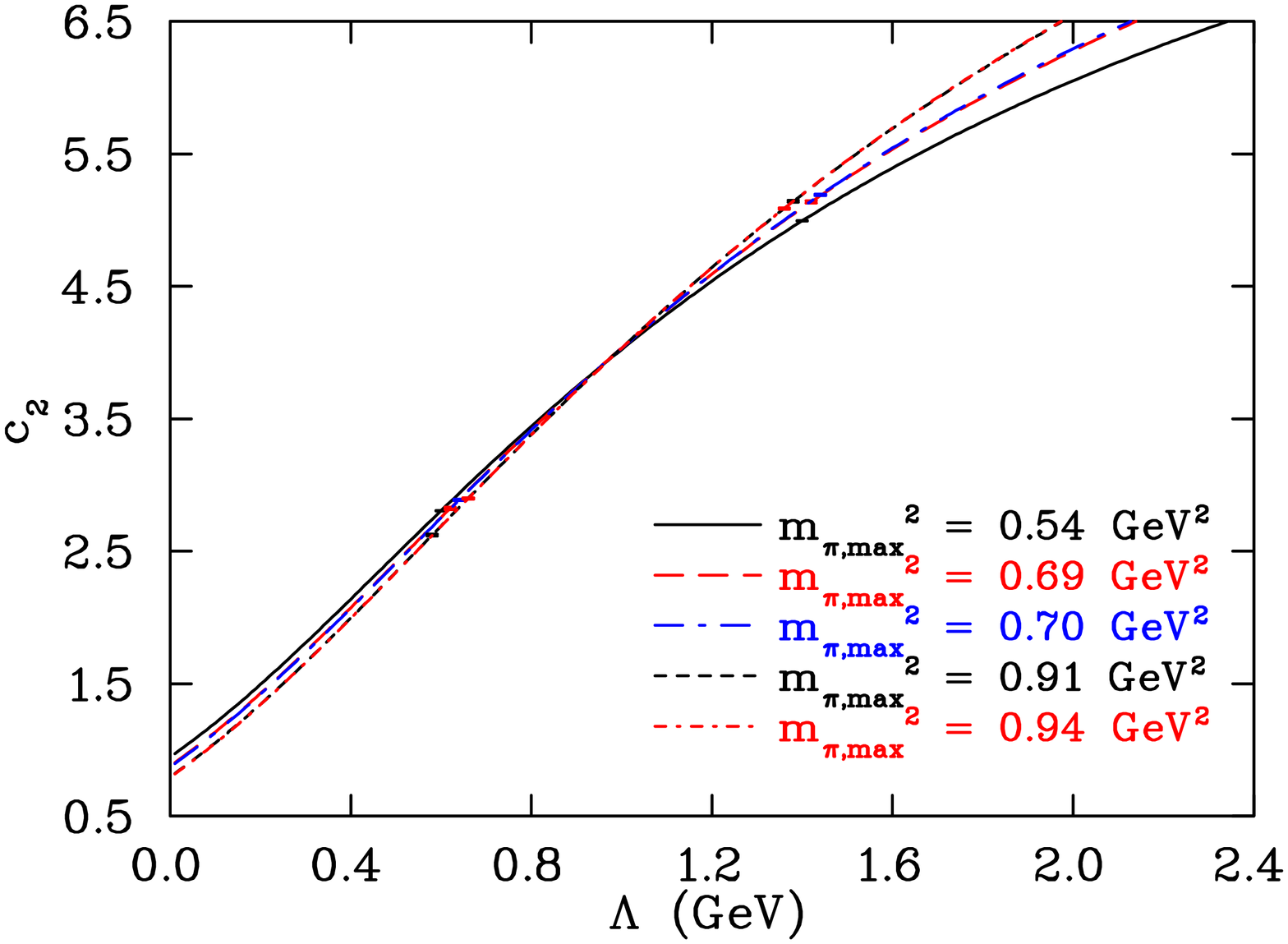}
\vspace{-12pt}
\caption{\footnotesize{(color online). Behaviour of $c_2$ vs.\ $\La$, based on CP-PACS data. The chiral expansion is taken to order $\ca{O}(m_\pi^3)$ and a double dipole regulator is used. A few points are selected to indicate the general size of the statistical error bars.}}
\label{fig:Youngc2truncDOUB}
\end{figure}

\begin{figure}[tp]
\includegraphics[height=0.76\hsize,angle=90]{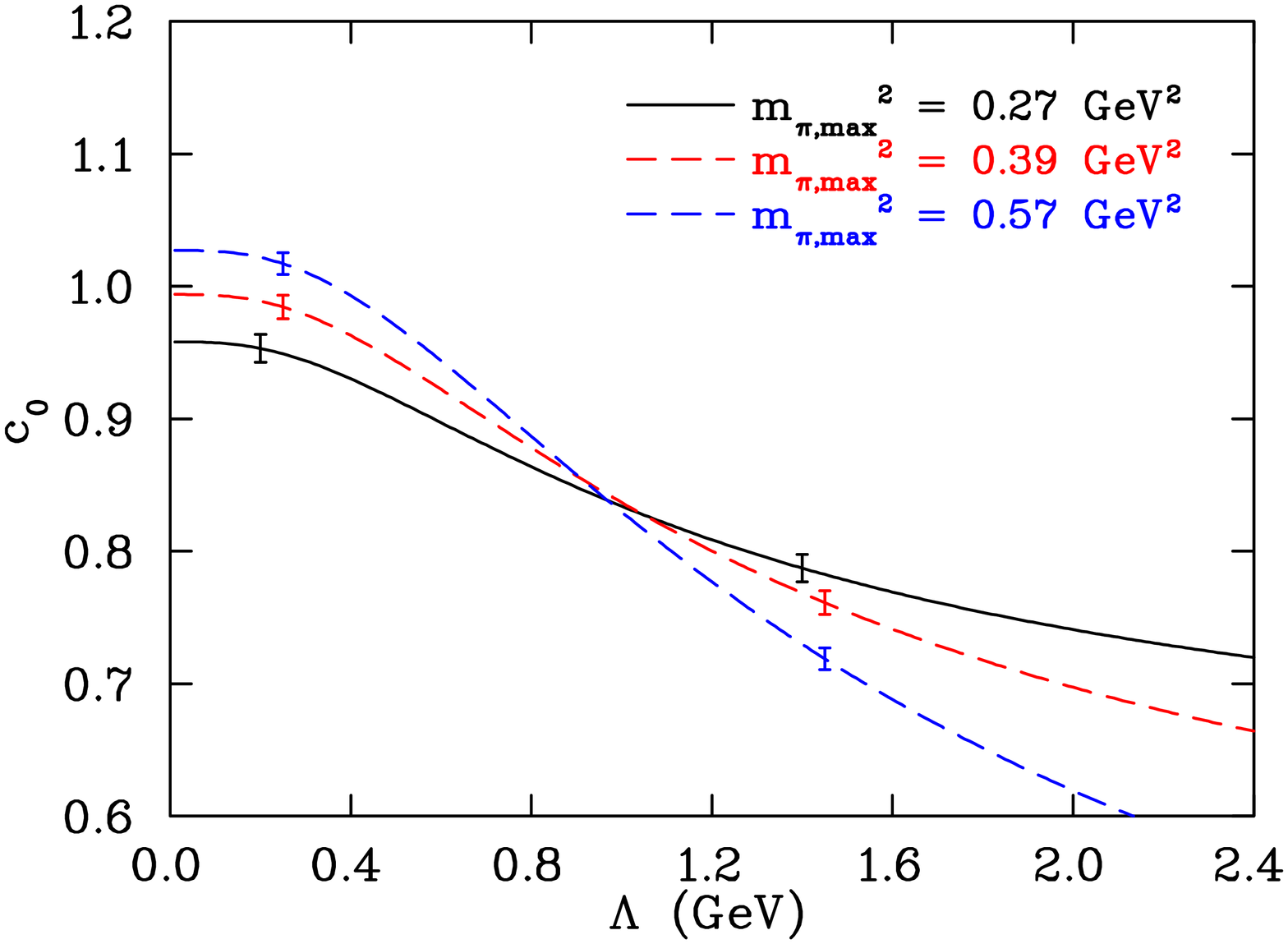}
\vspace{-12pt}
\caption{\footnotesize{(color online). Behaviour of $c_0$ vs.\ $\La$, based on JLQCD data. The chiral expansion is taken to order $\ca{O}(m_\pi^3)$ and a triple dipole regulator is used. A few points are selected to indicate the general size of the statistical error bars.}}
\label{fig:Ohkic0truncTRIP}
\includegraphics[height=0.76\hsize,angle=90]{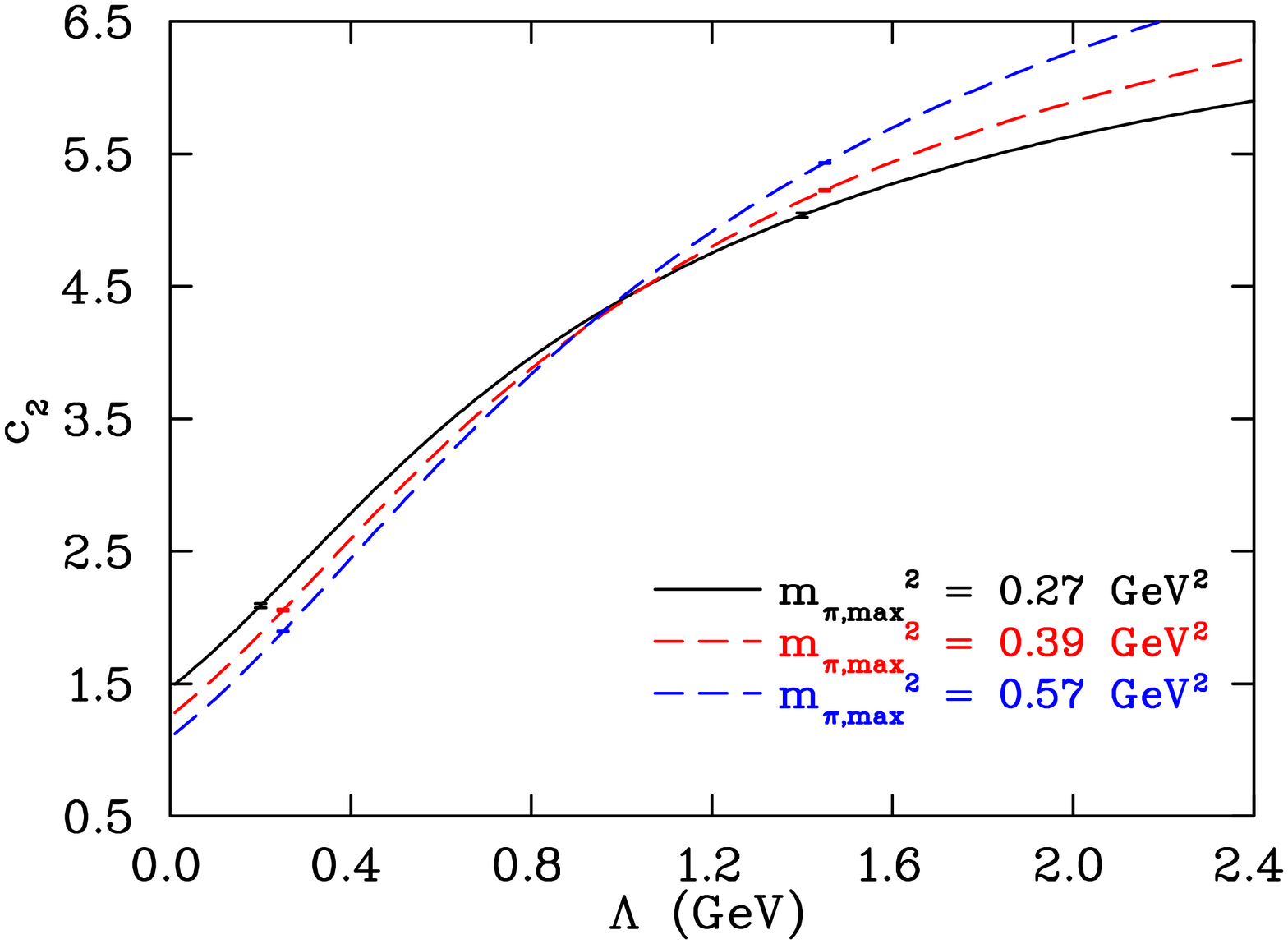}
\vspace{-12pt}
\caption{\footnotesize{(color online). Behaviour of $c_2$ vs.\ $\La$, based on JLQCD data. The chiral expansion is taken to order $\ca{O}(m_\pi^3)$ and a triple dipole regulator is used. A few points are selected to indicate the general size of the statistical error bars.}}
\label{fig:Ohkic2truncTRIP}
\includegraphics[height=0.76\hsize,angle=90]{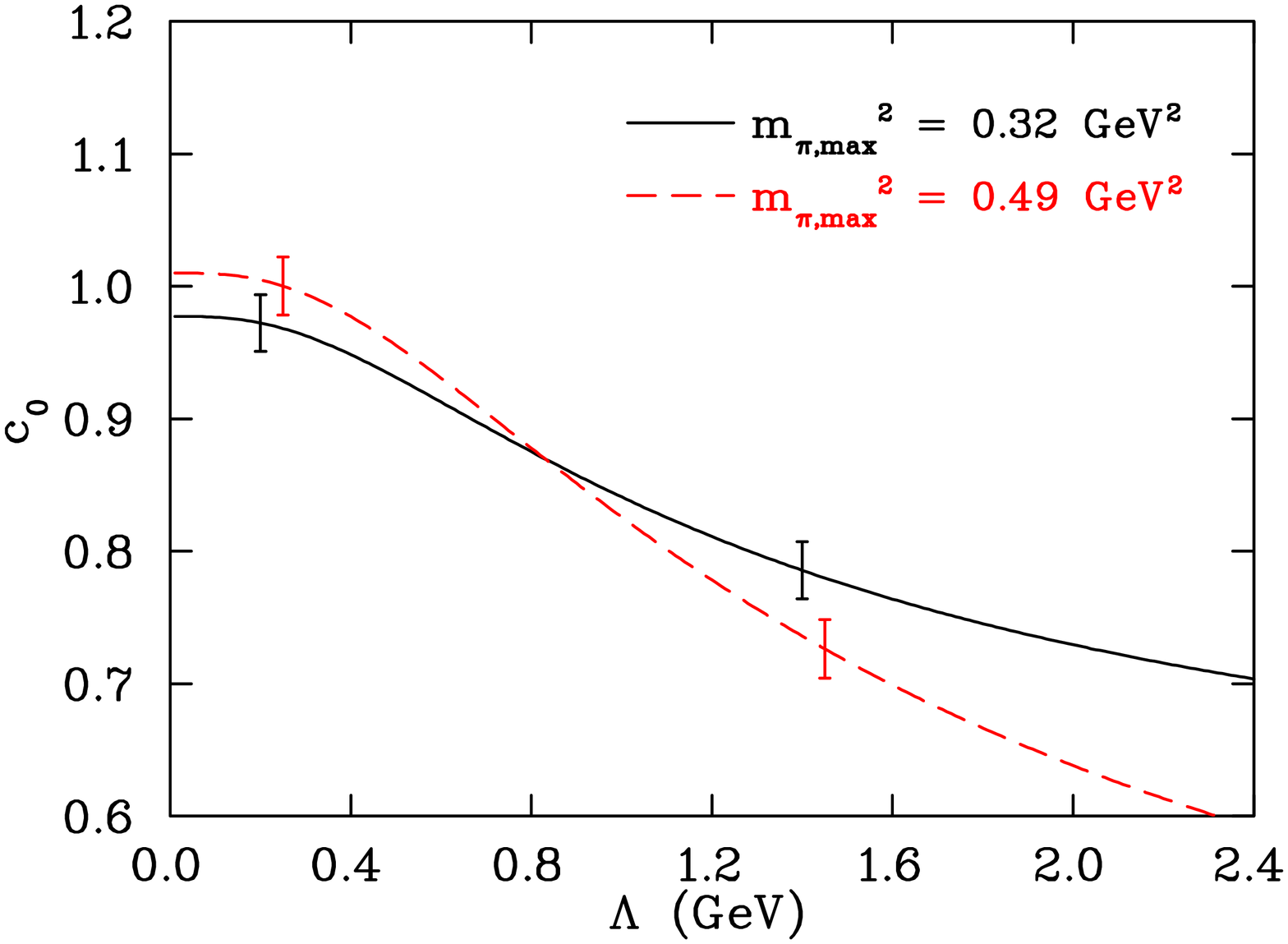}
\vspace{-12pt}
\caption{\footnotesize{(color online). Behaviour of $c_0$ vs.\ $\La$, based on PACS-CS data. The chiral expansion is taken to order $\ca{O}(m_\pi^3)$ and a triple dipole regulator is used. A few points are selected to indicate the general size of the statistical error bars.}}
\label{fig:Aokic0truncTRIP}
\includegraphics[height=0.76\hsize,angle=90]{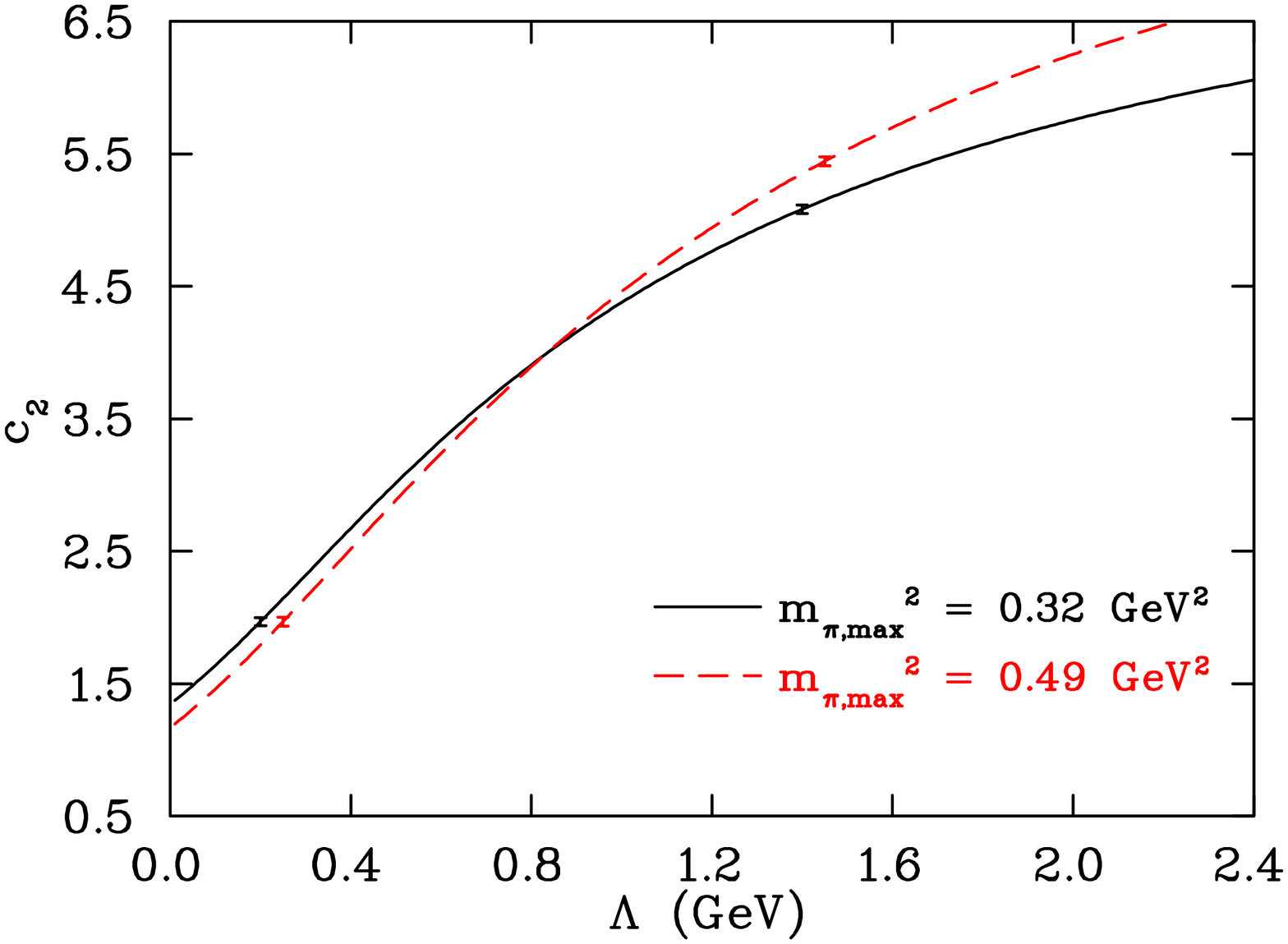}
\vspace{-12pt}
\caption{\footnotesize{(color online). Behaviour of $c_2$ vs.\ $\La$, based on PACS-CS data. The chiral expansion is taken to order $\ca{O}(m_\pi^3)$ and a triple dipole regulator is used. A few points are selected to indicate the general size of the statistical error bars.}}
\label{fig:Aokic2truncTRIP}
\end{figure}

\begin{figure}[tp]
\includegraphics[height=0.76\hsize,angle=90]{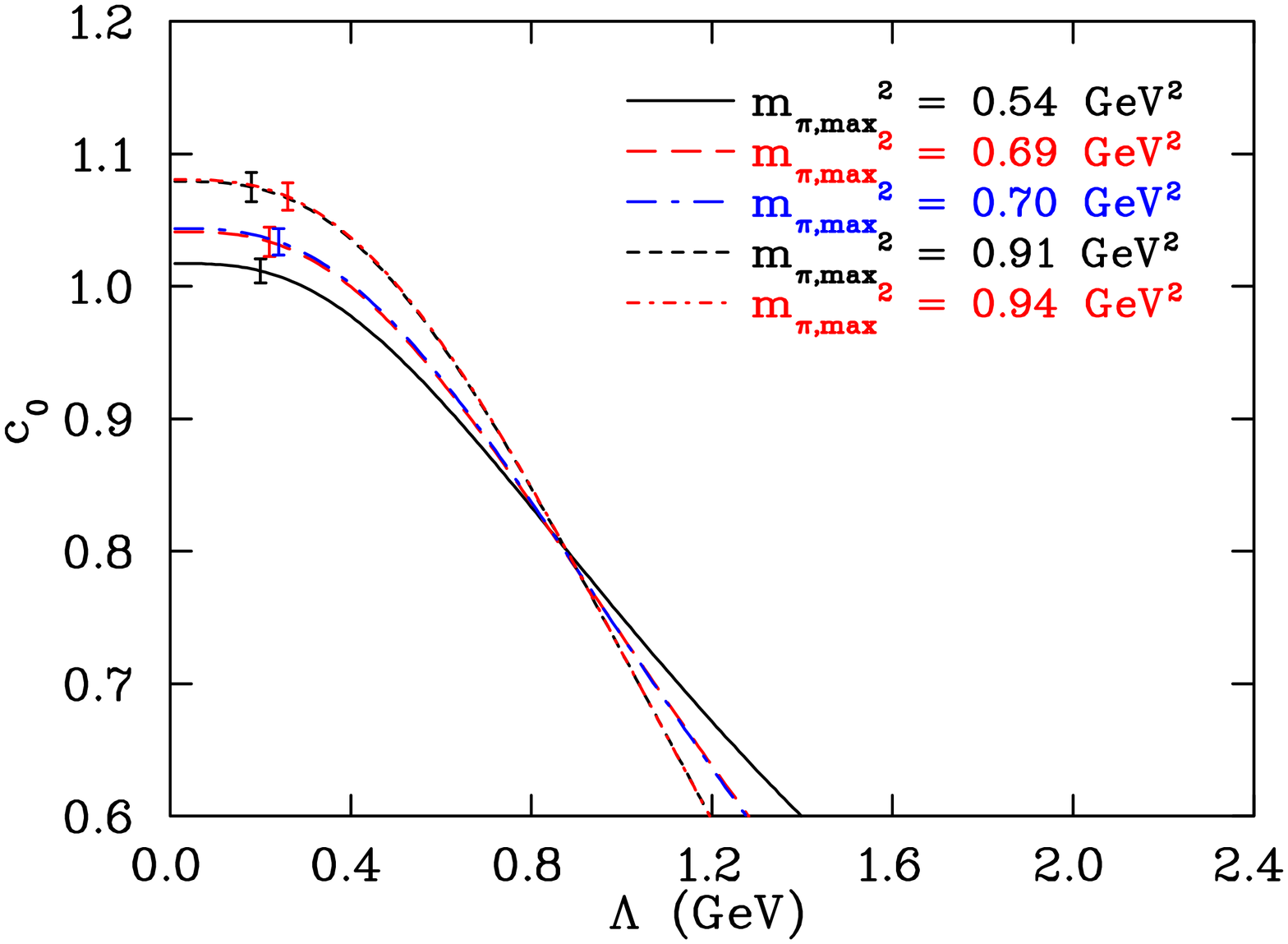}
\vspace{-12pt}
\caption{\footnotesize{(color online). Behaviour of $c_0$ vs.\ $\La$, based on CP-PACS data. The chiral expansion is taken to order $\ca{O}(m_\pi^3)$ and a triple dipole regulator is used. A few points are selected to indicate the general size of the statistical error bars.}}
\label{fig:Youngc0truncTRIP}
\includegraphics[height=0.76\hsize,angle=90]{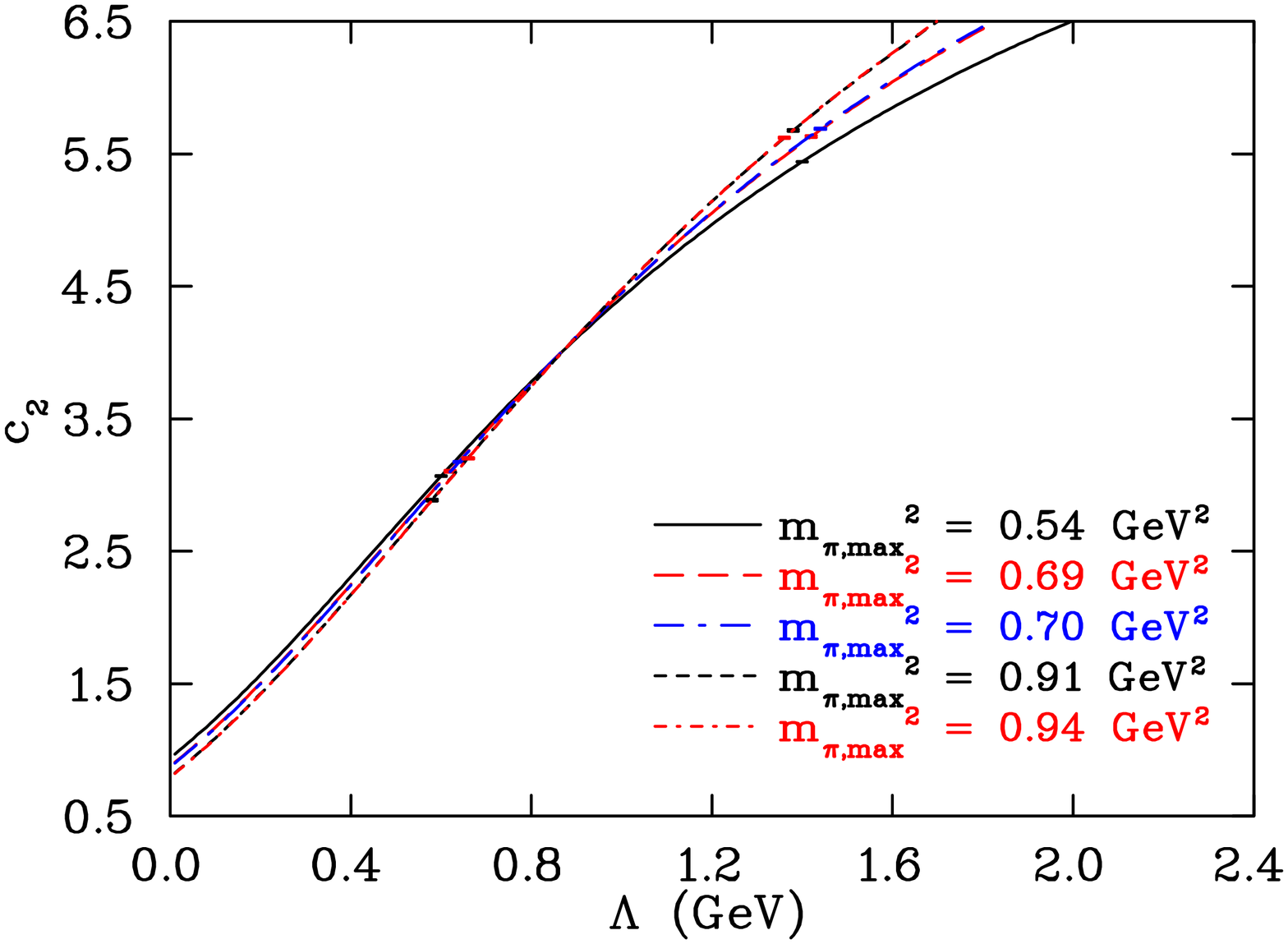}
\vspace{-12pt}
\caption{\footnotesize{(color online). Behaviour of $c_2$ vs.\ $\La$, based on CP-PACS data. The chiral expansion is taken to order $\ca{O}(m_\pi^3)$ and a triple dipole regulator is used. A few points are selected to indicate the general size of the statistical error bars.}}
\label{fig:Youngc2truncTRIP}
\end{figure}

The chiral expansion is first used to chiral order $\ca{O}(m_\pi^3)$.
In this case, the fit parameters are $c_0$ and $c_2$ only.
The results for a dipole regulator are shown in
Figures \ref{fig:Ohkic0truncDIP} through \ref{fig:Youngc2truncDIP},
the results for the double dipole case are shown in
Figures \ref{fig:Ohkic0truncDOUB} through \ref{fig:Youngc2truncDOUB}
and the results for the triple dipole are shown in
Figures \ref{fig:Ohkic0truncTRIP} through \ref{fig:Youngc2truncTRIP}. 
 To estimate the statistical error in the renormalized
constants $\de c$, a bootstrap technique of $200$ configurations of 
nucleon mass data is used. 
The configurations differ by the statistical error in
the data, with values generated by a Gaussian distribution. In each plot, 
the same configurations are used
 for a variety of values of $\La$ considered. 
A few points are selected in Figures \ref{fig:Ohkic0truncDIP} through
 \ref{fig:Youngc2truncTRIP} to indicate the general size of the statistical 
error bars.  

 It should be noted that none of these curves is flat to 
within $1\%$ accuracy.
 All fits have included data beyond the commonly accepted PCR.
Clearly, there is a well-defined intersection point in the 
renormalization flow curves. Also, the value of $\La$ at which the
intersection point occurs is the same even for different data sets,
and for different $c_i$. The tight groupings of the curve crossings lend 
credence to the \emph{ansatz} of an intrinsic scale associated with the finite
size of the source of the pion dressings of the nucleon. This is a central 
result of this analysis.

An intrinsic scale of $\La^\ro{scale}_\ro{dip} \approx 1.3$ GeV
 was obtained for the dipole, $\La^\ro{scale}_\ro{doub} \approx 1.0$ GeV 
for the double
dipole and $\La^\ro{scale}_\ro{trip} \approx 0.9$ GeV for the triple dipole.
These values differ because the regulators have different shapes, and different
 values of $\La^\ro{scale}$ are required to create a similar suppression of
large loop momenta.

%
%





\subsection{Statistical Errors}
\label{subsect:stat}

On each renormalization plot in Figures \ref{fig:Ohkic0truncDIP} through 
\ref{fig:Youngc2truncTRIP} there are many curves, each corresponding
to different values of $m_{\pi,\ro{max}}^2$. It is of primary interest to
what extent these curves match. Therefore,
 a $\chi^2_{dof}$ should be constructed, 
where $dof$ equals the number of curves 
on each plot minus one for the best fit value of $c_0$ or $c_2$, denoted
 by $c^T$ in the following.
 This also serves to quantify the constraint on the intrinsic
scale $\La^\ro{scale}$.
 The $\chi^2_{dof}$ is evaluated separately
 for each renormalized constant $c$ (with error $\de c$)
 and regulator value $\La$:
\eqb
\chi^2_{dof} = \f{1}{n-1} \sum_{i=1}^{n} \f{{(c_i(\La) - c^T(\La))}^2}
{{(\de c_i(\La))}^2}\,,
\eqe
for $i$ corresponding to data sets with differing 
$m_{\pi,\ro{max}}^2$. The theoretical value $c^T$ is given by the
weighted mean:
\eqb
c^T(\La) = \f{\sum_{i=1}^{n}c_i(\La)/{{(\de c_i(\La))}^2}}
{\sum_{j=1}^{n} 1 / {(\de c_j(\La))}^2}\,.
\eqe

\begin{figure}[tp]
\includegraphics[height=0.76\hsize,angle=90]{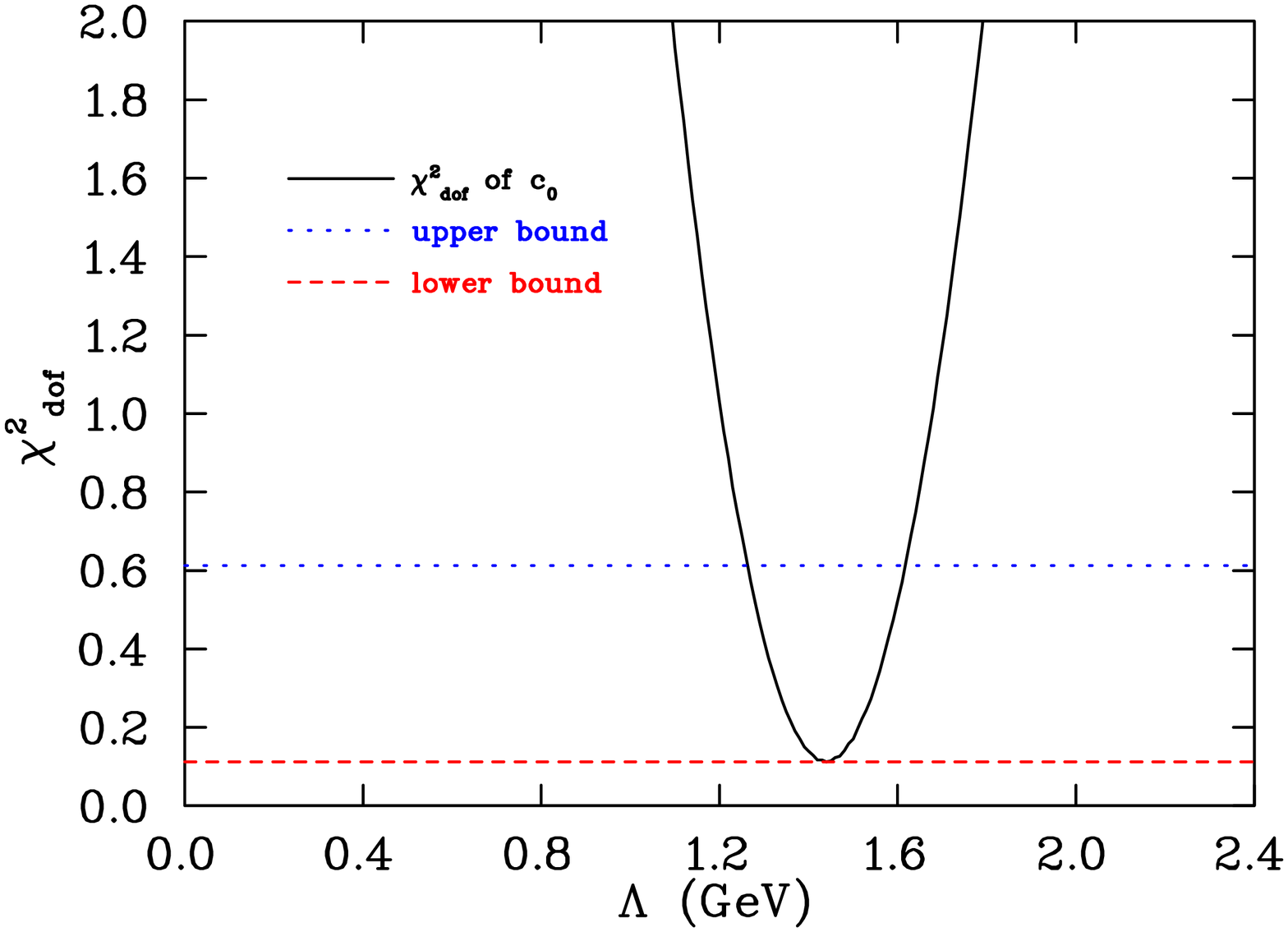}
\vspace{-12pt}
\caption{\footnotesize{(color online). Behaviour of $\chi^2_{dof}$ for $c_0$ vs.\ $\La$, based on JLQCD data. The chiral expansion is taken to order $\ca{O}(m_\pi^3)$, and a dipole regulator is used.}}
\label{fig:Ohkic0truncDIPchisqdof}
\includegraphics[height=0.76\hsize,angle=90]{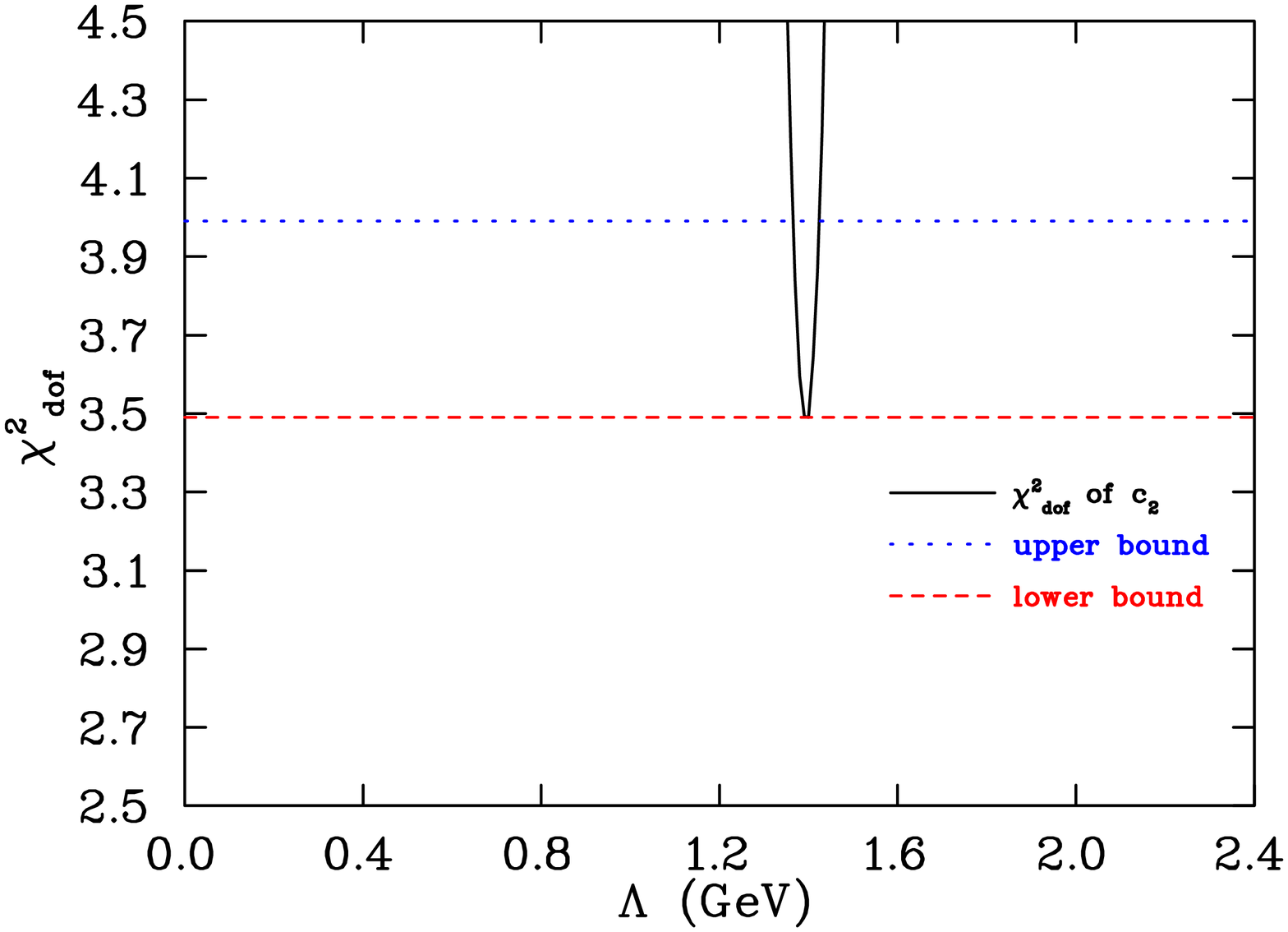}
\vspace{-12pt}
\caption{\footnotesize{(color online). Behaviour of $\chi^2_{dof}$ for $c_2$ vs.\ $\La$, based on JLQCD data. The chiral expansion is taken to order $\ca{O}(m_\pi^3)$, and a dipole regulator is used.}}
\label{fig:Ohkic2truncDIPchisqdof}
\includegraphics[height=0.76\hsize,angle=90]{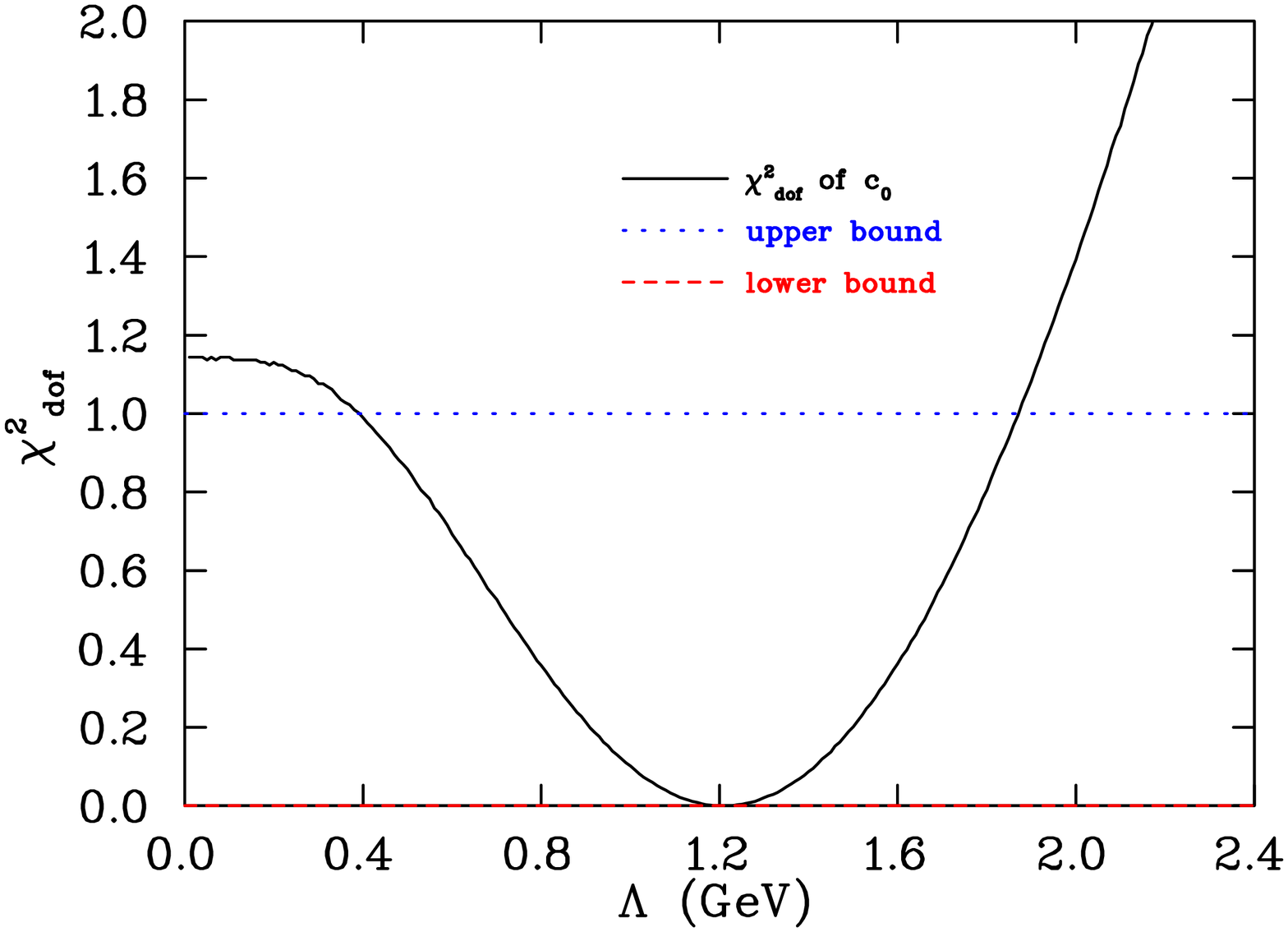}
\vspace{-12pt}
\caption{\footnotesize{(color online). Behaviour of $\chi^2_{dof}$ for $c_0$ vs.\ $\La$, based on PACS-CS data. The chiral expansion is taken to order $\ca{O}(m_\pi^3)$, and a dipole regulator is used.}}
\label{fig:Aokic0truncDIPchisqdof}
\includegraphics[height=0.76\hsize,angle=90]{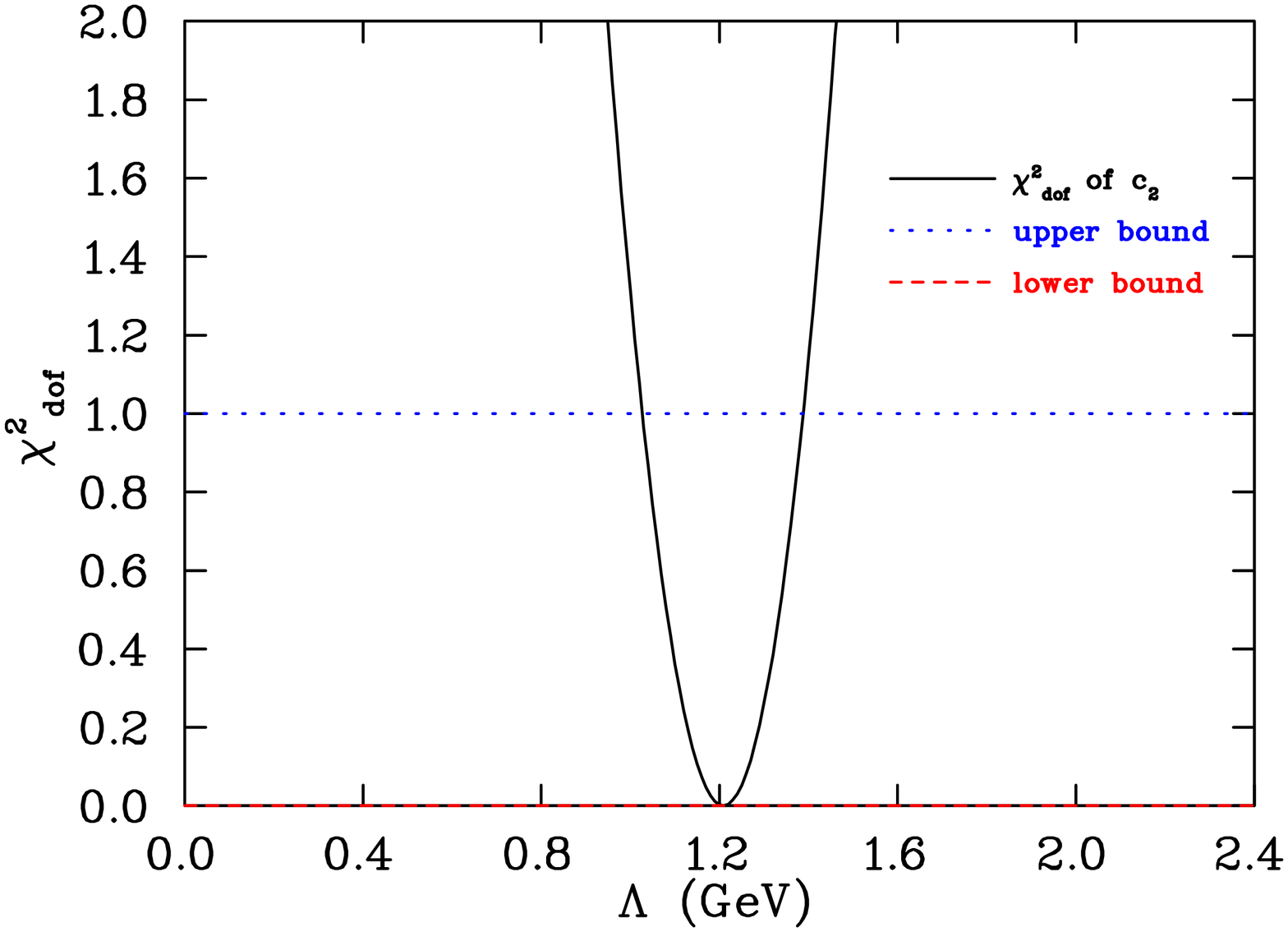}
\vspace{-12pt}
\caption{\footnotesize{(color online). Behaviour of $\chi^2_{dof}$ for $c_2$ vs.\ $\La$, based on PACS-CS data. The chiral expansion is taken to order $\ca{O}(m_\pi^3)$, and a dipole regulator is used.}}
\label{fig:Aokic2truncDIPchisqdof}
\end{figure}

\begin{figure}[tp]
\includegraphics[height=0.76\hsize,angle=90]{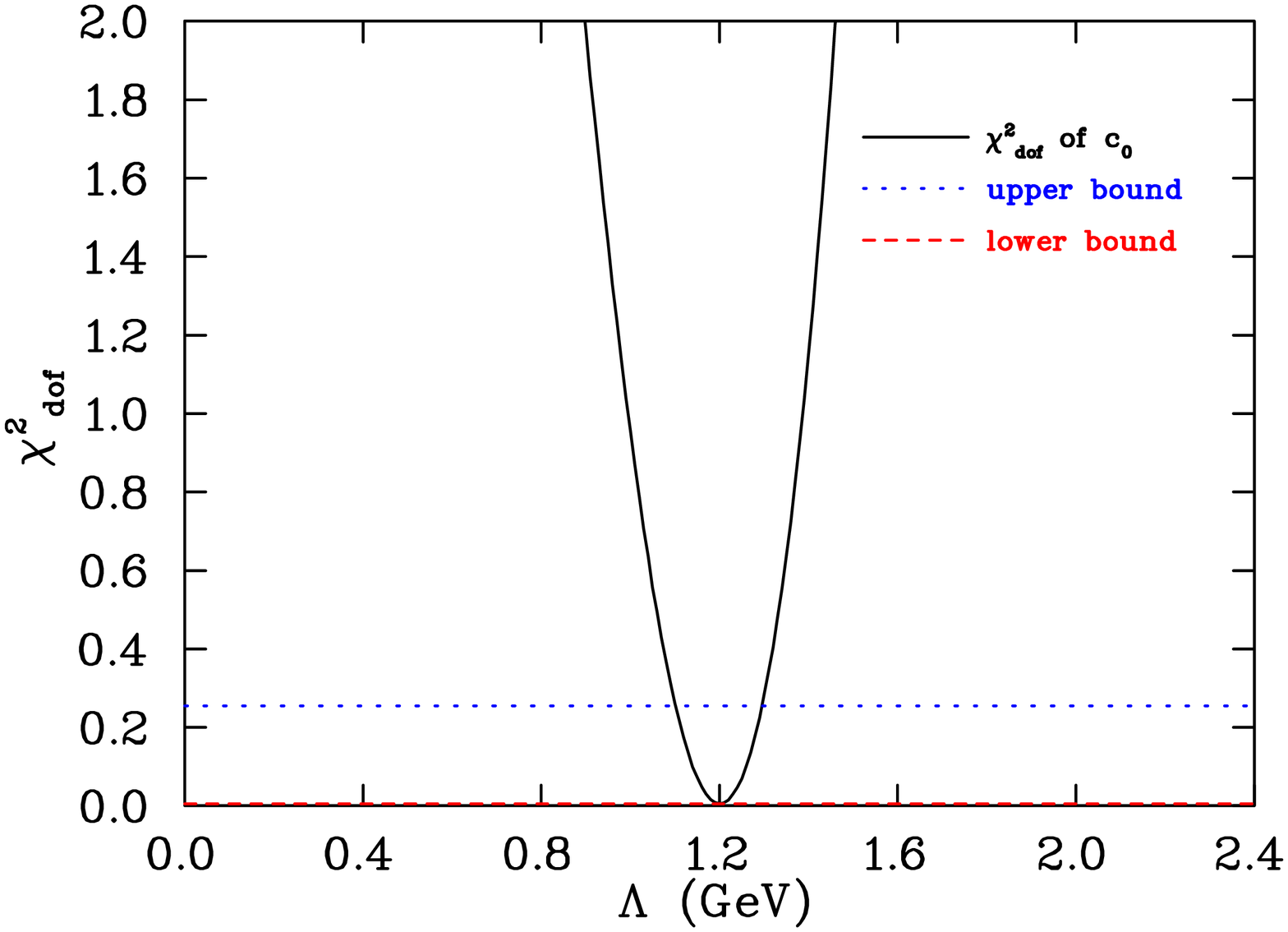}
\vspace{-12pt}
\caption{\footnotesize{(color online). Behaviour of $\chi^2_{dof}$ for $c_0$ vs.\ $\La$, based on CP-PACS data. The chiral expansion is taken to order $\ca{O}(m_\pi^3)$, and a dipole regulator is used.}}
\label{fig:Youngc0truncDIPchisqdof}
\includegraphics[height=0.76\hsize,angle=90]{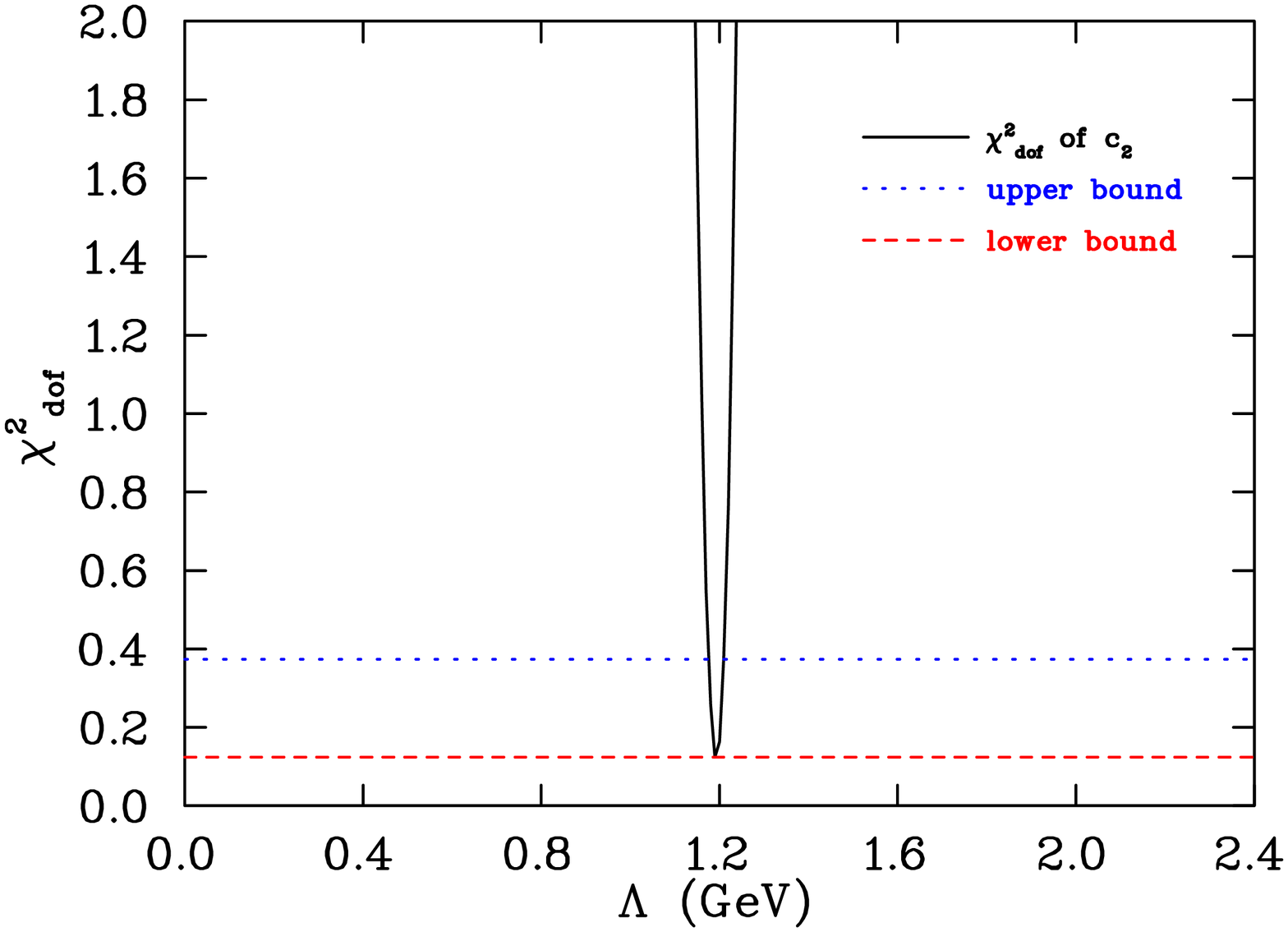}
\vspace{-12pt}
\caption{\footnotesize{(color online). Behaviour of $\chi^2_{dof}$ for $c_2$ vs.\ $\La$, based on CP-PACS data. The chiral expansion is taken to order $\ca{O}(m_\pi^3)$, and a dipole regulator is used.}}
\label{fig:Youngc2truncDIPchisqdof}
\end{figure}

\begin{figure}[tp]
\includegraphics[height=0.76\hsize,angle=90]{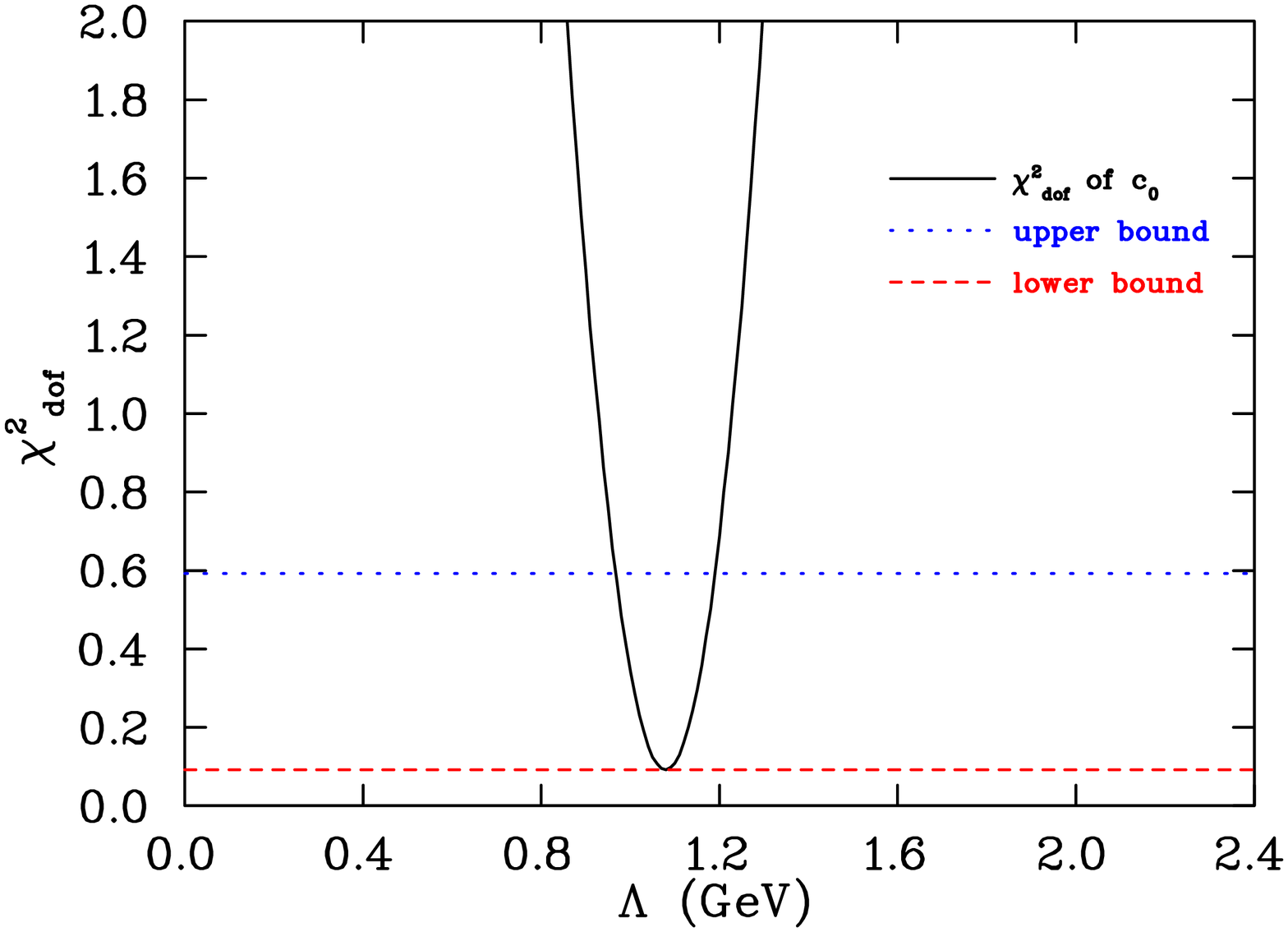}
\vspace{-12pt}
\caption{\footnotesize{(color online). Behaviour of $\chi^2_{dof}$ for $c_0$ vs.\ $\La$, based on JLQCD data. The chiral expansion is taken to order $\ca{O}(m_\pi^3)$, and a double dipole regulator is used.}}
\label{fig:Ohkic0truncDOUBchisqdof}
\includegraphics[height=0.76\hsize,angle=90]{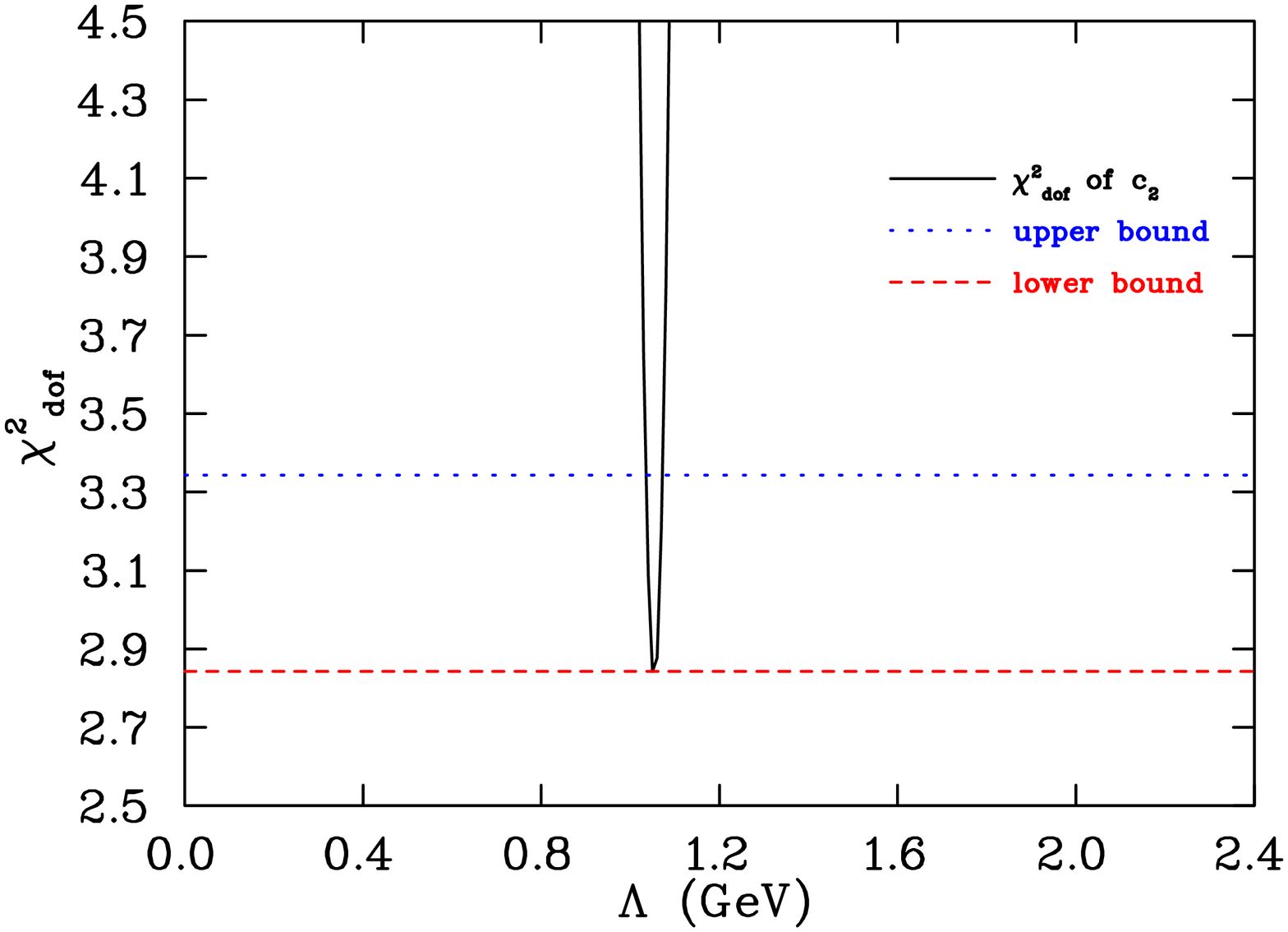}
\vspace{-12pt}
\caption{\footnotesize{(color online). Behaviour of $\chi^2_{dof}$ for $c_2$ vs.\ $\La$, based on JLQCD data. The chiral expansion is taken to order $\ca{O}(m_\pi^3)$, and a double dipole regulator is used.}}
\label{fig:Ohkic2truncDOUBchisqdof}
\includegraphics[height=0.76\hsize,angle=90]{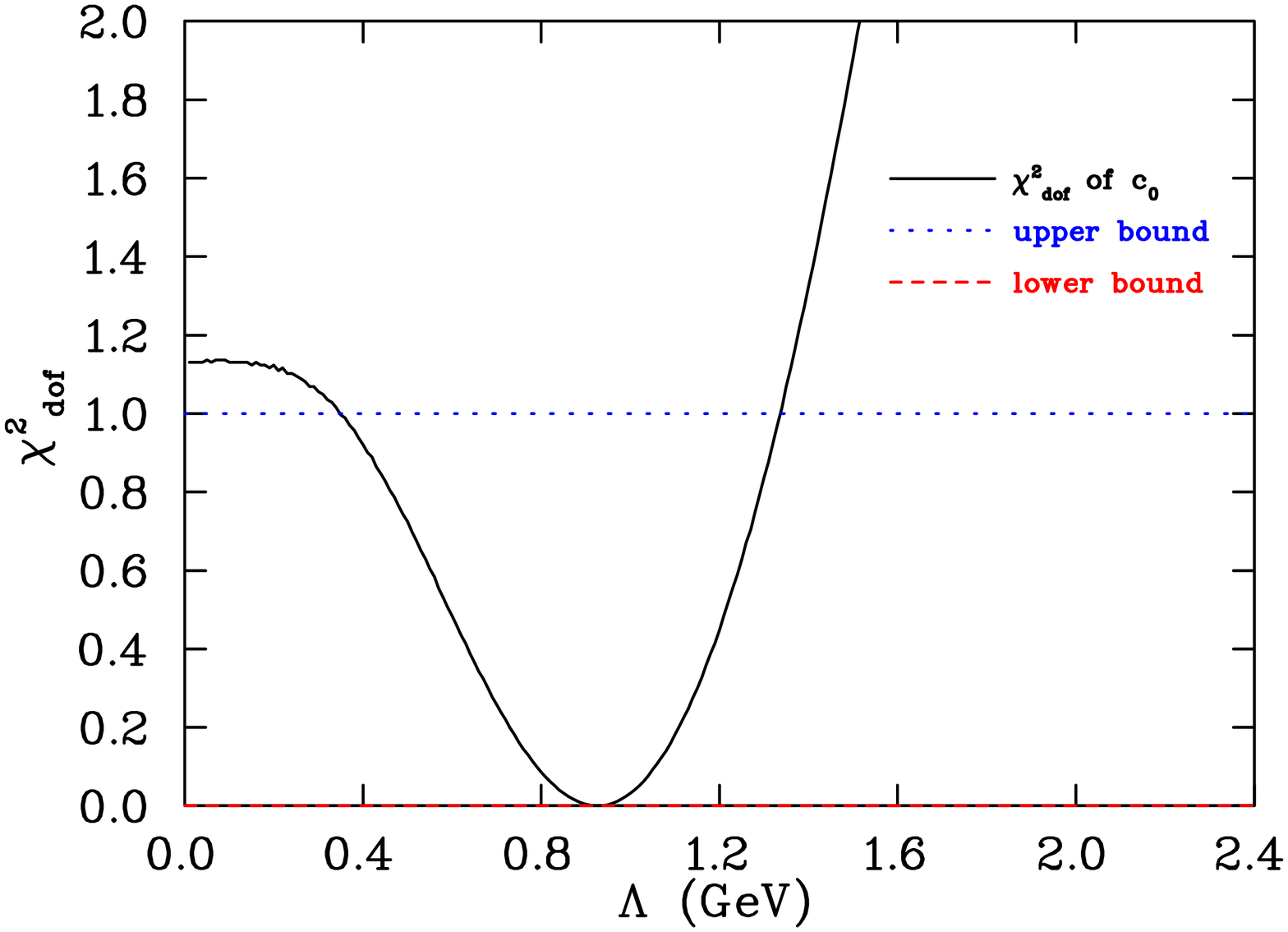}
\vspace{-12pt}
\caption{\footnotesize{(color online). Behaviour of $\chi^2_{dof}$ for $c_0$ vs.\ $\La$, based on PACS-CS data. The chiral expansion is taken to order $\ca{O}(m_\pi^3)$, and a double dipole regulator is used.}}
\label{fig:Aokic0truncDOUBchisqdof}
\includegraphics[height=0.76\hsize,angle=90]{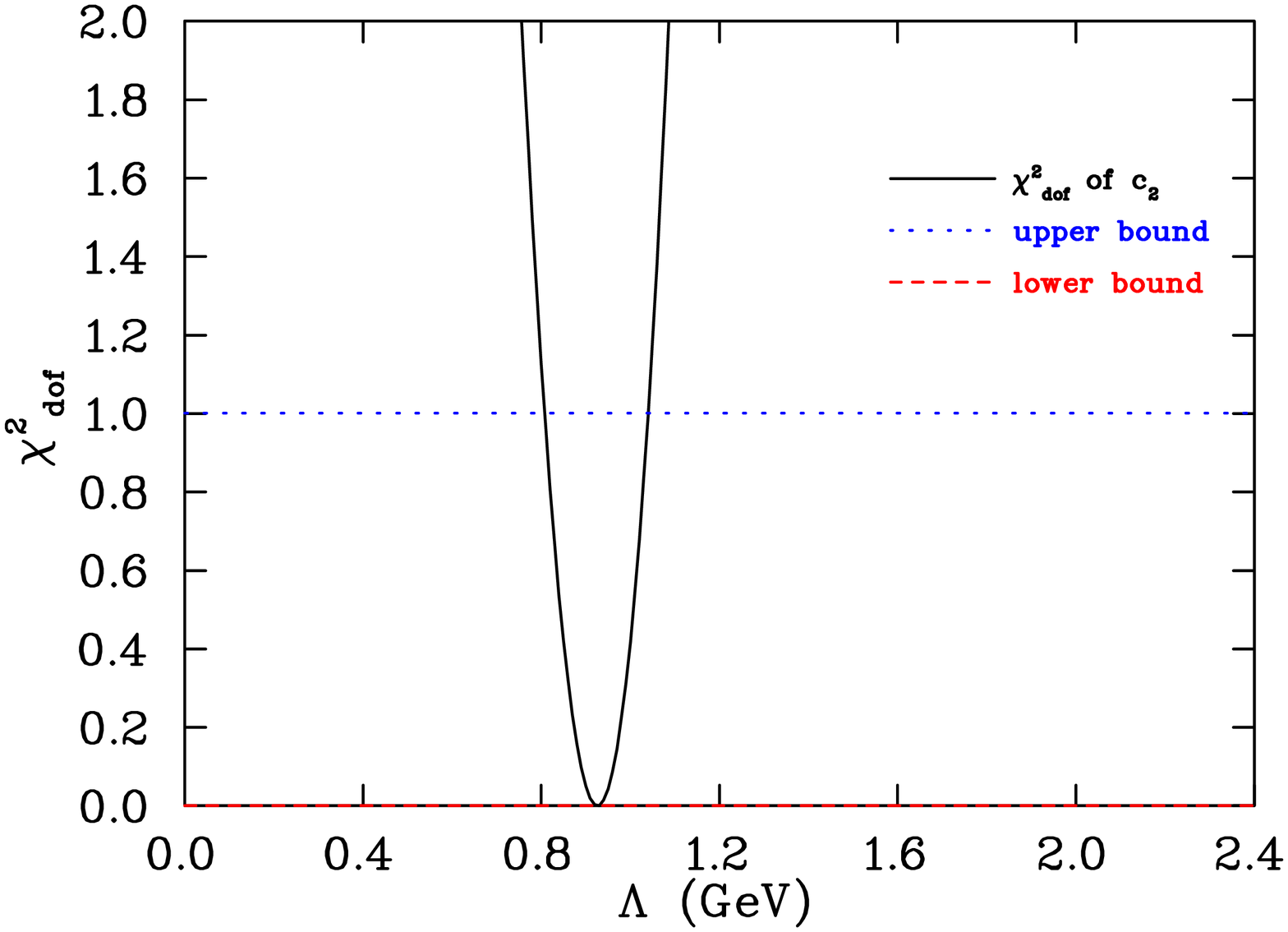}
\vspace{-12pt}
\caption{\footnotesize{(color online). Behaviour of $\chi^2_{dof}$ for $c_2$ vs.\ $\La$, based on PACS-CS data. The chiral expansion is taken to order $\ca{O}(m_\pi^3)$, and a double dipole regulator is used.}}
\label{fig:Aokic2truncDOUBchisqdof}
\end{figure}

\begin{figure}[tp]
\includegraphics[height=0.76\hsize,angle=90]{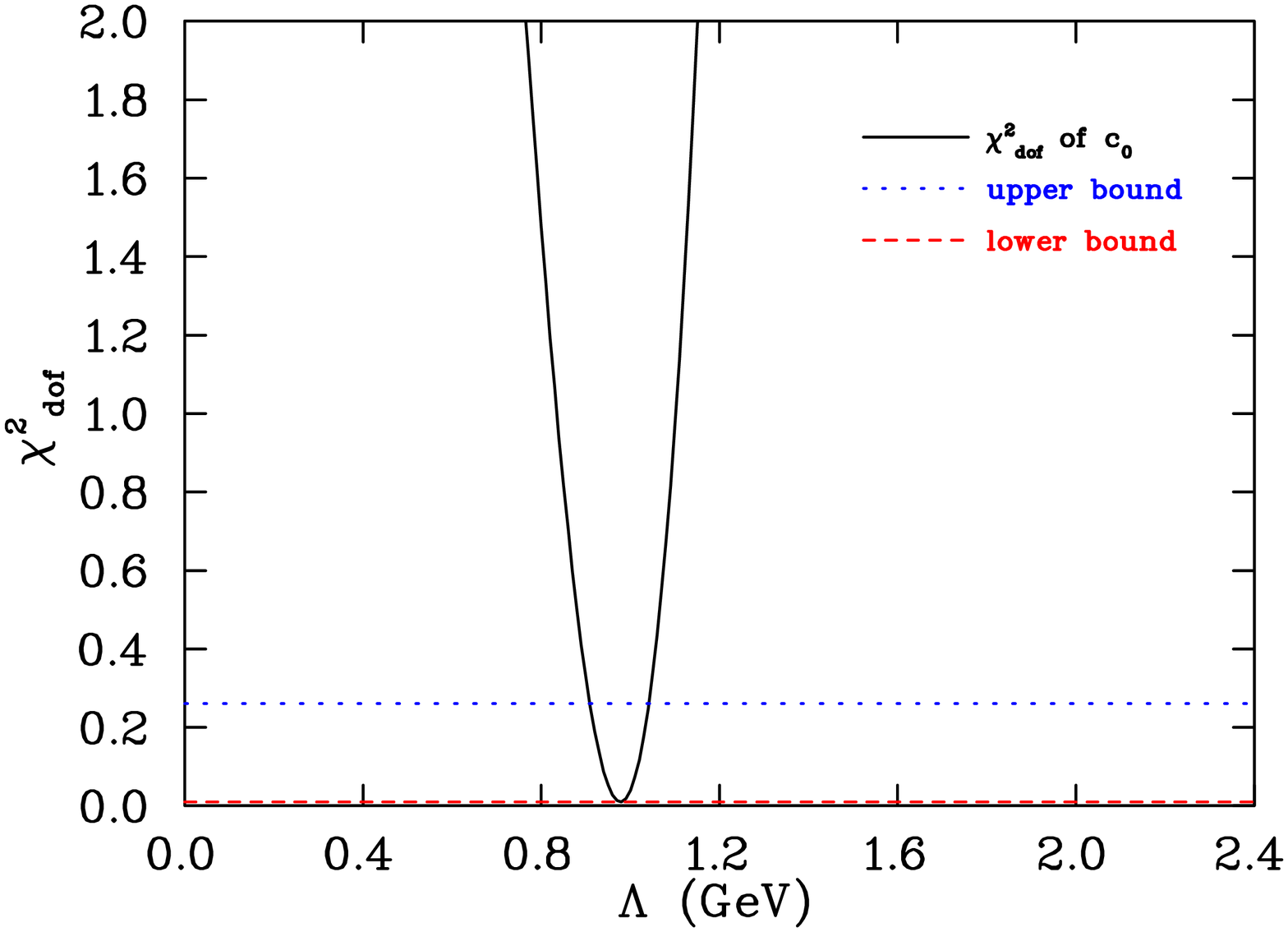}
\vspace{-12pt}
\caption{\footnotesize{(color online). Behaviour of $\chi^2_{dof}$ for $c_0$ vs.\ $\La$, based on CP-PACS data. The chiral expansion is taken to order $\ca{O}(m_\pi^3)$, and a double dipole regulator is used.}}
\label{fig:Youngc0truncDOUBchisqdof}
\includegraphics[height=0.76\hsize,angle=90]{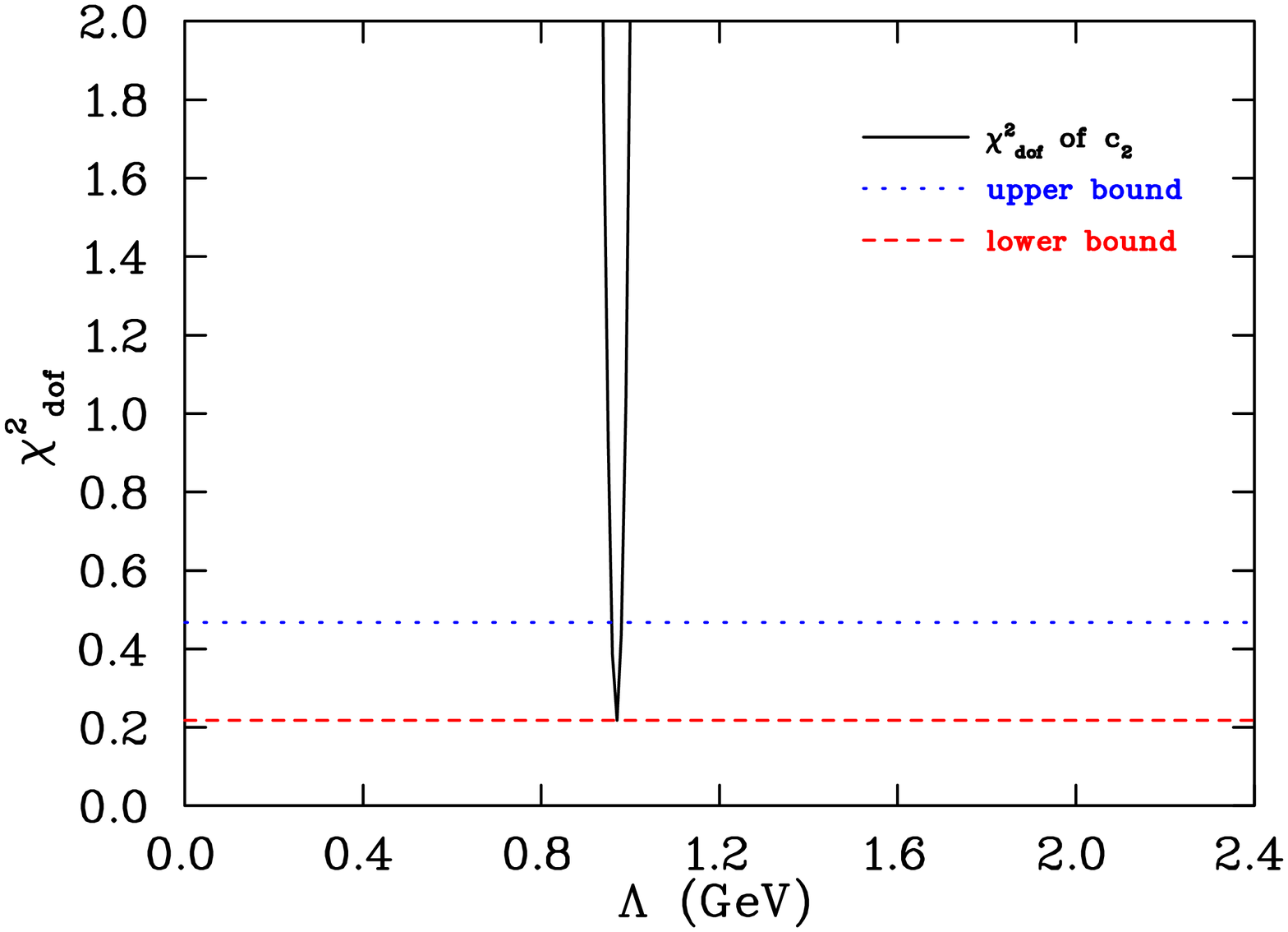}
\vspace{-12pt}
\caption{\footnotesize{(color online). Behaviour of $\chi^2_{dof}$ for $c_2$ vs.\ $\La$, based on CP-PACS data. The chiral expansion is taken to order $\ca{O}(m_\pi^3)$, and a double dipole regulator is used.}}
\label{fig:Youngc2truncDOUBchisqdof}
\end{figure}

\begin{figure}[tp]
\includegraphics[height=0.76\hsize,angle=90]{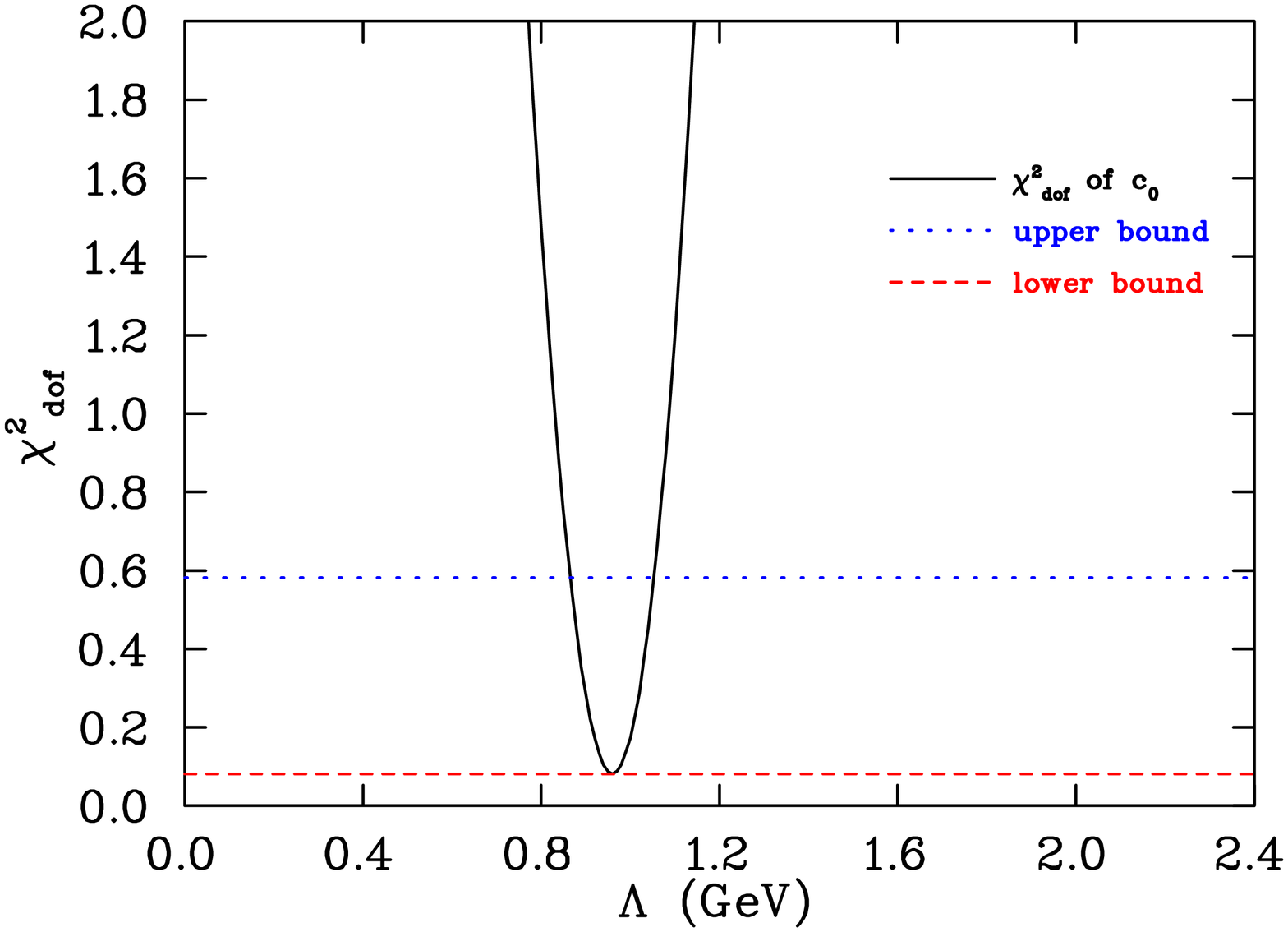}
\vspace{-12pt}
\caption{\footnotesize{(color online). Behaviour of $\chi^2_{dof}$ for $c_0$ vs.\ $\La$, based on JLQCD data. The chiral expansion is taken to order $\ca{O}(m_\pi^3)$, and a triple dipole regulator is used.}}
\label{fig:Ohkic0truncTRIPchisqdof}
\includegraphics[height=0.76\hsize,angle=90]{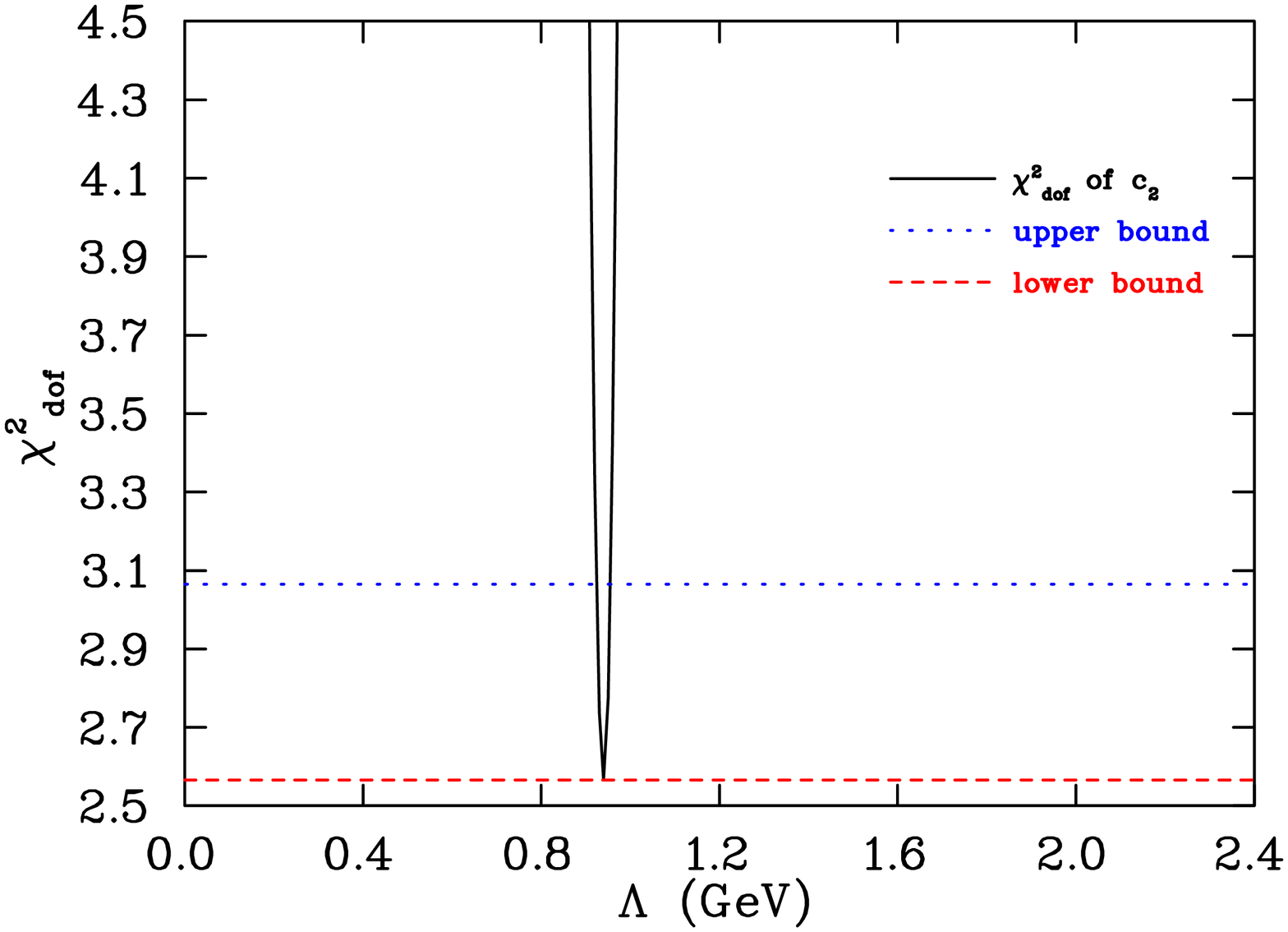}
\vspace{-12pt}
\caption{\footnotesize{(color online). Behaviour of $\chi^2_{dof}$ for $c_2$ vs.\ $\La$, based on JLQCD data. The chiral expansion is taken to order $\ca{O}(m_\pi^3)$, and a triple dipole regulator is used.}}
\label{fig:Ohkic2truncTRIPchisqdof}
\includegraphics[height=0.76\hsize,angle=90]{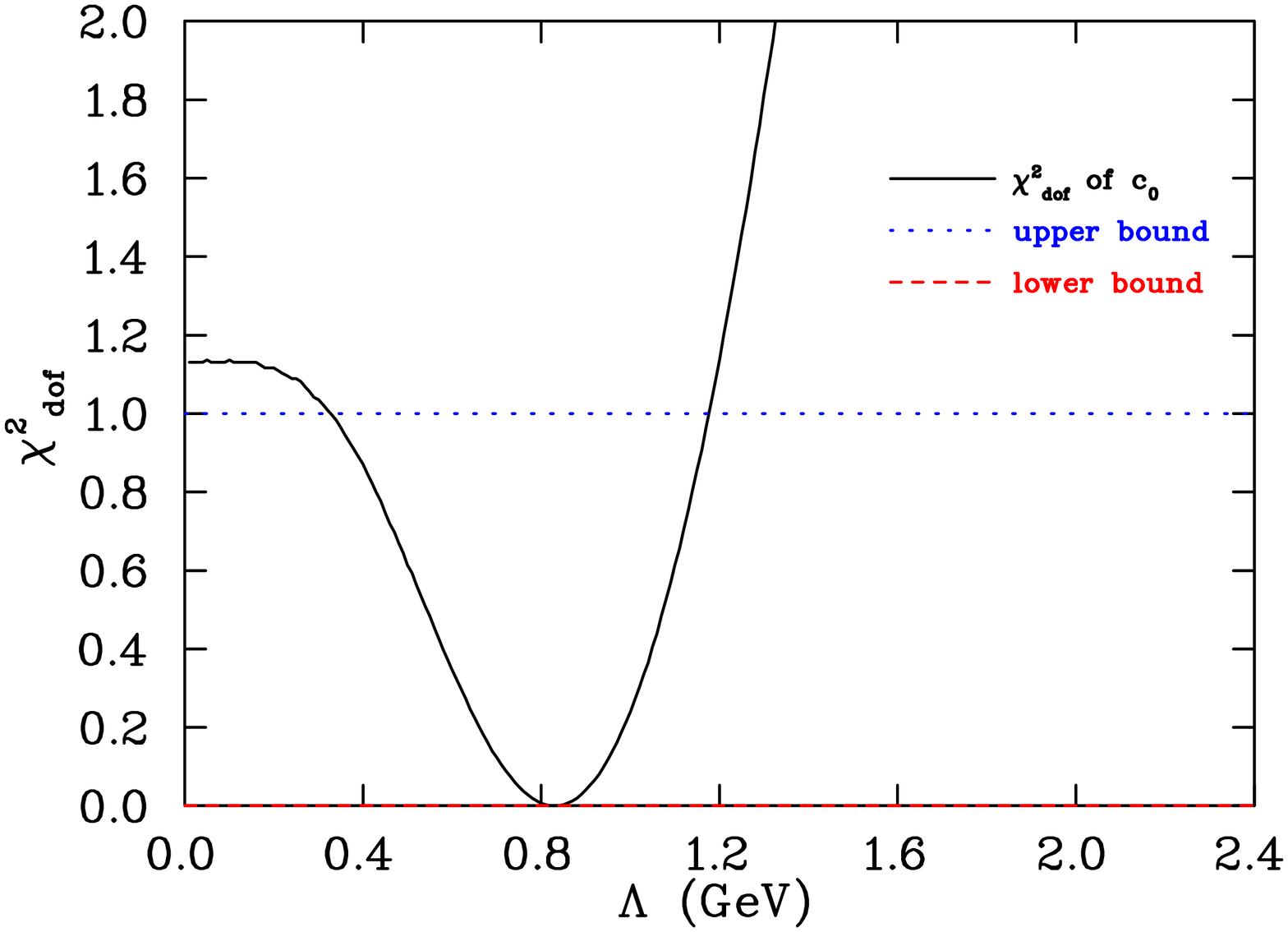}
\vspace{-12pt}
\caption{\footnotesize{(color online). Behaviour of $\chi^2_{dof}$ for $c_0$ vs.\ $\La$, based on PACS-CS data. The chiral expansion is taken to order $\ca{O}(m_\pi^3)$, and a triple dipole regulator is used.}}
\label{fig:Aokic0truncTRIPchisqdof}
\includegraphics[height=0.76\hsize,angle=90]{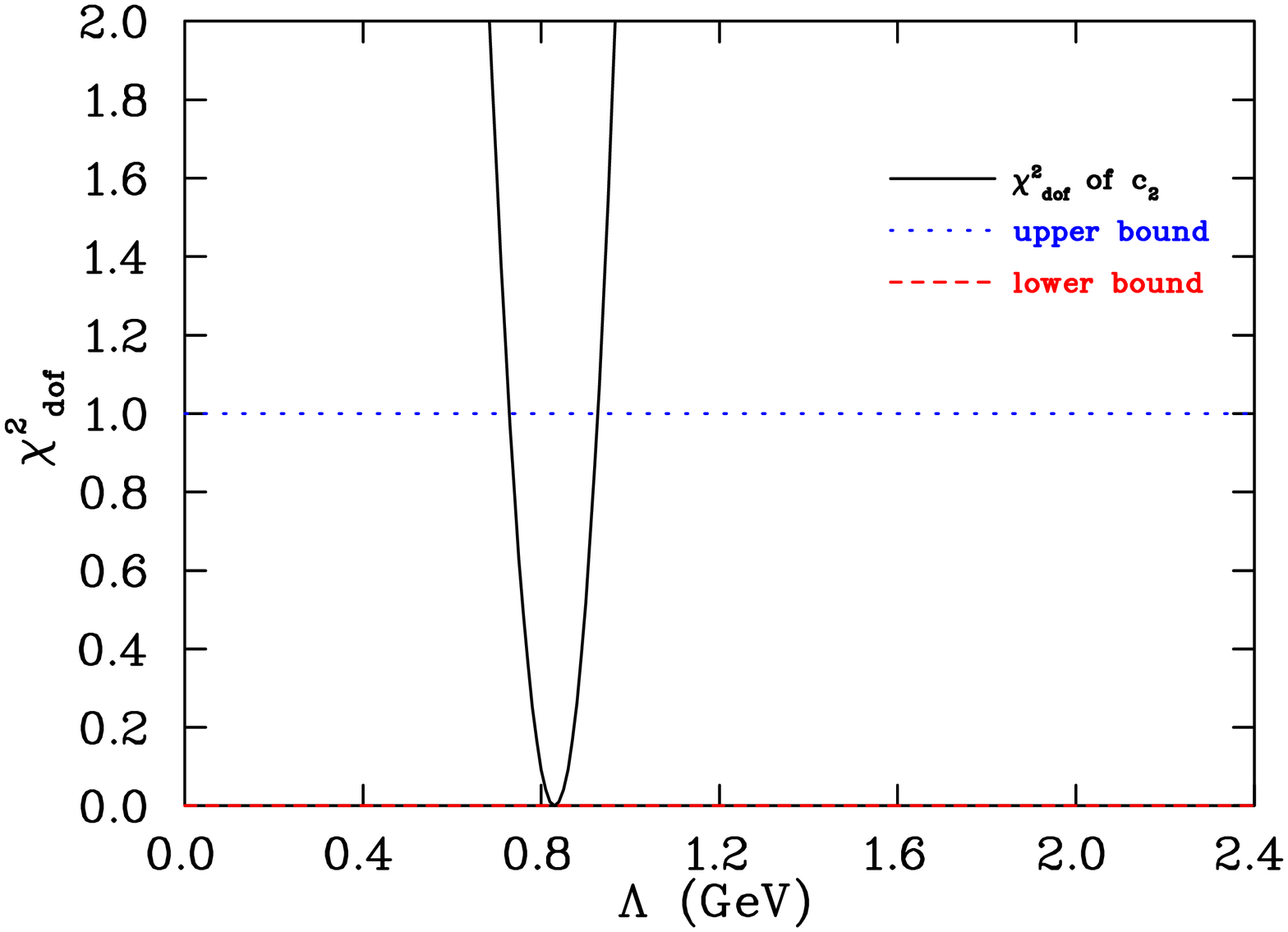}
\vspace{-12pt}
\caption{\footnotesize{(color online). Behaviour of $\chi^2_{dof}$ for $c_2$ vs.\ $\La$, based on PACS-CS data. The chiral expansion is taken to order $\ca{O}(m_\pi^3)$, and a triple dipole regulator is used.}}
\label{fig:Aokic2truncTRIPchisqdof}
\end{figure}

\begin{figure}[tp]
\includegraphics[height=0.76\hsize,angle=90]{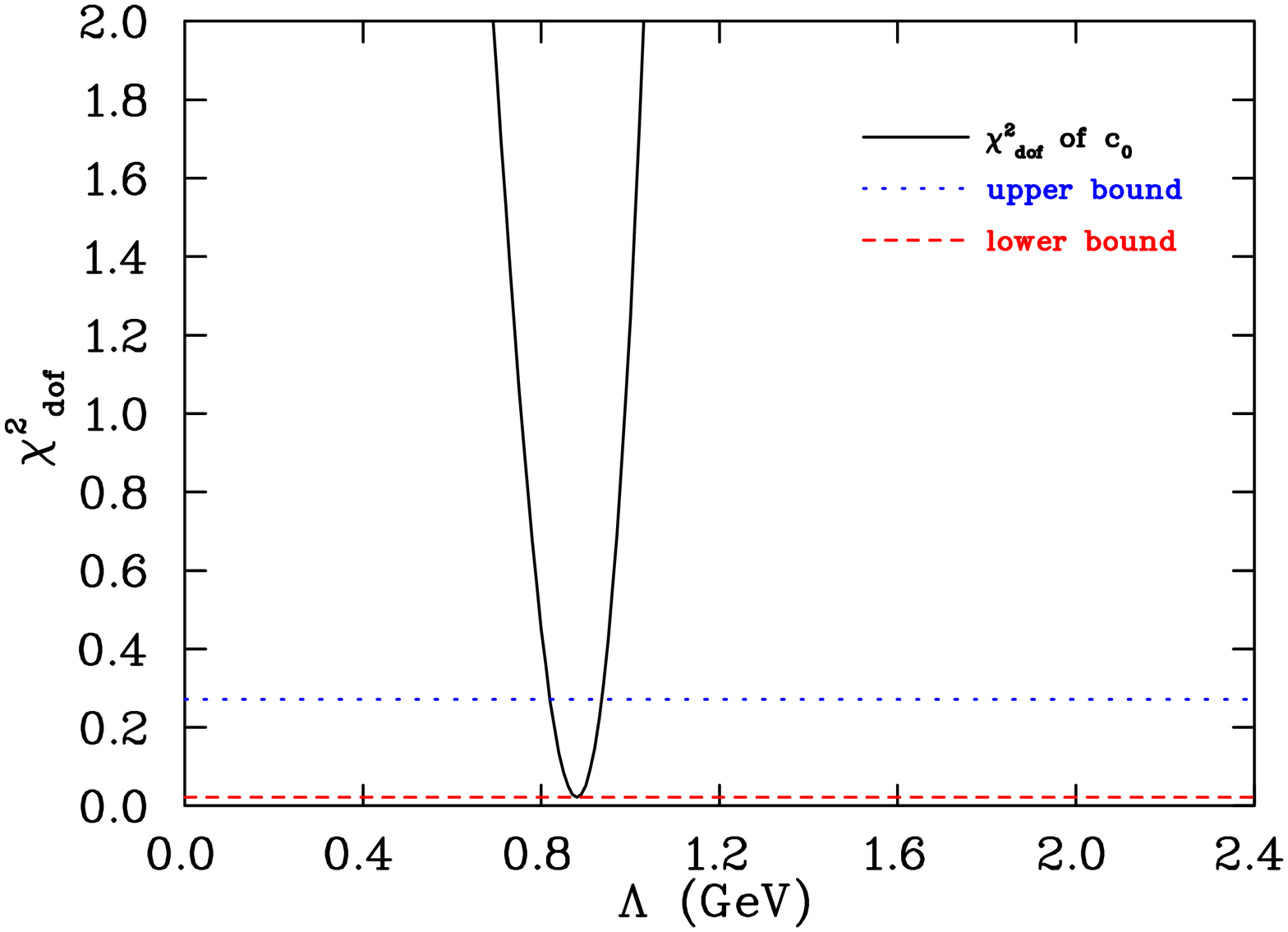}
\vspace{-12pt}
\caption{\footnotesize{(color online). Behaviour of $\chi^2_{dof}$ for $c_0$ vs.\ $\La$, based on CP-PACS data. The chiral expansion is taken to order $\ca{O}(m_\pi^3)$, and a triple dipole regulator is used.}}
\label{fig:Youngc0truncTRIPchisqdof}
\includegraphics[height=0.76\hsize,angle=90]{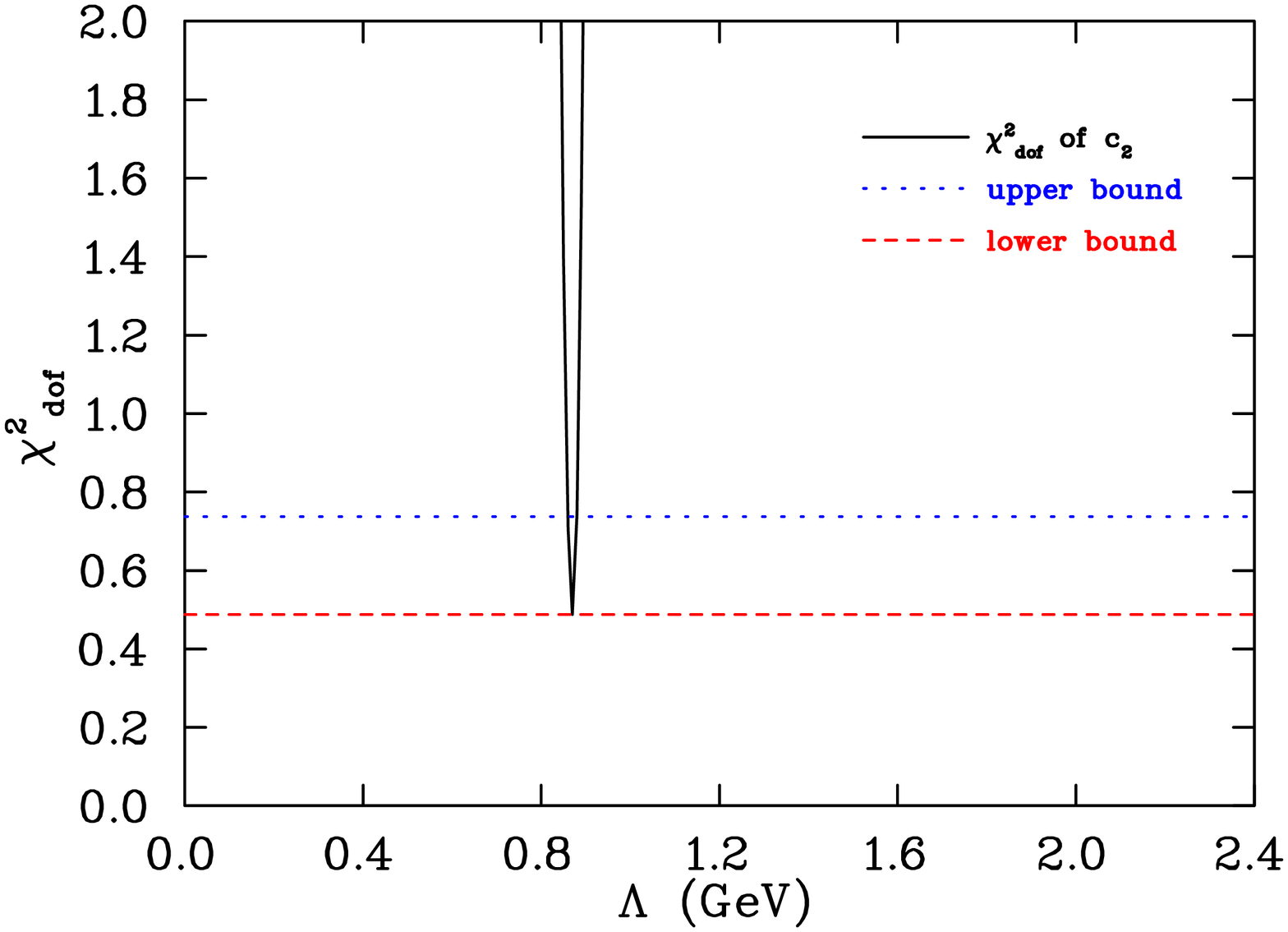}
\vspace{-12pt}
\caption{\footnotesize{(color online). Behaviour of $\chi^2_{dof}$ for $c_2$ vs.\ $\La$, based on CP-PACS data. The chiral expansion is taken to order $\ca{O}(m_\pi^3)$, and a triple dipole regulator is used.}}
\label{fig:Youngc2truncTRIPchisqdof}
\end{figure}

The $\chi^2_{dof}$ can be calculated as a function of 
the regulator parameter $\La$ for each of the renormalization plots of
 Figures \ref{fig:Ohkic0truncDIP} through 
\ref{fig:Youngc2truncTRIP}.
  In the case of the PACS-CS data,
 the minimum of the $\chi^2_{dof}$
curve will be centred at the intersection point. In the case of the 
JLQCD and CP-PACS data, 
there appears to be a single intersection point on each plot,
but in fact there are multiple intersections over a very small window of $\La$.
The results for $\chi^2_{dof}$ will indicate the `best' central value of $\La$.
This central value of $\La$ will be taken to be the intrinsic scale.
The $\chi^2_{dof}$ curves for a dipole regulator are shown in
Figures \ref{fig:Ohkic0truncDIPchisqdof}
 through \ref{fig:Youngc2truncDIPchisqdof},
the $\chi^2_{dof}$ curves for the double dipole case are shown in
Figures \ref{fig:Ohkic0truncDOUBchisqdof}
 through \ref{fig:Youngc2truncDOUBchisqdof}
and the $\chi^2_{dof}$ curves for the triple dipole are shown in
Figures \ref{fig:Ohkic0truncTRIPchisqdof}
 through \ref{fig:Youngc2truncTRIPchisqdof}.
%





\subsection{Higher Chiral Order}
\label{subsect:higher}
\begin{figure}[tp]
\includegraphics[height=0.76\hsize,angle=90]{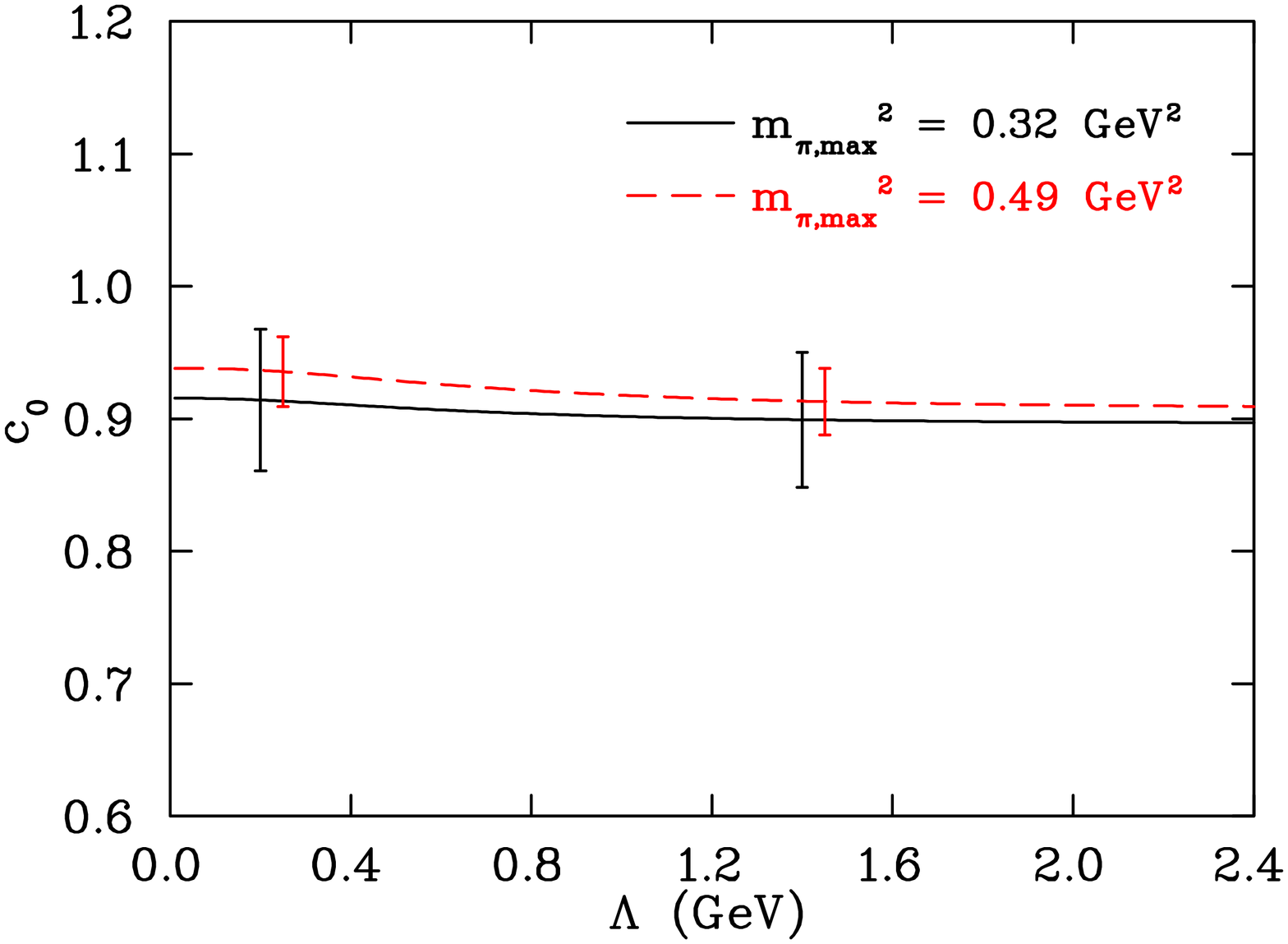}
\vspace{-12pt}
\caption{\footnotesize{(color online). Behaviour of $c_0$ vs.\ $\La$, based on PACS-CS data. The chiral expansion taken to order $\ca{O}(m_\pi^4\,\ro{log}\,m_\pi)$ and a dipole regulator is used. A few points are selected to indicate the general size of the statistical error bars.}}
\label{fig:Aokic0DIP}
\includegraphics[height=0.76\hsize,angle=90]{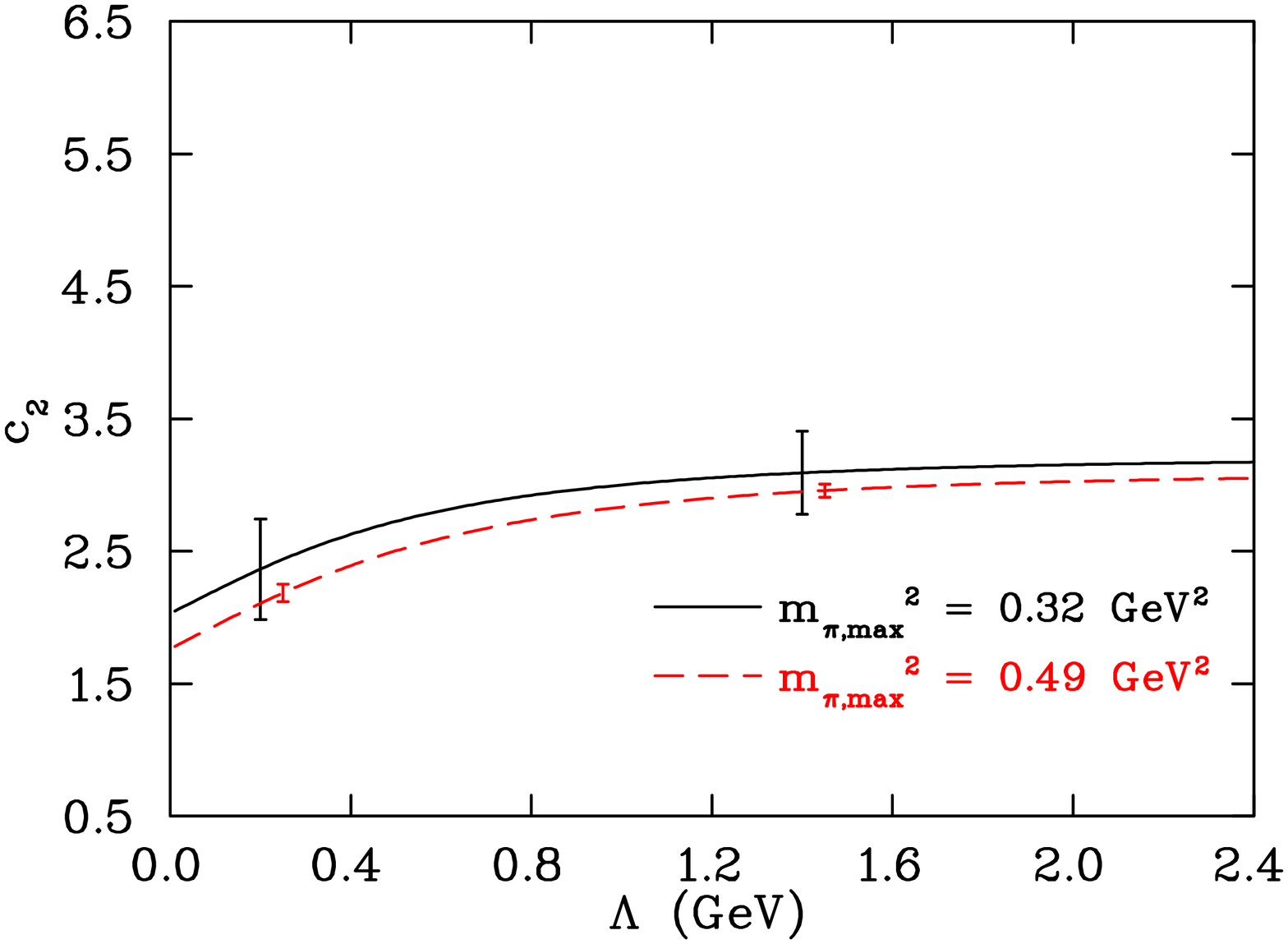}
\vspace{-12pt}
\caption{\footnotesize{(color online). Behaviour of $c_2$ vs.\ $\La$, based on PACS-CS data. The chiral expansion taken to order $\ca{O}(m_\pi^4\,\ro{log}\,m_\pi)$ and a dipole regulator is used. A few points are selected to indicate the general size of the statistical error bars.}}
\label{fig:Aokic2DIP}
\includegraphics[height=0.76\hsize,angle=90]{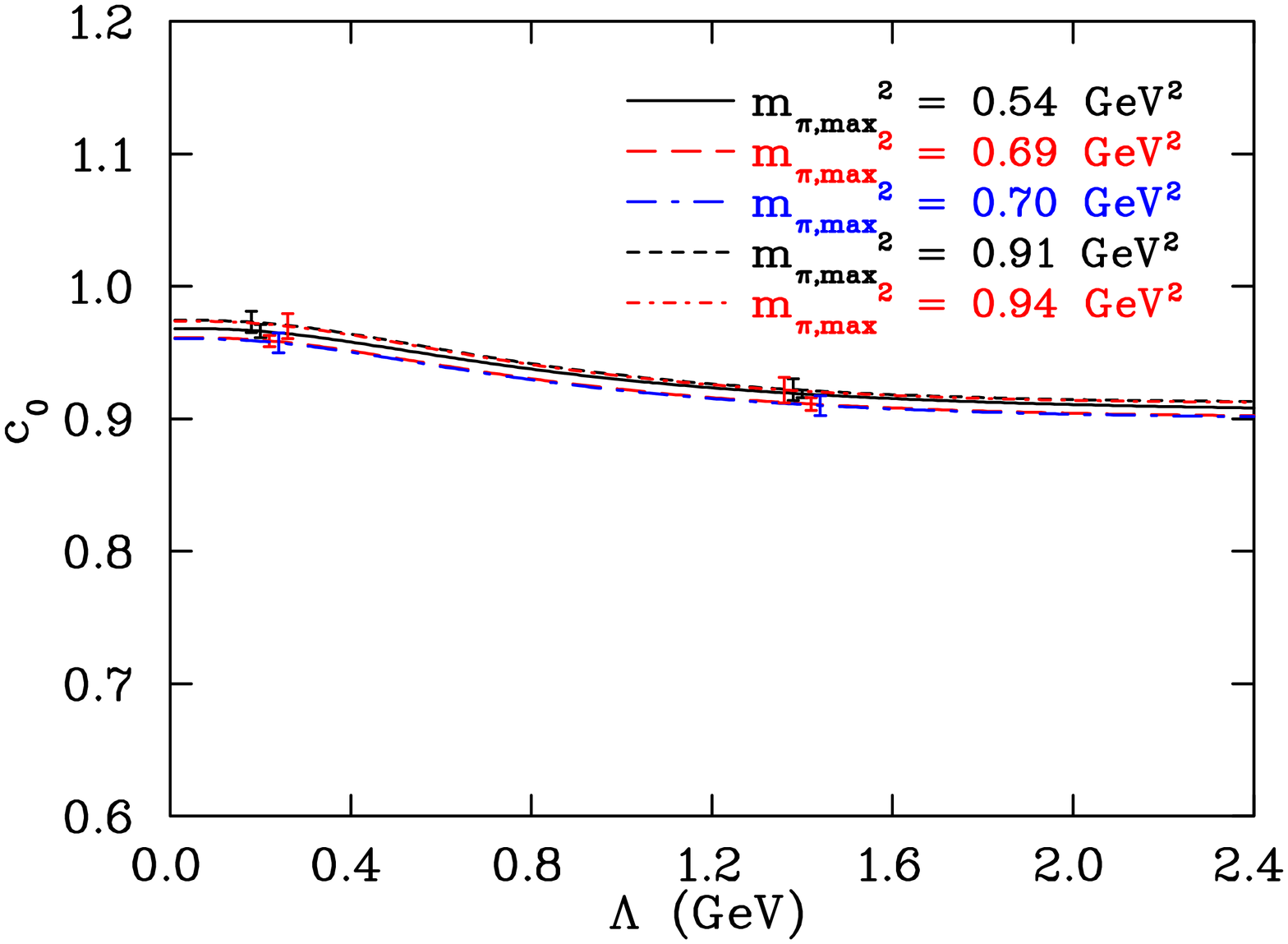}
\vspace{-12pt}
\caption{\footnotesize{(color online). Behaviour of $c_0$ vs.\ $\La$, based on CP-PACS data. The chiral expansion taken to order $\ca{O}(m_\pi^4\,\ro{log}\,m_\pi)$ and a dipole regulator is used. A few points are selected to indicate the general size of the statistical error bars.}}
\label{fig:Youngc0DIP}
\includegraphics[height=0.76\hsize,angle=90]{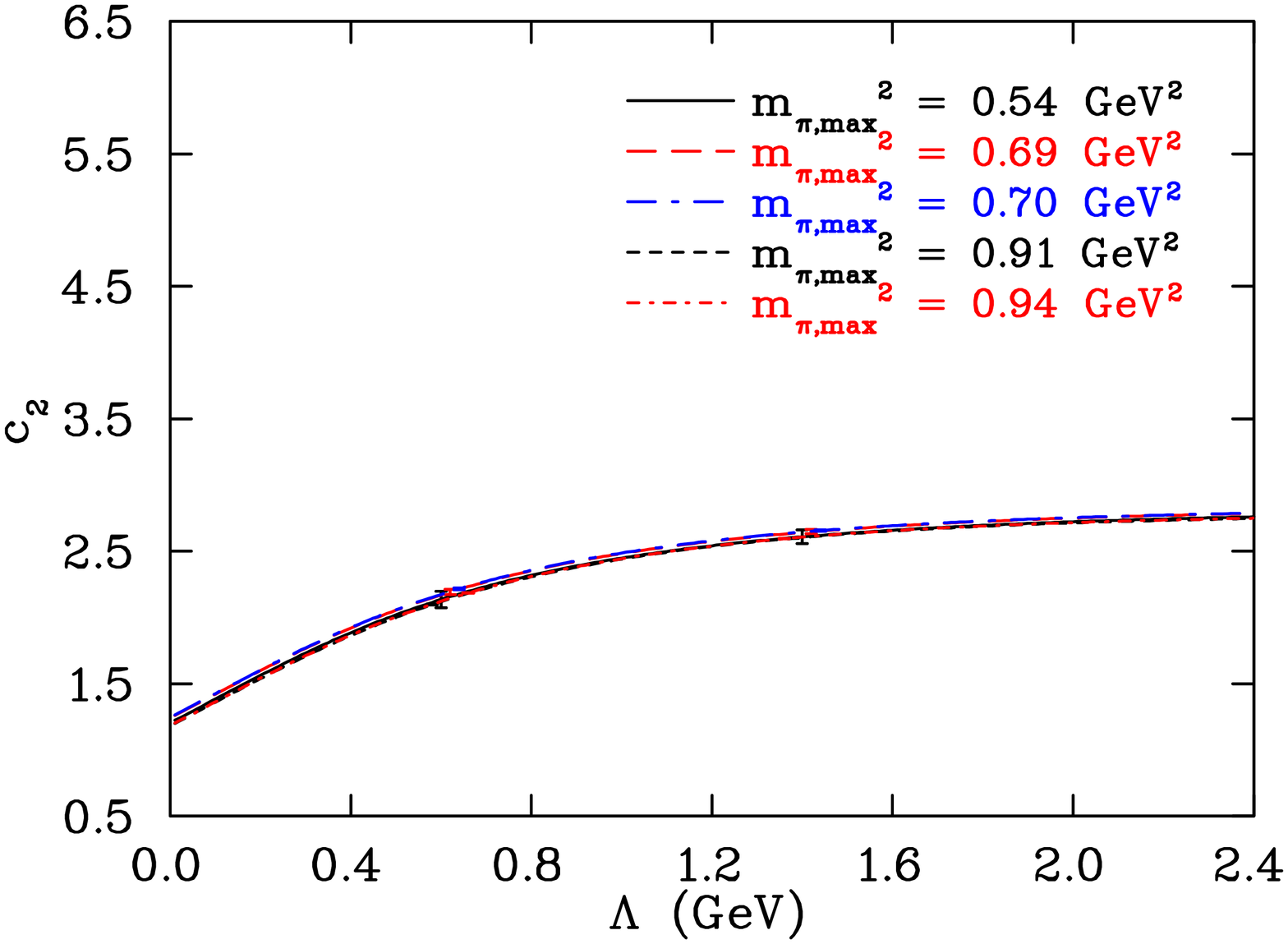}
\vspace{-12pt}
\caption{\footnotesize{(color online). Behaviour of $c_2$ vs.\ $\La$, based on CP-PACS data. The chiral expansion taken to order $\ca{O}(m_\pi^4\,\ro{log}\,m_\pi)$ and a dipole regulator is used. A few points are selected to indicate the general size of the statistical error bars.}}
\label{fig:Youngc2DIP}
\end{figure}

Consider the renormalization of $c_0$ and $c_2$ as a function of $\La$,
for chiral order $\ca{O}(m_\pi^4\,\ro{log}\,m_\pi)$.
The results for PACS-CS and CP-PACS data are shown in 
Figures \ref{fig:Aokic0DIP} through \ref{fig:Youngc2DIP}, as an example.
In this case, no clear intersection points in the renormalization
 curves can be found, and so one is unable to specify an intrinsic scale.
 This certainly should be the case 
when working with data entirely within the PCR,  
 because all renormalization procedures would be equivalent
 (to a prescribed level of accuracy) 
and so there would be no optimal regulator parameter.
 It is known that this is not the case for the data sets used in this study. 
 This is verified 
 by considering the evident scale dependence of $c_0$ and $c_2$ in 
 Figures \ref{fig:Aokic0DIP} through \ref{fig:Youngc2DIP}.
 The fact that $c_0$ and $c_2$ change over the range of $\La$ values indicates
 that the data are not inside the PCR.
Further, since no preferred scale is revealed, any choice of $\La$ appears 
 equivalent at this order. While this is encouraging that the scheme dependence
 is being weakened by working to higher order, it must be recognized that 
there is a systematic error associated with the choice of $\La$. In 
 the case of 
the CP-PACS results shown in Figures \ref{fig:Youngc0DIP} 
and \ref{fig:Youngc2DIP}, it can be seen that the 
statistical errors are substantially smaller than the systematic error 
associated with a characteristic range, $\La_\ro{lower}<\La<\infty$, where
 $\La_\ro{lower}$ is the lowest reasonable value of $\La$.

Since it is difficult to identify the intrinsic scale
at this chiral order, the results for chiral order $\ca{O}(m_\pi^3)$ will be
 chosen to demonstrate the process of handling the existence of an optimal 
regulator scale in lattice QCD data. 
 The results for the calculation of the intrinsic 
 scales $\La^\ro{scale}$
 for different data sets and regulators 
are given in Table \ref{table:scales}.
This table simply summarizes the central values from Figures 
\ref{fig:Ohkic0truncDIPchisqdof} through \ref{fig:Youngc2truncTRIPchisqdof}.
 \emph{
Such excellent agreement between the $c_0$ analysis and the $c_2$ analysis
   is remarkable, and indicative of the
 existence of an instrinsic scale in the data.}
 There is also consistency among independent data sets.
 It is important to realize that the value of 
$\La^\ro{scale}$ is always the order of $\sim 1$ GeV, not 
 $10$ GeV, nor $100$ GeV; nor is it infinity.
\begin{table}[tp]
  \newcommand\T{\rule{0pt}{2.8ex}}
  \newcommand\B{\rule[-1.4ex]{0pt}{0pt}}
  \begin{center}
    \begin{tabular}{llll}
      \hline
      \hline
       \T\B                  \qquad &\multicolumn{3}{c}{regulator form} \\ 
      optimal scale \qquad & dipole $\,\,\,\,\,$& double $\,\,$& triple   \\
      \hline
      $\La^\ro{scale}_{c_0,\ro{JLQCD}}$   &\T $1.44$ & $1.08$ & $0.96$ \\
      $\La^\ro{scale}_{c_2,\ro{JLQCD}}$   &\T $1.40$ & $1.05$ & $0.94$ \\
      $\La^\ro{scale}_{c_0,\ro{PACS-CS}}$ &\T $1.21$ & $0.93$ & $0.83$ \\
      $\La^\ro{scale}_{c_2,\ro{PACS-CS}}$ &\T $1.21$ & $0.93$ & $0.83$ \\
      $\La^\ro{scale}_{c_0,\ro{CP-PACS}}$ &\T $1.20$ & $0.98$ & $0.88$ \\
      $\La^\ro{scale}_{c_2,\ro{CP-PACS}}$ &\T $1.19$ & $0.97$ & $0.87$ \\
      \hline
    \end{tabular}
  \end{center}
\vspace{-6pt}
  \caption{\footnotesize{Values of the central $\La$ value in GeV, taken from the $\chi^2_{dof}$ analysis for $c_0$ and $c_2$, based on JLQCD, PACS-CS and CP-PACS data.}}
  \label{table:scales}
\end{table}

In calculating the systematic 
uncertainty in the observables $c_0$, $c_2$ and the nucleon mass at the 
physical point 
  due to the intrinsic scale
 at order $\ca{O}(m_\pi^4\,\ro{log}\,m_\pi)$, 
two methods are provided. Firstly, the upper and lower bounds from the 
$\chi^2_{dof}$ analysis at order $\ca{O}(m_\pi^3)$ will be used to constrain 
$\La$, 
and taken to be
 an accurate estimate of the systematic uncertainty
 in the contributions of higher 
order terms. Secondly, variation of the observables  
across the characteristic range of scale values, 
$\La_\ro{lower}<\La<\infty$ will be used, where $\La_\ro{lower}$ takes the 
value of $0.6$, $0.4$ and $0.3$ GeV for the dipole, double dipole and triple
 dipole regulator, respectively.
The results from both of these methods 
are displayed in
Table \ref{table:lerr}.

The final results for the calculation of the renormalized constants $c_0$,
 $c_2$ and the nucleon mass  extrapolated to the physical point 
($m_{\pi,\ro{phys}} = 140$ MeV) are summarized in Table \ref{table:cs}. 
 The lightest four data points from
 each of JLQCD, PACS-CS and CP-PACS lattice QCD data are used.
 The nucleon mass is calculated at the scale determined by 
  the data.
\begin{table*}[tp]
  \newcommand\T{\rule{0pt}{2.8ex}}
  \newcommand\B{\rule[-1.4ex]{0pt}{0pt}}
  \begin{center}
    \begin{tabular}{llllllll}
      \hline
      \hline
      \T\B            \quad  &\multicolumn{6}{c}{regulator form} \\
       sys. err. \quad  & \multicolumn{2}{c}{dipole} $\,\,\,$& \multicolumn{2}{c}{double} $\,\,\,$& \multicolumn{2}{c}{triple}   \\
      \hline

      $\de^\La\! c_0^\ro{JLQCD}$   &\T $0.001$, &$0.009$\quad & $0.001$, &$0.013$ & $0.001$, &$0.016$ \\
      $\de^\La\! c_0^\ro{PACS-CS}$ &\T $0.005$, &$0.006$ & $0.005$, &$0.010$ & $0.006$, &$0.012$ \\
      $\de^\La\! c_0^\ro{CP-PACS}$ &\T $0.002$, &$0.024$ & $0.002$, &$0.037$ & $0.002$, &$0.045$ \\
      $\de^\La\! c_2^\ro{JLQCD}$   &\T $0.02$, &$0.31$ & $0.03$, &$0.38$ & $0.01$, &$0.48$ \\
      $\de^\La\! c_2^\ro{PACS-CS}$ &\T $0.18$, &$0.25$ & $0.16$, &$0.33$ & $0.14$, &$0.43$ \\
      $\de^\La\! c_2^\ro{CP-PACS}$ &\T $0.02$, &$0.40$ & $0.02$, &$0.58$ & $0.02$, &$0.73$  \\
      $\de^\La\! M_{N,\ro{phys}}^\ro{JLQCD}$  &\T $0.0004$, &$0.0051$ & $0.0003$, &$0.0073$ & $0.0003$, &$0.0090$  \\
      $\de^\La\! M_{N,\ro{phys}}^\ro{PACS-CS}$ &\T $0.0022$, &$0.0030$ & $0.0025$, &$0.0046$ & $0.0025$, &$0.0058$ \\
      $\de^\La\! M_{N,\ro{phys}}^\ro{CP-PACS}$ &\T $0.0012$, &$0.0175$ & $0.0013$, &$0.0270$ & $0.0014$, &$0.0326$   \\
      \hline
    \end{tabular}
  \end{center}
\vspace{-6pt}
  \caption{\footnotesize{Results at $\ca{O}(m_\pi^4\,\ro{log}\,m_\pi)$ for the systematic error due to the intrinsic scale, calculated using two methods,
 for the 
 values of $c_0$ (GeV), $c_2$ (GeV$^{-1}$) and the nucleon mass $M_N$ (GeV) extrapolated to the physical point ($m_{\pi,\ro{phys}} = 140$ MeV). The first number in each column is the systematic error due to the intrinsic scale using the upper and lower bound from the $\chi^2_{dof}$ analysis at order $\ca{O}(m_\pi^3)$. The second number is the systematic error due to the instrinsic scale across the whole range of $\La$ values from the lowest reasonable value ($\La = \La_\ro{lower}$) obtained from the pseudodata analysis, to the asymptotic value ($\La = \infty$).}}
  \label{table:lerr}
\end{table*}
\begin{table*}[tp]
  \newcommand\T{\rule{0pt}{2.8ex}}
  \newcommand\B{\rule[-1.4ex]{0pt}{0pt}}
  \begin{center}
    \begin{tabular}{llllll}
      \hline
      \hline
      \T\B            \quad  &\multicolumn{3}{c}{regulator form} \\
       parameter \quad  & dipole $\,\,$& double $\,\,$& triple $\,\,$& WM(1) $\,\,$& WM(2) \\
      \hline

      $c_0^\ro{JLQCD}$               &\T $0.873(18)(16)$ & $0.875(17)(16)$ & $0.891(17)(16)$ & $0.880(29)$ & $0.879(32)$ \\
      $c_0^\ro{PACS-CS}$             &\T $0.900(51)(15)$ & $0.899(51)(14)$ & $0.898(51)(14)$ & $0.899(53)$ & $0.899(55)$ \\
      $c_0^\ro{CP-PACS}$             &\T $0.924(3)(8)$ & $0.914(3)(7)$ & $0.918(3)(7)$ & $0.918(13)$ & $0.920(37)$\\
      $c_2^\ro{JLQCD}$               &\T $3.09(9)(11)$ & $3.18(9)(12)$ & $3.20(9)(11)$ & $3.16(18)$ & $3.14(43)$\\
      $c_2^\ro{PACS-CS}$             &\T $3.06(32)(15)$ & $3.15(31)(14)$ & $3.17(31)(14)$ & $3.13(39)$ & $3.12(49)$ \\
      $c_2^\ro{CP-PACS}$             &\T $2.54(5)(4)$ & $2.70(5)(2)$ & $2.71(5)(3)$ & $2.66(18)$ & $2.61(60)$ \\
      $M_{N,\ro{phys}}^\ro{JLQCD}$   &\T $1.02(2)(9)$ & $1.02(2)(9)$ & $1.02(2)(9)$ & $1.02(9)$ & $1.02(9)$\\
      $M_{N,\ro{phys}}^\ro{PACS-CS}$ &\T $0.967(45)(43)$ & $0.966(45)(43)$ & $0.966(45)(43)$ & $0.966(62)$ & $0.966(62)$ \\
      $M_{N,\ro{phys}}^\ro{CP-PACS}$ &\T $0.982(2)(40)$ & $0.975(2)(43)$ & $0.978(2)(42)$ & $0.979(43)$ & $0.979(50)$\\
      \hline
    \end{tabular}
  \end{center}
\vspace{-6pt}
  \caption{\footnotesize{Results at $\ca{O}(m_\pi^4\,\ro{log}\,m_\pi)$ for the values of $c_0$ (GeV), $c_2$ (GeV$^{-1}$) and the nucleon mass $M_N$ (GeV) extrapolated to the physical point ($m_{\pi,\ro{phys}} = 140$ MeV). WM is the weighted mean of each row.
The nucleon mass is calculated at the optimal scale $\La^\ro{scale}$, which
is the average of $\La^\ro{scale}_{c_0}$ and $\La^\ro{scale}_{c_2}$ for each
 data set. The extrapolations are performed at box sizes relevant to each data set: $L_{\ro{extrap}}^\ro{JLQCD} = 1.9$ fm, $L_{\ro{extrap}}^\ro{PACS-CS} = 2.9$ fm and $L_{\ro{extrap}}^\ro{CP-PACS} = 2.8$ fm.
 The errors are quoted as the estimate of the statistical error first (based on random bootstrap configurations), and the systematic error obtained from the number of $m_\pi^2$ values used second.
Two seperate weighted means are calculated for each row. WM(1) incorporates the systematic error in the intrinsic scale using the upper and lower bound from the $\chi^2_{dof}$ analysis at order $\ca{O}(m_\pi^3)$. The WM(2) incorporates the systematic error due to the intrinsic scale across the whole range of $\La$ values from the lowest reasonable value ($\La = \La_\ro{lower}$) obtained from the pseudodata analysis, to the asymptotic value ($\La = \infty$).
The weighted means also include an estimate of the systematic error in the choice of regulator. All errors are added in quadrature. Note that any order $\ca{O}(a)$ errors have not been incorporated into the total error analysis.}} 
  \label{table:cs}
\end{table*}

%
\section{Conclusion}
\label{sect:conc}
In conclusion, it has been demonstrated that
chiral effective field theory is an important tool
for investigating the chiral properties of hadrons, and for
extrapolating lattice QCD results. 
Because the chiral expansion is only convergent within a
 PCR, a renormalization scheme such as 
finite-range regularization should be used for current lattice QCD results,
 and into the foreseeable future.
Renormalization scheme dependence occurs
when lattice QCD data extending outside the PCR are used in 
the extrapolation. This provides a new quantitative test for determining 
 when lattice QCD data lie within the PCR.
 As most lattice data extend beyond the PCR, a formalism was developed 
to determine if
there is an optimal regularization scale $\La^\ro{scale}$
 in the finite-range regulator,
 and to calculate it if it exists.
 It was concluded that such an optimal scale can be obtained from
 the data itself by analyzing the renormalization flow curves of the 
low energy coefficients in the chiral expansion.
 The optimal scale is selected by the value for which the renormalized 
constants are independent of the upper bound of the fit domain.
 This also means that the renormalized constants are not to be identified 
with their asymptotic values at large $\La$.

 It was revealed that a preferred regularization scheme 
exists only for data sets
extending outside the PCR. Such a preferred 
regularization scheme 
 is associated with an \emph{intrinsic scale}
 for the size of the pion dressings of the nucleon.
 By working to sufficiently high chiral order, it was discovered that the 
scale dependence was weakened. Nevertheless, the residual scale dependence 
persists as a significant component of the systematic uncertainty.
 For efficient propagation of this uncertainty, an interesting future direction
 would be to consider marginalization over the scale 
dependence \cite{Schindler:2008fh}.
 The described procedure was
 used to calculate the nucleon mass at the physical point,
 the low energy coefficients $c_0$ and $c_2$ and their associated statistical 
 and systematic errors.  Several different
functional forms of regulator were considered, and lattice QCD data from
 JLQCD, PACS-CS 
 and CP-PACS were used. 
 An optimal cutoff scale $\La^\ro{scale}$
for each set of lattice QCD data was obtained, and 
the systematic error in the choice of 
renormalization scheme was calculated. 

In summary, the existence of a well defined intrinsic 
scale has been discovered.  
It has also been illustrated how its value can be determined from lattice
 QCD results.

\bibliographystyle{apsrev} \bibliography{references,refs2}

\begin{thebibliography}{29}
\expandafter\ifx\csname natexlab\endcsname\relax\def\natexlab#1{#1}\fi
\expandafter\ifx\csname bibnamefont\endcsname\relax
  \def\bibnamefont#1{#1}\fi
\expandafter\ifx\csname bibfnamefont\endcsname\relax
  \def\bibfnamefont#1{#1}\fi
\expandafter\ifx\csname citenamefont\endcsname\relax
  \def\citenamefont#1{#1}\fi
\expandafter\ifx\csname url\endcsname\relax
  \def\url#1{\texttt{#1}}\fi
\expandafter\ifx\csname urlprefix\endcsname\relax\def\urlprefix{URL }\fi
\providecommand{\bibinfo}[2]{#2}
\providecommand{\eprint}[2][]{\url{#2}}

\bibitem[{\citenamefont{McGovern and Birse}(1999)}]{McGovern:1998tm}
\bibinfo{author}{\bibfnamefont{J.~A.} \bibnamefont{McGovern}} \bibnamefont{and}
  \bibinfo{author}{\bibfnamefont{M.~C.} \bibnamefont{Birse}},
  \bibinfo{journal}{Phys. Lett.} \textbf{\bibinfo{volume}{B446}},
  \bibinfo{pages}{300} (\bibinfo{year}{1999}), \eprint{hep-ph/9807384}.

\bibitem[{\citenamefont{McGovern and Birse}(2006)}]{McGovern:2006fm}
\bibinfo{author}{\bibfnamefont{J.~A.} \bibnamefont{McGovern}} \bibnamefont{and}
  \bibinfo{author}{\bibfnamefont{M.~C.} \bibnamefont{Birse}},
  \bibinfo{journal}{Phys. Rev.} \textbf{\bibinfo{volume}{D74}},
  \bibinfo{pages}{097501} (\bibinfo{year}{2006}), \eprint{hep-lat/0608002}.

\bibitem[{\citenamefont{Schindler et~al.}(2007)\citenamefont{Schindler,
  Djukanovic, Gegelia, and Scherer}}]{Schindler:2006ha}
\bibinfo{author}{\bibfnamefont{M.~R.} \bibnamefont{Schindler}},
  \bibinfo{author}{\bibfnamefont{D.}~\bibnamefont{Djukanovic}},
  \bibinfo{author}{\bibfnamefont{J.}~\bibnamefont{Gegelia}}, \bibnamefont{and}
  \bibinfo{author}{\bibfnamefont{S.}~\bibnamefont{Scherer}},
  \bibinfo{journal}{Phys. Lett.} \textbf{\bibinfo{volume}{B649}},
  \bibinfo{pages}{390} (\bibinfo{year}{2007}), \eprint{hep-ph/0612164}.

\bibitem[{\citenamefont{Beane}(2004{\natexlab{a}})}]{Beane:2004ks}
\bibinfo{author}{\bibfnamefont{S.~R.} \bibnamefont{Beane}},
  \bibinfo{journal}{Nucl. Phys.} \textbf{\bibinfo{volume}{B695}},
  \bibinfo{pages}{192} (\bibinfo{year}{2004}{\natexlab{a}}),
  \eprint{hep-lat/0403030}.

\bibitem[{\citenamefont{Leinweber
  et~al.}(2005{\natexlab{a}})\citenamefont{Leinweber, Thomas, and
  Young}}]{Leinweber:2005xz}
\bibinfo{author}{\bibfnamefont{D.~B.} \bibnamefont{Leinweber}},
  \bibinfo{author}{\bibfnamefont{A.~W.} \bibnamefont{Thomas}},
  \bibnamefont{and} \bibinfo{author}{\bibfnamefont{R.~D.} \bibnamefont{Young}},
  \bibinfo{journal}{Nucl. Phys.} \textbf{\bibinfo{volume}{A755}},
  \bibinfo{pages}{59} (\bibinfo{year}{2005}{\natexlab{a}}),
  \eprint{hep-lat/0501028}.

\bibitem[{\citenamefont{Gell-Mann et~al.}(1968)\citenamefont{Gell-Mann, Oakes,
  and Renner}}]{GellMann:1968rz}
\bibinfo{author}{\bibfnamefont{M.}~\bibnamefont{Gell-Mann}},
  \bibinfo{author}{\bibfnamefont{R.~J.} \bibnamefont{Oakes}}, \bibnamefont{and}
  \bibinfo{author}{\bibfnamefont{B.}~\bibnamefont{Renner}},
  \bibinfo{journal}{Phys. Rev.} \textbf{\bibinfo{volume}{175}},
  \bibinfo{pages}{2195} (\bibinfo{year}{1968}).

\bibitem[{\citenamefont{Leinweber et~al.}(2000)\citenamefont{Leinweber, Thomas,
  and Wright}}]{Leinweber:2000sa}
\bibinfo{author}{\bibfnamefont{D.~B.} \bibnamefont{Leinweber}},
  \bibinfo{author}{\bibfnamefont{A.~W.} \bibnamefont{Thomas}},
  \bibnamefont{and} \bibinfo{author}{\bibfnamefont{S.~V.}
  \bibnamefont{Wright}}, \bibinfo{journal}{Phys. Lett.}
  \textbf{\bibinfo{volume}{B482}}, \bibinfo{pages}{109} (\bibinfo{year}{2000}),
  \eprint{hep-lat/0001007}.

\bibitem[{\citenamefont{Wright et~al.}(2000)\citenamefont{Wright, Leinweber,
  and Thomas}}]{Wright:2000gg}
\bibinfo{author}{\bibfnamefont{S.~V.} \bibnamefont{Wright}},
  \bibinfo{author}{\bibfnamefont{D.~B.} \bibnamefont{Leinweber}},
  \bibnamefont{and} \bibinfo{author}{\bibfnamefont{A.~W.}
  \bibnamefont{Thomas}}, \bibinfo{journal}{Nucl. Phys.}
  \textbf{\bibinfo{volume}{A680}}, \bibinfo{pages}{137} (\bibinfo{year}{2000}),
  \eprint{nucl-th/0005003}.

\bibitem[{\citenamefont{Holl et~al.}(2006)\citenamefont{Holl, Maris, Roberts,
  and Wright}}]{Holl:2005st}
\bibinfo{author}{\bibfnamefont{A.}~\bibnamefont{Holl}},
  \bibinfo{author}{\bibfnamefont{P.}~\bibnamefont{Maris}},
  \bibinfo{author}{\bibfnamefont{C.~D.} \bibnamefont{Roberts}},
  \bibnamefont{and} \bibinfo{author}{\bibfnamefont{S.~V.}
  \bibnamefont{Wright}}, \bibinfo{journal}{Nucl. Phys. Proc. Suppl.}
  \textbf{\bibinfo{volume}{161}}, \bibinfo{pages}{87} (\bibinfo{year}{2006}),
  \eprint{nucl-th/0512048}.

\bibitem[{\citenamefont{Donoghue et~al.}(1999)\citenamefont{Donoghue, Holstein,
  and Borasoy}}]{Donoghue:1998bs}
\bibinfo{author}{\bibfnamefont{J.~F.} \bibnamefont{Donoghue}},
  \bibinfo{author}{\bibfnamefont{B.~R.} \bibnamefont{Holstein}},
  \bibnamefont{and} \bibinfo{author}{\bibfnamefont{B.}~\bibnamefont{Borasoy}},
  \bibinfo{journal}{Phys. Rev.} \textbf{\bibinfo{volume}{D59}},
  \bibinfo{pages}{036002} (\bibinfo{year}{1999}), \eprint{hep-ph/9804281}.

\bibitem[{\citenamefont{Young et~al.}(2002)\citenamefont{Young, Leinweber,
  Thomas, and Wright}}]{Young:2002cj}
\bibinfo{author}{\bibfnamefont{R.~D.} \bibnamefont{Young}},
  \bibinfo{author}{\bibfnamefont{D.~B.} \bibnamefont{Leinweber}},
  \bibinfo{author}{\bibfnamefont{A.~W.} \bibnamefont{Thomas}},
  \bibnamefont{and} \bibinfo{author}{\bibfnamefont{S.~V.}
  \bibnamefont{Wright}}, \bibinfo{journal}{Phys. Rev.}
  \textbf{\bibinfo{volume}{D66}}, \bibinfo{pages}{094507}
  (\bibinfo{year}{2002}), \eprint{hep-lat/0205017}.

\bibitem[{\citenamefont{Young et~al.}(2003)\citenamefont{Young, Leinweber, and
  Thomas}}]{Young:2002ib}
\bibinfo{author}{\bibfnamefont{R.~D.} \bibnamefont{Young}},
  \bibinfo{author}{\bibfnamefont{D.~B.} \bibnamefont{Leinweber}},
  \bibnamefont{and} \bibinfo{author}{\bibfnamefont{A.~W.}
  \bibnamefont{Thomas}}, \bibinfo{journal}{Prog. Part. Nucl. Phys.}
  \textbf{\bibinfo{volume}{50}}, \bibinfo{pages}{399} (\bibinfo{year}{2003}),
  \eprint{hep-lat/0212031}.

\bibitem[{\citenamefont{Borasoy et~al.}(2002)\citenamefont{Borasoy, Holstein,
  Lewis, and Ouimet}}]{Borasoy:2002jv}
\bibinfo{author}{\bibfnamefont{B.}~\bibnamefont{Borasoy}},
  \bibinfo{author}{\bibfnamefont{B.~R.} \bibnamefont{Holstein}},
  \bibinfo{author}{\bibfnamefont{R.}~\bibnamefont{Lewis}}, \bibnamefont{and}
  \bibinfo{author}{\bibfnamefont{P.~P.~A.} \bibnamefont{Ouimet}},
  \bibinfo{journal}{Phys. Rev.} \textbf{\bibinfo{volume}{D66}},
  \bibinfo{pages}{094020} (\bibinfo{year}{2002}), \eprint{hep-ph/0210092}.

\bibitem[{\citenamefont{Leinweber et~al.}(2004)\citenamefont{Leinweber, Thomas,
  and Young}}]{Leinweber:2003dg}
\bibinfo{author}{\bibfnamefont{D.~B.} \bibnamefont{Leinweber}},
  \bibinfo{author}{\bibfnamefont{A.~W.} \bibnamefont{Thomas}},
  \bibnamefont{and} \bibinfo{author}{\bibfnamefont{R.~D.} \bibnamefont{Young}},
  \bibinfo{journal}{Phys. Rev. Lett.} \textbf{\bibinfo{volume}{92}},
  \bibinfo{pages}{242002} (\bibinfo{year}{2004}), \eprint{hep-lat/0302020}.

\bibitem[{\citenamefont{Bernard et~al.}(2004)\citenamefont{Bernard, Hemmert,
  and Meissner}}]{Bernard:2003rp}
\bibinfo{author}{\bibfnamefont{V.}~\bibnamefont{Bernard}},
  \bibinfo{author}{\bibfnamefont{T.~R.} \bibnamefont{Hemmert}},
  \bibnamefont{and} \bibinfo{author}{\bibfnamefont{U.-G.}
  \bibnamefont{Meissner}}, \bibinfo{journal}{Nucl. Phys.}
  \textbf{\bibinfo{volume}{A732}}, \bibinfo{pages}{149} (\bibinfo{year}{2004}),
  \eprint{hep-ph/0307115}.

\bibitem[{\citenamefont{Djukanovic et~al.}(2005)\citenamefont{Djukanovic,
  Schindler, Gegelia, and Scherer}}]{Djukanovic:2004px}
\bibinfo{author}{\bibfnamefont{D.}~\bibnamefont{Djukanovic}},
  \bibinfo{author}{\bibfnamefont{M.~R.} \bibnamefont{Schindler}},
  \bibinfo{author}{\bibfnamefont{J.}~\bibnamefont{Gegelia}}, \bibnamefont{and}
  \bibinfo{author}{\bibfnamefont{S.}~\bibnamefont{Scherer}},
  \bibinfo{journal}{Phys. Rev.} \textbf{\bibinfo{volume}{D72}},
  \bibinfo{pages}{045002} (\bibinfo{year}{2005}), \eprint{hep-ph/0407170}.

\bibitem[{\citenamefont{Young et~al.}(2005)\citenamefont{Young, Leinweber, and
  Thomas}}]{Young:2004tb}
\bibinfo{author}{\bibfnamefont{R.~D.} \bibnamefont{Young}},
  \bibinfo{author}{\bibfnamefont{D.~B.} \bibnamefont{Leinweber}},
  \bibnamefont{and} \bibinfo{author}{\bibfnamefont{A.~W.}
  \bibnamefont{Thomas}}, \bibinfo{journal}{Phys. Rev.}
  \textbf{\bibinfo{volume}{D71}}, \bibinfo{pages}{014001}
  (\bibinfo{year}{2005}), \eprint{hep-lat/0406001}.

\bibitem[{\citenamefont{Leinweber
  et~al.}(2005{\natexlab{b}})}]{Leinweber:2004tc}
\bibinfo{author}{\bibfnamefont{D.~B.} \bibnamefont{Leinweber}}
  \bibnamefont{et~al.}, \bibinfo{journal}{Phys. Rev. Lett.}
  \textbf{\bibinfo{volume}{94}}, \bibinfo{pages}{212001}
  (\bibinfo{year}{2005}{\natexlab{b}}), \eprint{hep-lat/0406002}.

\bibitem[{\citenamefont{Jenkins}(1992)}]{Jenkins:1991ts}
\bibinfo{author}{\bibfnamefont{E.~E.} \bibnamefont{Jenkins}},
  \bibinfo{journal}{Nucl. Phys.} \textbf{\bibinfo{volume}{B368}},
  \bibinfo{pages}{190} (\bibinfo{year}{1992}).

\bibitem[{\citenamefont{Lebed}(1995)}]{Lebed:1994ga}
\bibinfo{author}{\bibfnamefont{R.~F.} \bibnamefont{Lebed}},
  \bibinfo{journal}{Phys. Rev.} \textbf{\bibinfo{volume}{D51}},
  \bibinfo{pages}{5039} (\bibinfo{year}{1995}), \eprint{hep-ph/9411204}.

\bibitem[{\citenamefont{Beane}(2004{\natexlab{b}})}]{Beane:2004tw}
\bibinfo{author}{\bibfnamefont{S.~R.} \bibnamefont{Beane}},
  \bibinfo{journal}{Phys. Rev.} \textbf{\bibinfo{volume}{D70}},
  \bibinfo{pages}{034507} (\bibinfo{year}{2004}{\natexlab{b}}),
  \eprint{hep-lat/0403015}.

\bibitem[{\citenamefont{Armour et~al.}(2006)\citenamefont{Armour, Allton,
  Leinweber, Thomas, and Young}}]{Armour:2005mk}
\bibinfo{author}{\bibfnamefont{W.}~\bibnamefont{Armour}},
  \bibinfo{author}{\bibfnamefont{C.~R.} \bibnamefont{Allton}},
  \bibinfo{author}{\bibfnamefont{D.~B.} \bibnamefont{Leinweber}},
  \bibinfo{author}{\bibfnamefont{A.~W.} \bibnamefont{Thomas}},
  \bibnamefont{and} \bibinfo{author}{\bibfnamefont{R.~D.} \bibnamefont{Young}},
  \bibinfo{journal}{J. Phys.} \textbf{\bibinfo{volume}{G32}},
  \bibinfo{pages}{971} (\bibinfo{year}{2006}), \eprint{hep-lat/0510078}.

\bibitem[{\citenamefont{Aoki et~al.}(2008)}]{Aoki:2008sm}
\bibinfo{author}{\bibfnamefont{S.}~\bibnamefont{Aoki}} \bibnamefont{et~al.}
  (\bibinfo{collaboration}{PACS-CS}) (\bibinfo{year}{2008}),
  \eprint{0807.1661}.

\bibitem[{\citenamefont{Ali~Khan et~al.}(2004)}]{AliKhan:2003cu}
\bibinfo{author}{\bibfnamefont{A.}~\bibnamefont{Ali~Khan}} \bibnamefont{et~al.}
  (\bibinfo{collaboration}{QCDSF-UKQCD}), \bibinfo{journal}{Nucl. Phys.}
  \textbf{\bibinfo{volume}{B689}}, \bibinfo{pages}{175} (\bibinfo{year}{2004}),
  \eprint{hep-lat/0312030}.

\bibitem[{\citenamefont{Ohki et~al.}(2008)}]{Ohki:2008ff}
\bibinfo{author}{\bibfnamefont{H.}~\bibnamefont{Ohki}} \bibnamefont{et~al.},
  \bibinfo{journal}{Phys. Rev.} \textbf{\bibinfo{volume}{D78}},
  \bibinfo{pages}{054502} (\bibinfo{year}{2008}), \eprint{0806.4744}.

\bibitem[{\citenamefont{Ali~Khan et~al.}(2002)}]{AliKhan:2001tx}
\bibinfo{author}{\bibfnamefont{A.}~\bibnamefont{Ali~Khan}} \bibnamefont{et~al.}
  (\bibinfo{collaboration}{CP-PACS}), \bibinfo{journal}{Phys. Rev.}
  \textbf{\bibinfo{volume}{D65}}, \bibinfo{pages}{054505}
  (\bibinfo{year}{2002}), \eprint{hep-lat/0105015}.

\bibitem[{\citenamefont{Young et~al.}(2009)\citenamefont{Young, Hall, and
  Leinweber}}]{Young:2009ub}
\bibinfo{author}{\bibfnamefont{R.~D.} \bibnamefont{Young}},
  \bibinfo{author}{\bibfnamefont{J.~M.~M.} \bibnamefont{Hall}},
  \bibnamefont{and} \bibinfo{author}{\bibfnamefont{D.~B.}
  \bibnamefont{Leinweber}} (\bibinfo{year}{2009}), \eprint{0907.0408}.

\bibitem[{\citenamefont{Young and Thomas}(2009)}]{Young:2009zb}
\bibinfo{author}{\bibfnamefont{R.~D.} \bibnamefont{Young}} \bibnamefont{and}
  \bibinfo{author}{\bibfnamefont{A.~W.} \bibnamefont{Thomas}}
  (\bibinfo{year}{2009}), \eprint{0901.3310}.

\bibitem[{\citenamefont{Schindler and Phillips}(2009)}]{Schindler:2008fh}
\bibinfo{author}{\bibfnamefont{M.~R.} \bibnamefont{Schindler}}
  \bibnamefont{and} \bibinfo{author}{\bibfnamefont{D.~R.}
  \bibnamefont{Phillips}}, \bibinfo{journal}{Annals Phys.}
  \textbf{\bibinfo{volume}{324}}, \bibinfo{pages}{682} (\bibinfo{year}{2009}),
  \eprint{0808.3643}.

\end{thebibliography}


\end{document}